%% file: main.tex
\documentclass[a4paper,11pt]{article}
\pdfoutput=1
\usepackage{jinstpub}

\usepackage{amsmath}
\usepackage{bbm}
\usepackage{graphicx}
\usepackage{subcaption}
\usepackage{xcolor}
\newcommand{\eps}{\varepsilon}

\newcommand{\parnassus}{\textsc{Parnassus}}
\newcommand{\pandora}{\textsc{PandoraPFA}}
\newcommand{\hitpf}{\textsc{HitPF}}
\newcommand{\delphes}{\textsc{Delphes}}
\newcommand{\gfour}{\textsc{Geant4}}
\newcommand{\ddsim}{\textsc{ddsim}}

\newcommand{\keyfourhep}{\textsc{Key4hep}}
\newcommand{\fastjet}{\textsc{FastJet}}
\newcommand{\pT}{p_\mathrm{T}}
\newcommand{\ptrel}{p_\mathrm{T}^{\mathrm{rel}}}
\newcommand{\HT}{H_\mathrm{T}}
\newcommand{\dzero}{d_0}
\newcommand{\zzero}{z_0}
\newcommand{\Xzero}{X_0}
\newcommand{\bT}{\mathcal{T}}
\newcommand{\bR}{\mathcal{R}}
\newcommand{\bx}{\mathbf{x}}
\newcommand{\bv}{\mathbf{v}}
\newcommand{\bu}{\mathbf{u}}
\newcommand{\bc}{\mathbf{c}}
\newcommand{\bg}{\mathbf{g}}
\newcommand{\dd}{\mathrm{d}}

\title{\boldmath \textsc{Parnassus} for the CLD Detector: A Generative Machine-Learning Surrogate for Detector Simulation and Reconstruction at the FCC-ee}

\author[a,b,1]{Umar Sohail Qureshi\note{Corresponding author. },}
\author[b,c]{Benjamin Nachman,}
\author[b,c]{and Caterina Vernieri}
\affiliation[a]{Department of Physics, Stanford University,
Stanford, CA 94305, USA}
\affiliation[b]{Fundamental Physics Directorate, SLAC National Accelerator Laboratory,
Menlo Park, CA 94025, USA}
\emailAdd{uqureshi@cern.ch}
\affiliation[c]{Department of Particle Physics and Astrophysics, Stanford University,
Stanford, CA 94305, USA}

\abstract{Detector simulation and event reconstruction will be computationally expensive for future $e^+e^-$ collider programs and are currently critical bottlenecks for accurate feasibility studies. 
To address this challenge, we build a {\parnassus} model for the CLD detector concept. 
{\parnassus} is a framework for automatically tuning a surrogate model---in our case, a conditional flow matching neural network---to emulate a full detector simulation and reconstruction.  
We train on 
$e^+e^-\to Z\to q\bar q$ events at
$\sqrt{s}=91.2$~GeV processed through a {\gfour} simulation of the
CLD detector concept and the {\pandora} particle-flow reconstruction and reproduce single-particle kinematics, particle-IDs, and
impact-parameter distributions.  
We also examine jet- and event-level features, inclusively
and split by flavor, and find excellent fidelity, significantly better than the parameterized program {\delphes}.
Furthermore, we train a transformer-based flavor tagger on the reconstructed particle-flow constituents and show that the
surrogate preserves the $b/c/s/q$ discrimination of the full CLD reconstruction.
The {\parnassus} CLD model achieves generation cost of about $1.2$~ms (40~ms) per event on a single GPU (CPU), over three (two)
orders of magnitude faster than full simulation and reconstruction. 
Our model is publicly available for feasibility and design studies.
}
\begin{document}
\maketitle
\flushbottom

\section{Introduction}
\label{sec:intro}

The physics program of a future electron-positron collider such as the
FCC-ee~\cite{FCCee_FSR} rests on high-fidelity simulated samples of a size that is increasingly
difficult to produce. A single event must be propagated through a
{\gfour}~\cite{GEANT4:2002zbu} description of the detector and then through a
reconstruction chain whose cost is comparable to, and sometimes exceeds, that of
the simulation itself. At the $Z$-pole, where the FCC-ee anticipates
$\mathcal{O}(10^{12})$ hadronic $Z$ decays, the mismatch between the required and
attainable sample sizes is severe, and it is aggravated by the fact that both the
detector concept and the reconstruction algorithms are still under active
development.

Parameterized fast simulation tools
address the cost problem by replacing the detector with a small set of hand-tuned
resolution and efficiency functions. For example, the de-facto tool for fast FCC-ee simulations,  {\delphes}~\cite{deFavereau:2013fsa}, is fast, modular and transparent. However, the
parameterization is constructed by hand and
by design cannot express correlations that were not anticipated by its author.
\textbf{Par}ticle-flow \textbf{N}eural \textbf{As}sisted \textbf{S}im\textbf{u}lation\textbf{s}
({\parnassus})~\cite{Kobylianskii:2024sup,Dreyer:2024bhs,Dreyer:2025zhp,Elabd:2026qfw}
is a framework for automatically tuning the map from generator-level particles to reconstructed objects.  It has both parametric and deep generative model paths, trading off accuracy and speed/interpretability. %
%
%
%
Prior work has demonstrated and validated this method for
the CMS~\cite{Kobylianskii:2024sup,Dreyer:2024bhs,Elabd:2026poc} and ATLAS~\cite{ATLAS:2026parnassus} detectors at the LHC, the ALEPH detector at LEP~\cite{Lo:2026aleph}, and most recently, the SLD detector at SLAC~\cite{Lo:2026vbs}.

Generative machine learning is increasingly finding applications in studies for future colliders. Recent work has explored diffusion models for reconstructing projections of six-dimensional beam phase space~\cite{Scheinker:2024cDVAE}, mapping beam-loss measurements to two-dimensional phase-space projections~\cite{Scheinker:2025beamLoss}, incorporating physics constraints into six-dimensional beam diagnostics~\cite{Scheinker:2025superresolution}, reconstructing four-dimensional beam phase space from limited two-dimensional observations~\cite{Scheinker:2026PhaseFlow4D}, and simulating beam-induced backgrounds, in particular, incoherent pair creation at future $e^+e^-$ colliders~\cite{chahine2026deeffusioncapturingbeambeamphysics}. Deep generative models have also been developed to accelerate detector simulation by learning the calorimeter response to incident particles. Examples include CaloGAN, which uses generative adversarial networks to generate showers in calorimeters~\cite{Paganini:2018CaloGAN}, CaloFlow, which models shower energy deposits with normalizing flows~\cite{Krause:2021ilc}, and CaloDiffusion, which uses denoising diffusion with learned geometry mappings to accommodate irregular calorimeter segmentation~\cite{Amram:2023CaloDiffusion}. Hardware-aware algorithms have also been developed for fast calorimeter simulation on FPGAs~\cite{May:2026FPGA}. These developments also extend to experimental simulation frameworks, such as AtlFast3, which combines parameterized and machine-learning approaches for fast calorimeter simulation in ATLAS~\cite{ATLAS:2022AtlFast3}. Similarly, FlashSim \cite{Vaselli:2023FlashSim} uses normalizing flows to generate CMS  reconstructed analysis-level objects directly from generator-level information, bypassing the detector simulation and reconstruction chain entirely.

In this work, we apply {\parnassus} to the CLD detector
concept~\cite{Bacchetta:2019fmz} at the FCC-ee. We generate the particle flow constituents and their associated track impact parameters i.e.~the transverse ($\dzero$) and longitudinal ($\zzero$) distances of closest
approach and their significances $\dzero/\sigma_{\dzero}$,
$\zzero/\sigma_{\zzero}$ alongside the particle-flow kinematics. We then assess the fidelity of the surrogate output for a complex downstream analysis by training a transformer-based jet flavor tagger on the generated particle-flow constituents. We demonstrate that the resulting tagger achieves comparable performance to full simulation and reconstruction for heavy-flavor tagging.

Furthermore, since the surrogate conditions only on generator-level input and targets
whatever reconstruction it is trained against, the same framework applies
unchanged to alternative reconstruction algorithms. We describe the extension to
the hit-level machine-learned particle flow {\hitpf}~\cite{HitPF:2026} in
Sec.~\ref{sec:hitpf}. Throughout, we benchmark against {\delphes} with the CLD
card, the de-facto parameterized fast simulation recommended by the FCC-ee collaboration for users today. Our results demonstrate that {\parnassus} reproduces full-simulation levels of fidelity, thus significantly outperforming {\delphes}, while being more than three (two) orders of magnitude faster, running on a single GPU (CPU), than the full simulation and reconstruction chain. 

The rest of this paper is organized as follows. Section~\ref{sec:cld} describes the CLD
detector concept. Section~\ref{sec:chain} describes the simulation and reconstruction
chain, and Section~\ref{sec:dataset} the event representation. Section~\ref{sec:method}
develops the machine learning method in detail. Section~\ref{sec:results} presents the results and
the distribution-level analysis, Section~\ref{sec:ftag} the flavor-tagging study,
and Section~\ref{sec:hitpf} the {\hitpf} extension. Finally, we conclude with a short discussion in Section~\ref{sec:discussion}.

\section{The CLD detector}
\label{sec:cld}

\begin{figure}
\centering
\includegraphics[width=0.495\linewidth]{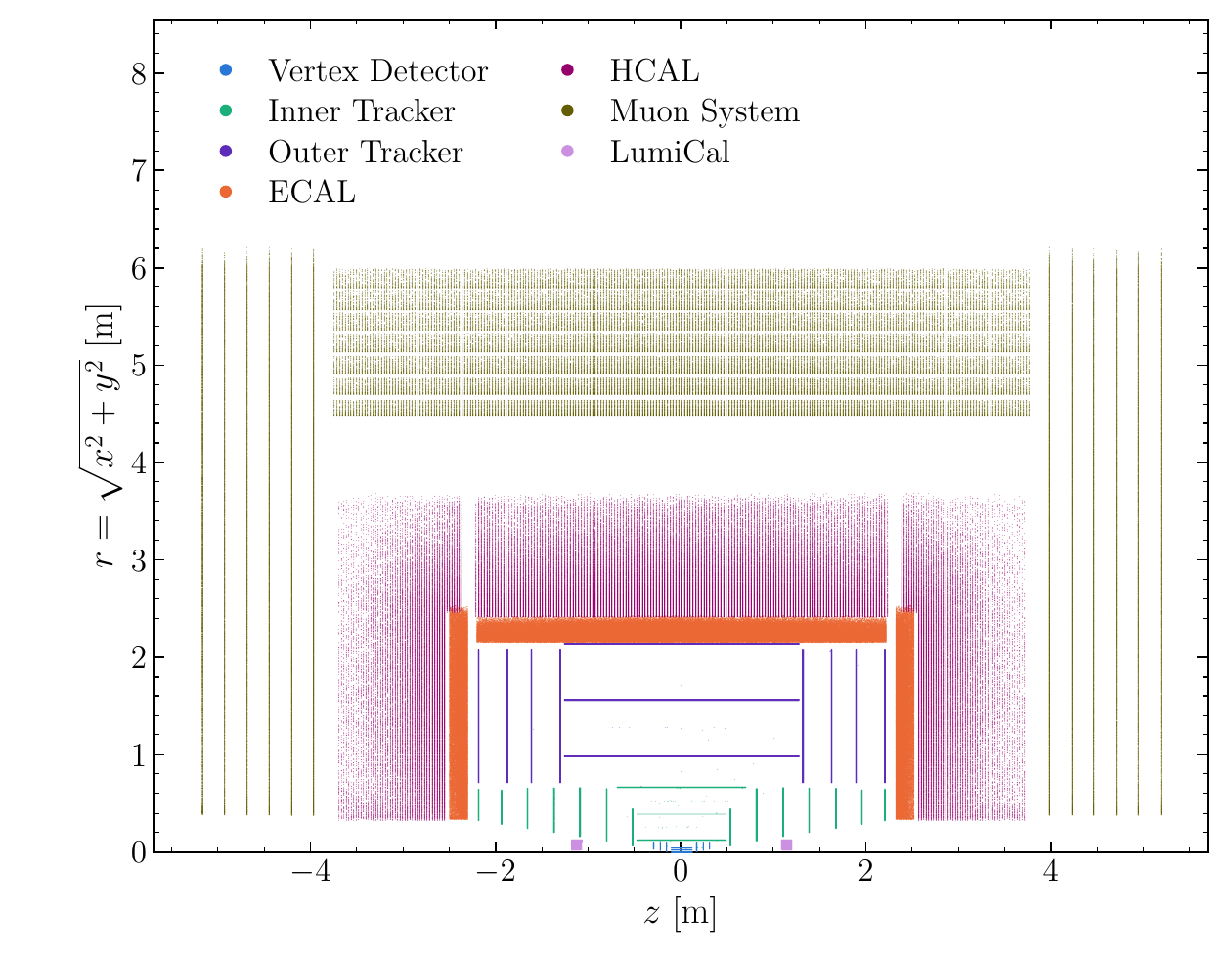}
\includegraphics[width=0.495\linewidth]{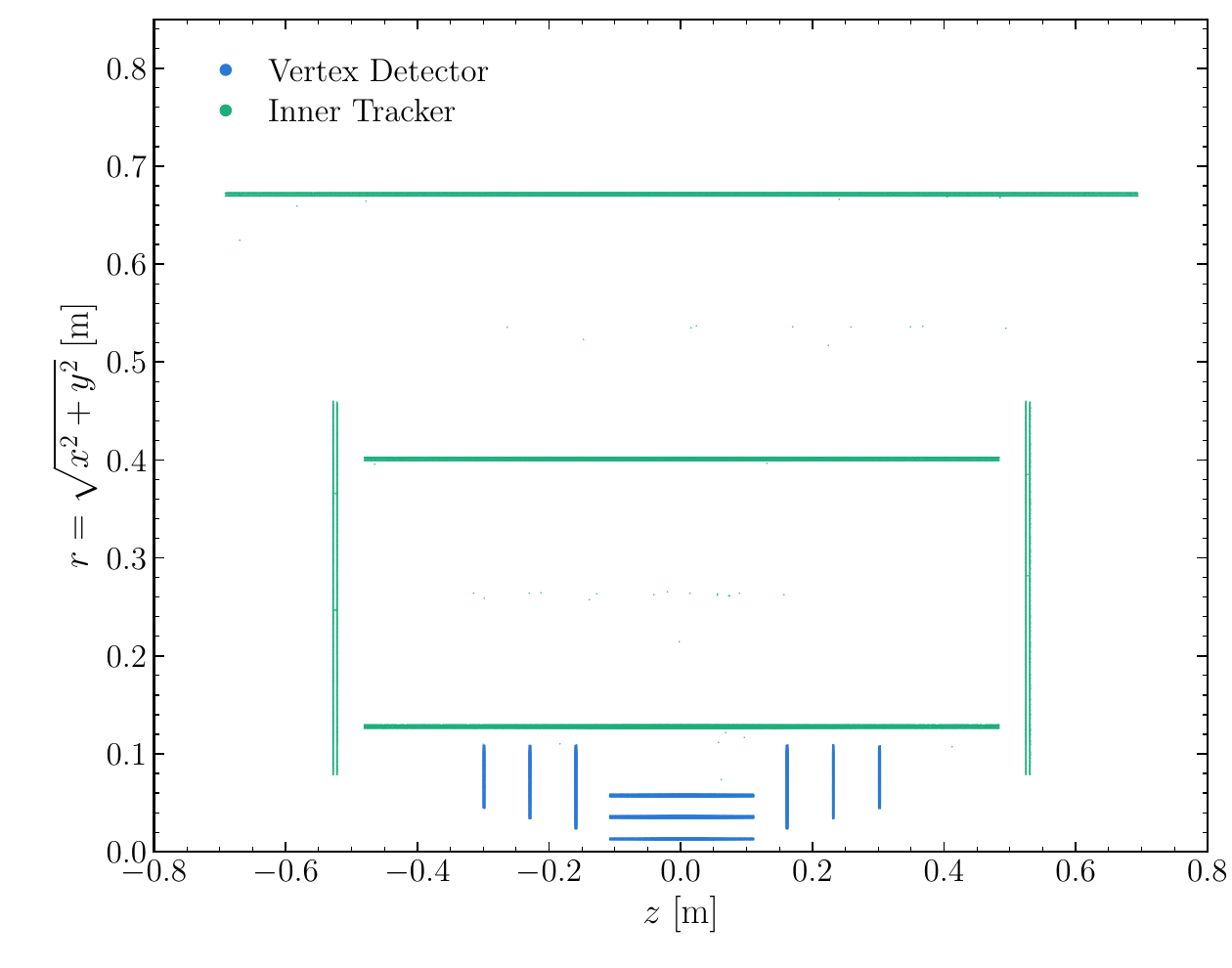}
\caption{Spatial distribution of {\gfour} simulated hits in the \texttt{CLD\_o2\_v07}
geometry, projected onto the $r$-$z$ plane and colored by subdetector,
accumulated over 100k $Z\to q\bar q$ events. Each point is an energy deposit recorded in a
sensitive layer. The left-hand side shows the full detector, including the silicon tracking
layers, the ECAL and HCAL barrels and endcaps, and the instrumented return
yoke. The right-hand side zooms into the vertex and inner-tracker
region.}
\label{fig:cld_rz}
\end{figure}

CLD (CLIC-Like Detector)~\cite{Bacchetta:2019fmz} is a general-purpose detector
concept for the FCC-ee, adapted from the CLIC detector model~\cite{CLICdet}. It
features full-silicon tracking inside highly granular calorimeters and a
2 Tesla superconducting solenoid, all enclosed by an iron return yoke instrumented
for muon identification. The high granularity characteristic of CLD is designed to enable accurate
particle-flow reconstruction \cite{Pata:2021MLPF,CMS:2026MLPF,Pata:2024MLPF,Mokhtar:2025Transfer,HitPF:2026,DiBello:2023HGPflow,Kakati:2025HGPflow,Suehara:2024ILD}, in which the momenta of charged particles are
measured in the tracker and the energies of neutral particles in the
calorimeters; CLD is therefore a natural target for a surrogate that must learn
the particle-flow response. This study uses the \texttt{CLD\_o2\_v07} geometry, which is
the most recent baseline in the {\keyfourhep} software stack at the time of conducting this study; the geometric
parameters quoted below are read directly from its DD4hep~\cite{Frank:2014zya}
compact description, and the design and performance figures follow the CLD note~\cite{Bacchetta:2019fmz}. The layout described below can be visualized directly in the simulated data. Fig.~\ref{fig:cld_rz} shows the spatial distribution of the {\gfour} hits in the $r$-$z$ plane for various subdetector systems integrated over 100k $Z\to q\bar{q}$ events.

The vertex detector is a scaled version of the CLIC design, consisting of
three cylindrical double layers in the barrel and three double disks in each
forward region. The barrel double layers are located at radii of approximately
$13$, $35$, and $57$~mm, the innermost sensitive surface sitting at
$12.5$~mm from the beam line, and extend to a half-length of $110$~mm. The
sensors are $25\times25~\mu\mathrm{m}^2$ silicon pixels, $50~\mu$m thick,
providing a single-point resolution of $\approx\!3~\mu$m; including support,
services, and cabling, the material budget is $0.6\%\,\Xzero$ per double layer. The fine pixels and small inner radius give the excellent impact-parameter resolution required for the flavor tagging of Section~\ref{sec:ftag}.

Surrounding the vertex detector is an all-silicon tracker divided into an inner
and an outer tracker, each with a barrel and forward disks, spanning radii from
$61$~mm to an outer radius of $2145$~mm, which is enlarged from the $1.5$~m of CLIC to
$2.15$~m to recover momentum resolution at the lower FCC-ee magnetic field. The
detection layers are built from silicon modules with strip-like sensors
($30\times30~\mathrm{mm}^2$ cells in the outer tracker, $15\times15~\mathrm{mm}^2$
in the inner disks) providing a single-point resolution of $\approx\!7~\mu$m in
the bending plane, with a per-layer material budget of $\approx\!1\%\,\Xzero$.
Together the vertex and tracker deliver an asymptotic transverse-momentum
resolution of $\sigma(\pT)/\pT^2\sim\!2\times10^{-5}~\mathrm{GeV}^{-1}$ and an
impact-parameter resolution well below $10~\mu$m for high-momentum central
tracks.

The electromagnetic calorimeter (ECAL) is a silicon-tungsten (SiW) sampling
calorimeter of $40$ layers, each a $1.9$~mm tungsten absorber followed by a
$0.5$~mm silicon sensor read out in $5.1\times5.1~\mathrm{mm}^2$ cells, for a
total depth of $\approx\!22\,\Xzero$; the barrel inner radius is $2150$~mm. The
hadronic calorimeter (HCAL) is a steel-scintillator sampling calorimeter of
$44$ layers, each a $19$~mm steel absorber followed by a $3$~mm polystyrene
scintillator tile of $30\times30~\mathrm{mm}^2$ read out with silicon
photomultipliers, for a total depth of $\approx\!5.5\,\lambda_\mathrm{I}$; the
barrel inner radius is $2400$~mm. The high transverse and longitudinal
granularity of both calorimeters is the enabling feature for particle flow. With
software compensation applied in reconstruction, the jet energy resolution is
$4.5$-$5\%$ for central jets at the $Z$ pole~\cite{Bacchetta:2019fmz}. A superconducting solenoid of inner radius $3719$~mm provides a 2 Tesla axial
field. The flux is returned by an iron yoke (barrel inner radius $4479$~mm)
instrumented with detector layers that, together with the tracker, provide muon
identification.

It is worth noting that the CLD geometry has evolved through a sequence of versions in
{\keyfourhep}~\cite{Bacchetta:2019fmz,CLICdet}. The original
\texttt{FCCee\_o2\_v01} model of the CLD note was updated in
\texttt{v02} (a fix to the tracker-endcap support) and \texttt{v04} (an alternate
vertex detector with a smaller radius preserving the barrel angular coverage).
The models were then renamed \texttt{CLD\_o2\_v05}, which adopts the latest
low-impedance beam-pipe design, adjusts the vertex-barrel layers to the new
beam-pipe constraints, and fixes tracker overlaps; \texttt{v06}, with minor
LumiCal changes; and \texttt{v07}, which speeds up loading by grouping tracker
staves into assemblies, adds a non-cylindrical (polycone) tracking volume
excluding the LumiCal, and fixes remaining LumiCal-related overlaps.
\texttt{CLD\_o2\_v07} is the version used throughout this work.

\section{Simulation and reconstruction chain}
\label{sec:chain}
Initial beam–beam and interactions and non-linear QED effects are first simulated with \textsc{GuineaPig++} \cite{4440556}, the modern C++
implementation of the \textsc{GuineaPig} \cite{schulte_1997_b8qnz-qvm60, Schulte:382453} particle-in-cell code to generate the $e^+e^-$ beam spread spectrum. For this, we use the official FCC-ee parameter set corresponding to the 2025 machine design. The generated spectra are subsequently converted to the \textsc{Circe2} format and interfaced with \textsc{Whizard} for convolution.

We generate the $e^+e^-\to Z\to j\bar j$ process at $\sqrt{s}=91.2$~GeV, with $j$ corresponding to four $Z$ decay channels i.e.~light
quarks ($q=u,d$), strange $s$, charm $c$, and bottom $b$. A total of 5M\footnote{While we simulate the 5M total events, only 2.75M are used in this study. The $c\bar{c}$ and $s\bar{s}$ channels are reserved for out-of-distribution testing; the remaining 2.25M events from these channels that would otherwise enter the training and validation sets are excluded from this study.} events are simulated with balanced flavor composition. As will be discussed in later sections, we train the model on $q\bar q$ and $b\bar b$ events and reserve $c\bar c$ and $s\bar s$ for out-of-distribution testing. Event generation including, initial state radiation, is performed using \textsc{Whizard} \cite{Kilian:2007gr, Moretti:2001zz} (v3.1.5) interfaced with \textsc{Pythia} \cite{Sjostrand:2006za} (v6.427) for parton showering and hadronization. Particle-level events are
written in the \textsc{HepMC3} format~\cite{Buckley:2019xhk} and propagated through a {\gfour}
simulation of the \texttt{CLD\_o2\_v07} geometry using {\ddsim} from the
{\keyfourhep} stack~\cite{Ganis:2021vgv,Gaede:2022leb} with the DD4hep geometry
toolkit~\cite{Frank:2014zya}.

The primary reconstruction target for this study is the tried and tested {\pandora} particle-flow
algorithm~\cite{Marshall:2015rfa,Thomson:2009rp} as configured in the standard
\texttt{CLDConfig} chain. {\pandora} associates calorimeter clusters with
extrapolated tracks and outputs a collection of particle-flow objects (PFOs),
each carrying a four-momentum, a charge, and a species hypothesis. For charged
PFOs we additionally read the perigee track parameters $(\dzero,\zzero)$ and their
covariances from the refitted silicon tracks at the point of closest approach to
the interaction point, from which the significances $\dzero/\sigma_{\dzero}$ and
$\zzero/\sigma_{\zzero}$ are formed. It is this augmented \texttt{PandoraPFOs}
collection that the surrogate learns to generate. As a fast-simulation baseline
we process the same generated events through {\delphes}~\cite{deFavereau:2013fsa}
with the CLD detector card, retaining its \texttt{ParticleFlowCandidate}
collection and its per-track $d_0$, $d_Z$ and their uncertainties. 

It is worth mentioning that the FCC-ee \textsc{Delphes} cards are more sophisticated than the ones employed by the LHC and LEP collaborations and leverage the latest advancements in \textsc{Delphes}. In particular, the \texttt{DetectorGeometry} and \texttt{TrackCovariance} modules, in which the the former takes as input, a geometric description of the detector's tracking system and the latter, using the geometry, provides an estimate for the track parameters, and associated covariance matrix, for each charged particle. A detailed description of these modules and their implementation is provided in Ref.~\cite{Bedeschi:2022rnj}.

\section{Dataset and representation}
\label{sec:dataset}

Let $\bT=\{\tau_i\}_{i=1}^{M}$ denote the generator-level (truth) particle set of
an event and $\bR=\{\rho_j\}_{j=1}^{N}$ the reconstructed set. Truth particles are
required to be stable, satisfy $|\eta|<3$, and
exclude neutrinos; reconstructed particles satisfy the same $|\eta|$ requirement.
Each particle is assigned to one of five particle-flow species,
$\kappa\in\{\text{charged\ hadron},\,e,\,\mu,\,\text{neutral\ hadron},\,\gamma\}$, by a
fixed map from its PDG identifier.

Kinematics are represented as $(\ptrel,\eta,\sin\phi,\cos\phi)$, where
$\ptrel=\pT/\HT$ with $\HT=\sum_k p_{\mathrm{T},k}$ removes the overall scale and
$(\sin\phi,\cos\phi)$ removes the $2\pi$ discontinuity. The reconstructed feature
vector augments this with the species one-hot and the four impact observables:
\begin{equation}
\rho_j = \big(\ptrel,\ \eta,\ \sin\phi,\ \cos\phi,\ \mathbbm{1}_\kappa,\
              \dzero,\ \zzero,\ \dzero/\sigma_{\dzero},\ \zzero/\sigma_{\zzero}\big)
         \in \mathbb{R}^{13},
\label{eq:fsfeat}
\end{equation}
where $\mathbbm{1}_\kappa\in\{0,1\}^5$ is the species one-hot and the four impact
features are zero for neutral PFOs, which carry no track.

The signed impact parameters follow heavy-tailed distributions with a
$\sim\!12~\mu$m prompt-resolution core (the median $|\dzero|$ over charged tracks)
coexists with millimeter-scale displaced tails from heavy-flavor decays, giving
a kurtosis of $\mathcal{O}(300)$. A naive standardization leaves a sharp cusp
at zero that the generative model cannot represent effectively. We therefore apply an
inverse-hyperbolic-sine transform:
\begin{equation}
\dzero \;\mapsto\; \operatorname{asinh}\!\big(\dzero/s\big),\qquad
s = 15~\mu\mathrm{m},
\label{eq:asinh}
\end{equation}
whose scale $s$ is matched to the prompt-resolution core. This places the core on
the bend of $\operatorname{asinh}$ rather than in its linear region, spreading the
cusp into a near-Gaussian target and reducing the transformed-space kurtosis from
$\approx\!15$ (at the CMS default $s=0.1$~mm) to $\approx\!6$, while the
logarithmic regime still accommodates the displaced tails. The scale was chosen
by a one-parameter scan minimizing the transformed-space kurtosis subject to
preserving tail dynamic range. Furthermore, in place of the resolutions $\sigma_{\dzero,\zzero}$
we target the significances $\dzero/\sigma_{\dzero}$, which are more
Gaussian in the prompt core (a pull of $\mathcal{O}(1)$) and are the more
tagging-relevant quantity. Moreover, the absolute resolution can be recovered downstream as
$\sigma_{\dzero}=\dzero/(\dzero/\sigma_{\dzero})$.

The truth conditioning vector carries the same relative kinematics together with
the truth charge and species one-hots and the MC truth production vertex:
\begin{equation}
\tau_i = \big(\ptrel,\ \eta,\ \sin\phi,\ \cos\phi,\ \mathbbm{1}_q,\ \mathbbm{1}_\kappa,\
              v_x,\ v_y,\ v_z,\ \dzero^{\mathrm{tr}},\ \zzero^{\mathrm{tr}}\big)
         \in \mathbb{R}^{17},
\label{eq:ctxtfeat}
\end{equation}
where $\mathbbm{1}_q\in\{0,1\}^3$ encodes $q\in\{-1,0,+1\}$, $(v_x,v_y,v_z)$ is
the MC production vertex, and $(\dzero^{\mathrm{tr}},\zzero^{\mathrm{tr}})$ are the
corresponding linear (straight-line) impact parameters. The vertex conditioning
enables the model to discriminate between a prompt track and a $b$-decay
daughter which would otherwise be indistinguishable to a surrogate
conditioned only on kinematics with identical $(\ptrel,\eta,\phi)$. In that case, the displaced tail would be conditionally invisible and regressed toward the core. Conditioning on the truth vertex makes the displacement a deterministic input, and the reconstructed impact
parameters then become a smeared function of it.

Each collection is sorted by descending $\pT$ and zero-padded to length $K=128$
with a Boolean mask; we note that all network operations are masked to remove any contributions from the padded entries. Continuous features are
standardized on the training split. The event-level conditioning vector $\bg$
collects the mask-weighted means of the continuous truth features and the truth
and reconstructed $\HT$, missing-momentum components
$E^{\mathrm{miss}}_{x,y}=\sum_k p_{\mathrm{T},k}\{\cos,\sin\}\phi_k$, and
multiplicities, so that $\bg\in\mathbb{R}^{16}$. The flavor of the generating $Z$
decay is retained per event to enable the downstream analysis of Section~\ref{sec:ftag} but is not used an input to the surrogate.

The events are partitioned into 2.125M training,
125k validation, and 500k test events, with the training and validation sets only containing $q\bar{q}$ and $b\bar{b}$ processes and the test set additionally including $s\bar{s}$ and $c\bar{c}$ events. All results presented in the later sections are computed on the test split. To ensure a robust comparison, the {\delphes}
baseline is produced from the exact same generator events.

\section{Method}
\label{sec:method}

\subsection{Conditional factorization}
{\parnassus} models the conditional density $p(\bR\mid\bT)$ of the reconstructed
set of particle flow candidates given the truth set, and factorizes it into an event-level and a
particle-level stage as follows:
\begin{equation}
p(\bR\mid\bT) = \underbrace{p(\mathcal{E}\mid\bT)}_{\text{Event Stage}}\,
                \underbrace{p(\bR\mid\mathcal{E},\bT)}_{\text{Particle Stage}},
\qquad
\mathcal{E}=\big(N_\bR,\ \HT^\bR,\ E^{\mathrm{miss}}_{x,\bR},\ E^{\mathrm{miss}}_{y,\bR}\big)\in\mathbb{R}^4 ,
\label{eq:factorization}
\end{equation}
where $\mathcal{E}$ is the vector of aggregate reconstructed quantities. The event
stage fixes the overall scale and multiplicity and the particle stage distributes
that budget over individual particles. In particular, given $\bT$ one
samples $\mathcal{E}\sim p(\cdot\mid\bT)$, rounds the multiplicity to an integer
$\widehat N_\bR=\max(0,\lfloor N+\tfrac12\rfloor)$, and samples the $\widehat N_\bR$
reconstructed particles from $p(\bR\mid\mathcal{E},\bT)$. No full-simulation
information enters at generation time, so the chain is a complete fast simulation
of the combined detector and reconstruction response.

While we find that the particle- and event-level observables adopted from the CMS studies \cite{Dreyer:2024bhs, Elabd:2026poc} perform well, future work could explore coordinates and observables more typical of $e^+e^-$ collisions, such as the polar angle $\theta$ for particle-level kinematics and thrust and sphericity for event-level variables.

\subsection{Flow matching}
\label{sec:fm}
Both stages are trained with a flow-matching
objective~\cite{lipman2022flow,Albergo2022StochasticInterpolants,Liu:2022rectified}.
With a Gaussian source $\bx_0\sim\mathcal{N}(0,\mathbb{I})$ and a target
$\bx_1\sim p_1$, a linear interpolant, given as:
\begin{equation}
\bx_t=\alpha(t)\,\bx_0+\sigma(t)\,\bx_1,\qquad t\in[0,1],
\label{eq:path}
\end{equation}
has a conditional velocity that is constant along the path,
$\bu_t=\dot\alpha(t)\,\bx_0+\dot\sigma(t)\,\bx_1$. The two stages adopt opposite
orientations, $(\alpha,\sigma)=(1-t,t)$ for the event stage and $(t,1-t)$ for the
particle stage, so that the data endpoint is $t=1$ in the former and $t=0$ in the
latter. The marginal velocity that transports $p_0$ into $p_1$ is the conditional
expectation $\bu_t(\bx)=\mathbb{E}[\bu_t\mid\bx_t=\bx]$, and is the objective the
network $\bv_\theta$ learns via the masked regression:
\begin{equation}
\mathcal{L}_{\mathrm{FM}}(\theta)=\mathbb{E}_{t,\bx_0,\bx_1}
  \big\|[\,\bv_\theta(\bx_t,t,\bT,\bg)-\bu_t\,]\odot m\big\|^2 ,
\label{eq:fmloss}
\end{equation}
where $m$ masks padded slots and is unity for the event stage. The particle stage
draws $t\sim\mathcal{U}(0,1)$; the event stage instead uses the power-law density
$p(t)=\alpha_p\,t^{\alpha_p-1}$ with $\alpha_p=3$ (obtained as $t=u^{1/\alpha_p}$,
$u\sim\mathcal{U}(0,1)$), which concentrates training near the data endpoint where
the event-level field varies most rapidly. For the particle stage we augment
Eq.~\eqref{eq:fmloss}, following Ref.~\cite{Kobylianskii:2024sup}, with a cosine
term that penalizes the angle between the predicted and target velocities:
\begin{equation}
\mathcal{L}(\theta)=\mathcal{L}_{\mathrm{FM}}(\theta)
  +\lambda\,\mathbb{E}\!\left[1-\frac{\langle\bv_\theta,\bu\rangle}{\|\bv_\theta\|\,\|\bu\|}\right],
\qquad \lambda=\mathcal{O}(1),
\label{eq:loss}
\end{equation}
which supplies a scale-free gradient direction where $\|\bu\|$ is small. At
generation, the field is integrated as the ordinary differential equation
$\dd\bx_t/\dd t=\bv_\theta(\bx_t,t,\bT,\bg)$ from the noise to the data endpoint,
using a DPM solver~\cite{Lu:2022dpm} with $10$ function evaluations for the
particle stage and a PNDM solver~\cite{Liu:2022pndm} with $25$ function evaluations for the event
stage. The species
one-hots, the impact features, and the multiplicity are transported as continuous
coordinates and discretized (species by $\arg\max$, multiplicity by rounding) at
the end of the integration; no separate categorical head or cross-entropy term is
used.

\begin{figure}
\centering
\includegraphics[width=\linewidth]{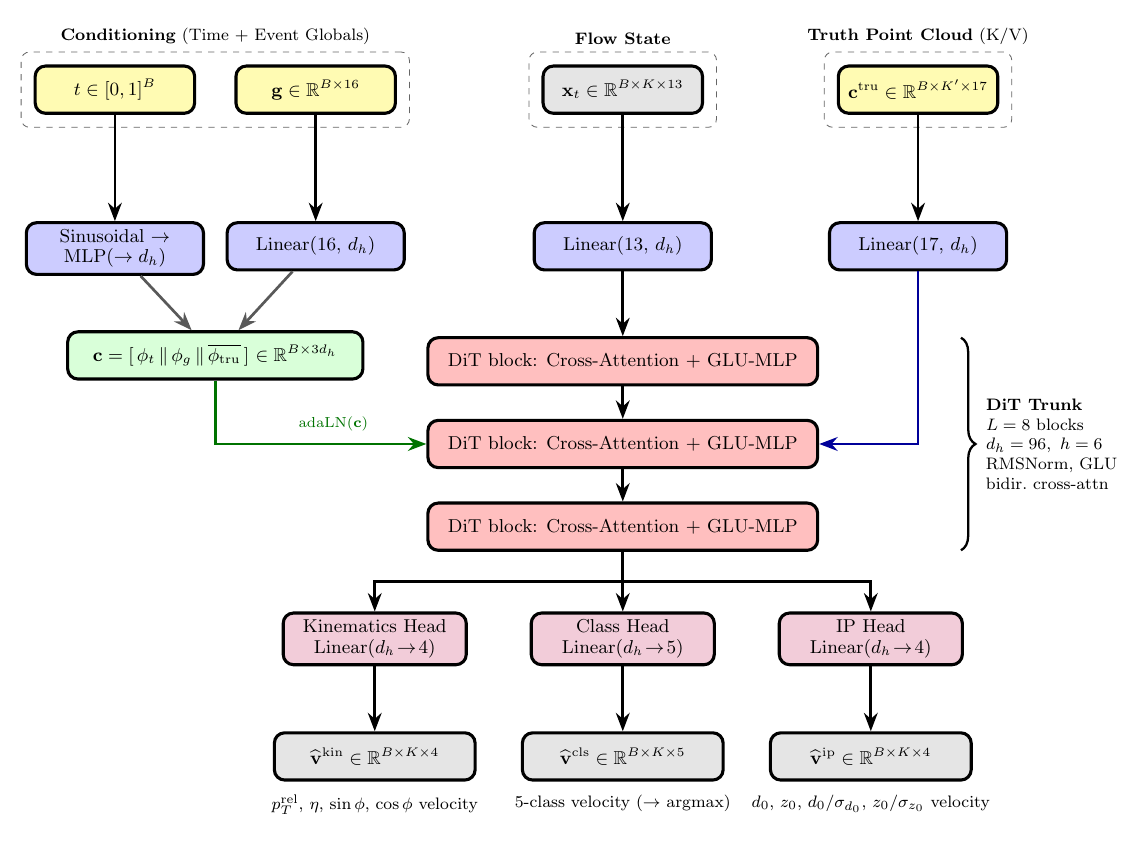}
\caption{Architecture of the particle-stage flow-matching network. The flow state
forms the query sequence of a cross-attention DiT trunk whose keys and values are
the embedded truth particle cloud; the flow time, the event-level globals, and
the pooled truth embedding are concatenated into the conditioning vector $\bc$ of
Eq.~\eqref{eq:cvec} that modulates every block and all output heads through
adaptive layer normalization.}
\label{fig:architecture}
\end{figure}

\subsection{Particle-stage architecture}
\label{sec:arch}
The velocity field is parameterized by a cross-attention diffusion transformer
(DiT)~\cite{Peebles:2022dit}, shown in Fig.~\ref{fig:architecture}. Three linear
maps embed the $13$-dimensional flow state $\bx_t$ (the target of
Eq.~\eqref{eq:fsfeat}), the $17$-dimensional truth particles
(Eq.~\eqref{eq:ctxtfeat}), and the $16$-dimensional global vector $\bg$ into a
common width $d_h=96$. Prior to being embedded by the neural network, the four categorical inputs (the
$\phi$ pair, and the truth charge and species one-hots) are expanded so that the
per-particle context carries $17$ features and $\bg$ the $16$ event-level
features (eight mask-weighted feature means and eight event scalars). The flow
time $t$ is embedded by a sinusoidal encoding followed by a two-layer perceptron,
and a single conditioning vector is formed by concatenation:
\begin{equation}
\bc \;=\; \big[\ \phi_t(t)\ \big\Vert\ \phi_g(\bg)\ \big\Vert\
                 \overline{\phi_\tau(\bT)}\ \big]\in\mathbb{R}^{3d_h},
\label{eq:cvec}
\end{equation}
where $\overline{\phi_\tau(\bT)}$ is the mask-weighted mean of the embedded truth
particles. The trunk consists of $L=8$ blocks, each updating the flow-state
sequence $h$ by cross-attention onto the embedded truth sequence $z$ followed by a
gated feed-forward layer, both modulated by $\bc$ through adaptive layer
normalization~\cite{Perez:2018film,Peebles:2022dit}:
\begin{align}
h &\leftarrow h + g_1\odot\mathrm{MHA}\big(\gamma_1\odot\mathrm{RMSNorm}(h)+\beta_1,\ \mathrm{RMSNorm}(z)\big), \label{eq:blockA}\\
h &\leftarrow h + g_2\odot\mathrm{MLP}\big(\gamma_2\odot\mathrm{RMSNorm}(h)+\beta_2\big), \label{eq:blockB}
\end{align}
where $(\beta_1,\gamma_1,g_1,\beta_2,\gamma_2,g_2)=\mathrm{Linear}(\mathrm{SiLU}(\bc))$
and the modulation is zero-initialized, so each block is the identity at
initialization. Attention uses $6$ heads with query-key normalization, and
RMSNorm~\cite{Zhang:2019rmsnorm} throughout; cross-attention is bidirectional
(the truth sequence also attends back onto the flow state) in all but the final
block. Because the architecture contains no positional encoding and all
reductions are masked means, the model is permutation-equivariant on the flow
state and permutation-invariant in the truth cloud. Three output heads, each a
perceptron modulated by $\bc$, produce the kinematic, species, and impact
components of the velocity:
\begin{equation}
\bv_\theta = \big[\ \bv^{\mathrm{kin}}_\theta\ (\in\mathbb{R}^4)\ \big\Vert\
                    \bv^{\mathrm{cls}}_\theta\ (\in\mathbb{R}^5)\ \big\Vert\
                    \bv^{\mathrm{ip}}_\theta\ (\in\mathbb{R}^4)\ \big]\odot m .
\label{eq:heads}
\end{equation}
The resulting particle-stage model has approximately 5M trainable
parameters, of which the DiT trunk
backbone accounts for about 4.7M parameters.

\subsection{Event-stage architecture}
The event stage models the four-dimensional density $p(\mathcal{E}\mid\bT)$ of
Eq.~\eqref{eq:factorization}. Because its target is a fixed-length vector rather
than a set, it requires neither attention nor permutation equivariance in the
target, and is a residual multilayer perceptron of three pre-normalized blocks of
three layers each at width $d_h^{\mathcal{E}}=128$ with GELU \cite{hendrycks2023gaussianerrorlinearunits} activations and
learned layer scaling. The truth cloud enters only through the mask-weighted mean
of the embedded truth particles, concatenated with the embedded truth-level event
quantities and the embedded flow time to form the conditioning vector, in direct
analogy with Eq.~\eqref{eq:cvec}. The event stage is trained with the vanilla
flow-matching loss of Eq.~\eqref{eq:fmloss}.

\subsection{Training and hyperparameters}
Both the particle-level and event-level stages are trained with the \textsc{Lion} optimizer~\cite{Chen:2023lion} using a learning
rate of $3\times10^{-5}$ and a weight decay $10^{-2}$, with a cosine annealing
schedule \cite{loshchilov2017sgdrstochasticgradientdescent}, and gradient-norm clipping at unity. We train using a batch size of $256$, for $15$
epochs over the full training set and select the best performing based on the validation set loss. The transformer hyperparameters, width $d_h=96$, depth $L=8$,
$6$ attention heads, are taken without modification from the CMS
configuration of Ref.~\cite{Kobylianskii:2024sup}, hence the present results
are obtained without a per-detector architecture search and represent a
conservative baseline. The only component that was for CLD is the
impact-parameter representation of Section~\ref{sec:dataset}. In particular, the
$\operatorname{asinh}$ scale $s$ was selected by the kurtosis-matching scan
described above, and the choice of significances over raw resolutions as the
learned target was validated on the reconstructed impact-parameter distributions.

\section{Results}
\label{sec:results}

All results are evaluated on the 500k held-out test events. All
distributions presented in this section are normalized to unit area with the ratio panels showing the
generated-to-reference ratio with the reference statistical uncertainty as a
band. {\pandora} full simulation is the reference (blue), and {\parnassus} (red)
and {\delphes} (green) are the surrogate and parameterized baseline.

\subsection{Particle-level observables}
\label{sec:res_particle}
\begin{figure}
\centering
\begin{subfigure}{0.32\linewidth}\includegraphics[width=\linewidth]{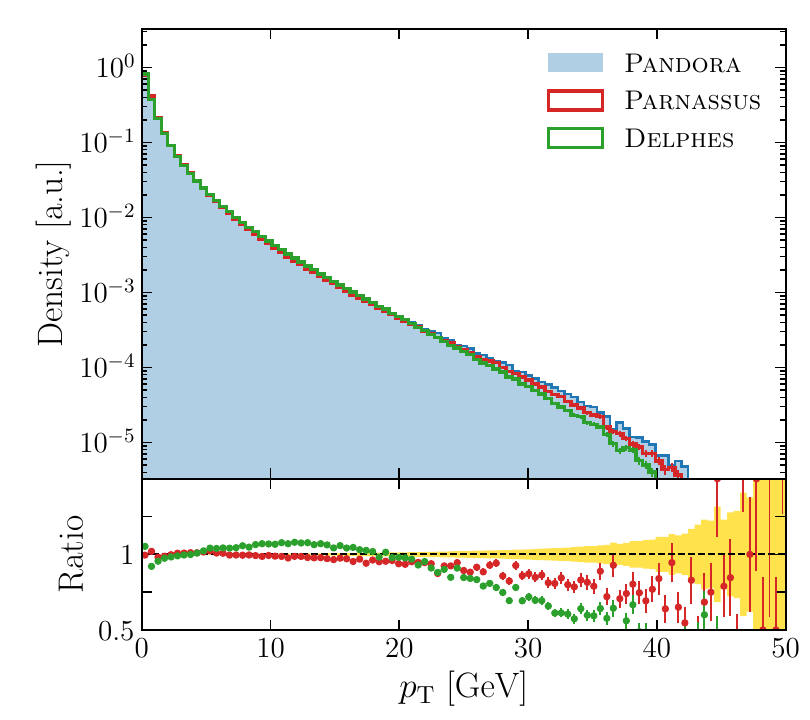}\end{subfigure}
\begin{subfigure}{0.32\linewidth}\includegraphics[width=\linewidth]{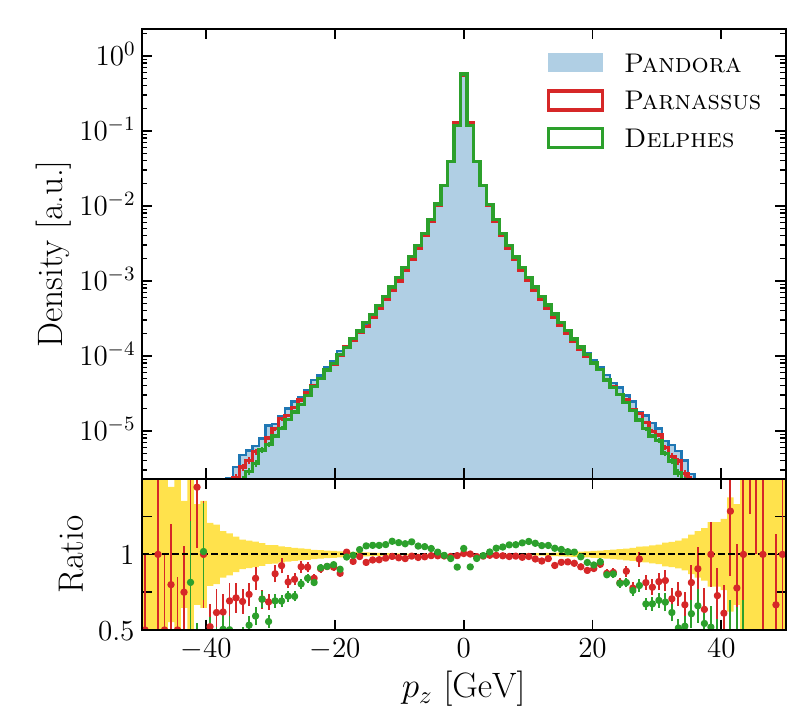}\end{subfigure}
\begin{subfigure}{0.32\linewidth}\includegraphics[width=\linewidth]{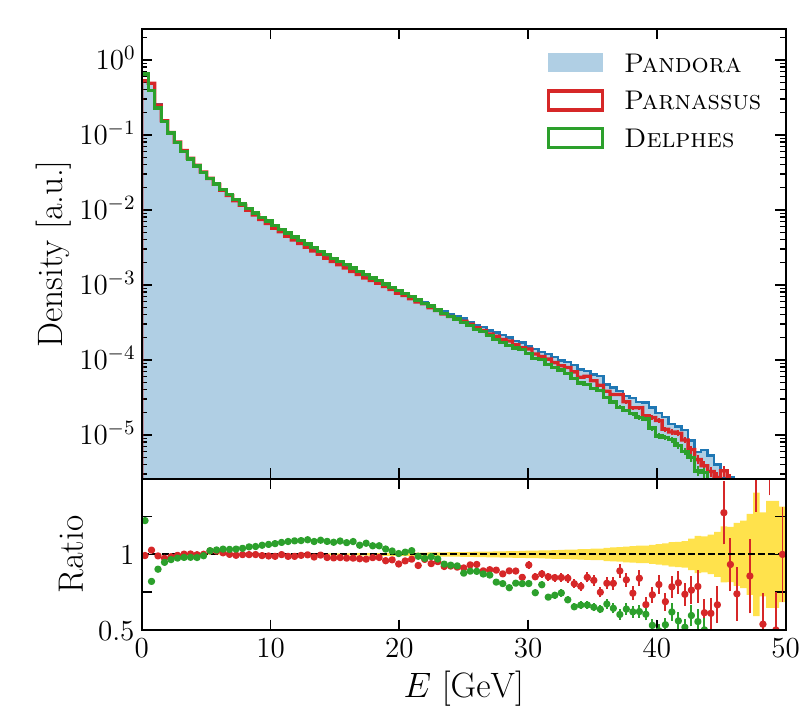}\end{subfigure}
\begin{subfigure}{0.32\linewidth}\includegraphics[width=\linewidth]{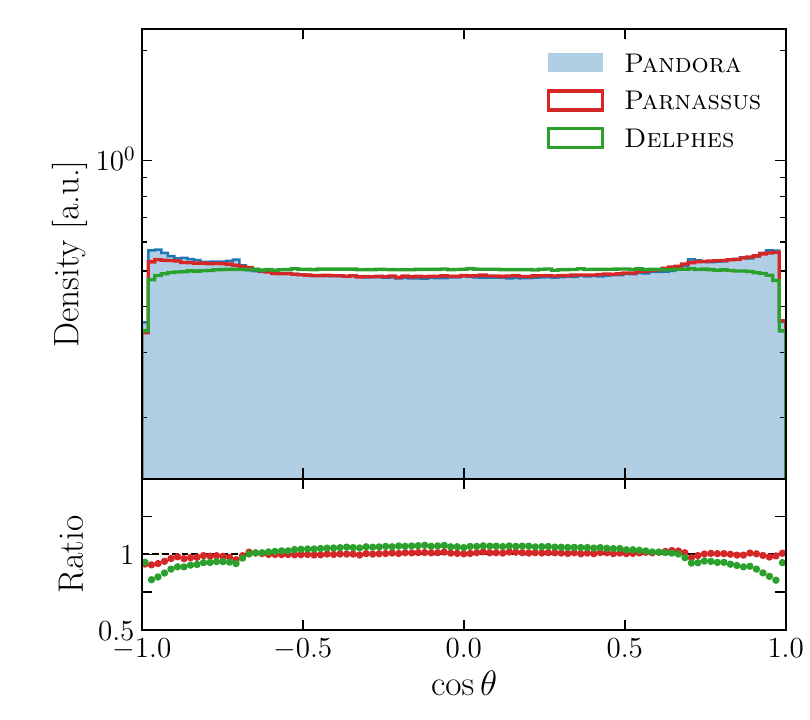}\end{subfigure}
\begin{subfigure}{0.32\linewidth}\includegraphics[width=\linewidth]{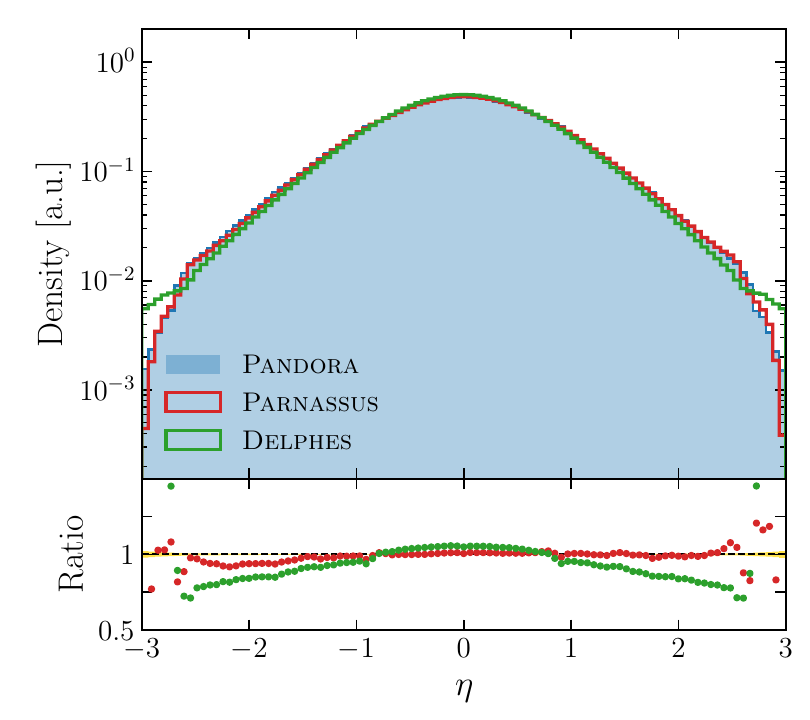}\end{subfigure}
\begin{subfigure}{0.32\linewidth}\includegraphics[width=\linewidth]{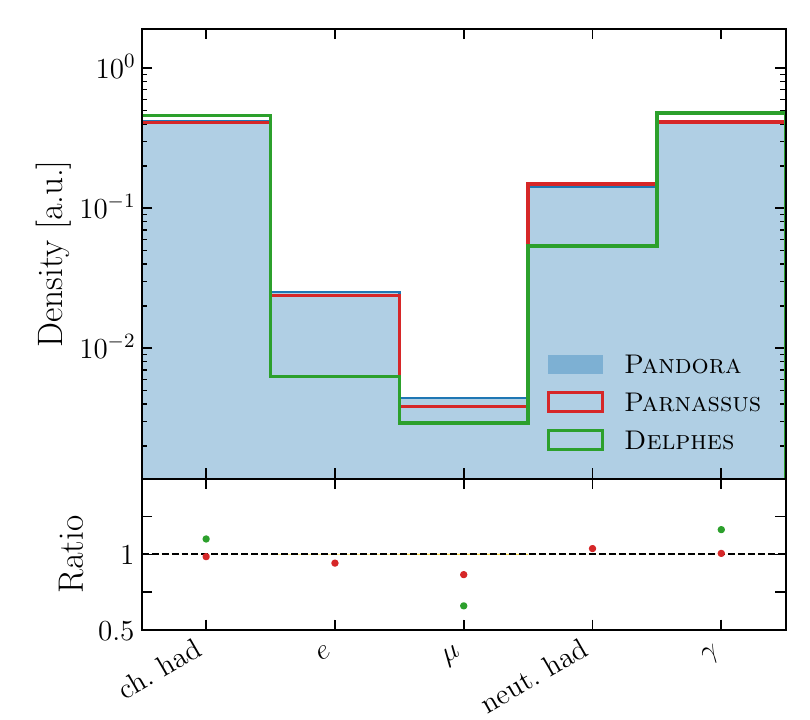}\end{subfigure}
\caption{Particle-level kinematics and species composition: {\pandora} (blue
filled), {\parnassus} (red), {\delphes} (green). Upper panels show
unit-normalized densities; lower panels the ratio to {\pandora} with its
statistical uncertainty as a band.}
\label{fig:particle_kin}
\end{figure}

\begin{figure}
\centering
\begin{subfigure}{0.48\linewidth}\includegraphics[width=\linewidth]{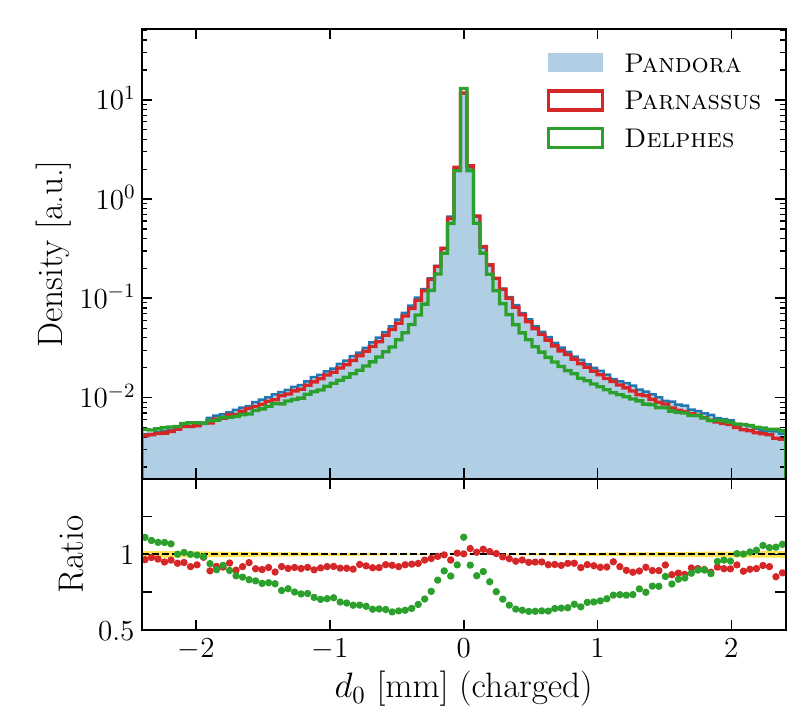}\end{subfigure}
\begin{subfigure}{0.48\linewidth}\includegraphics[width=\linewidth]{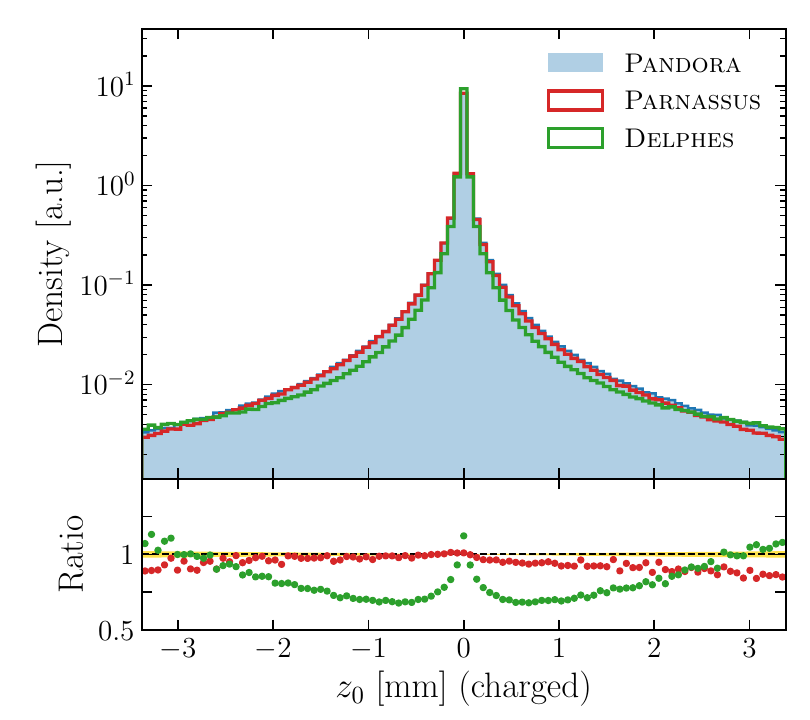}\end{subfigure}\\[2pt]
\begin{subfigure}{0.48\linewidth}\includegraphics[width=\linewidth]{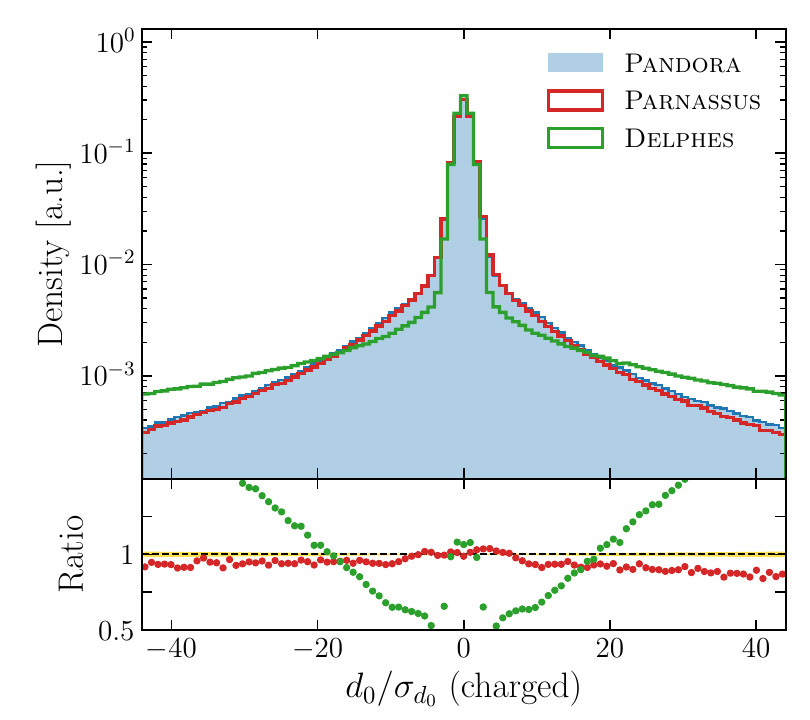}\end{subfigure}
\begin{subfigure}{0.48\linewidth}\includegraphics[width=\linewidth]{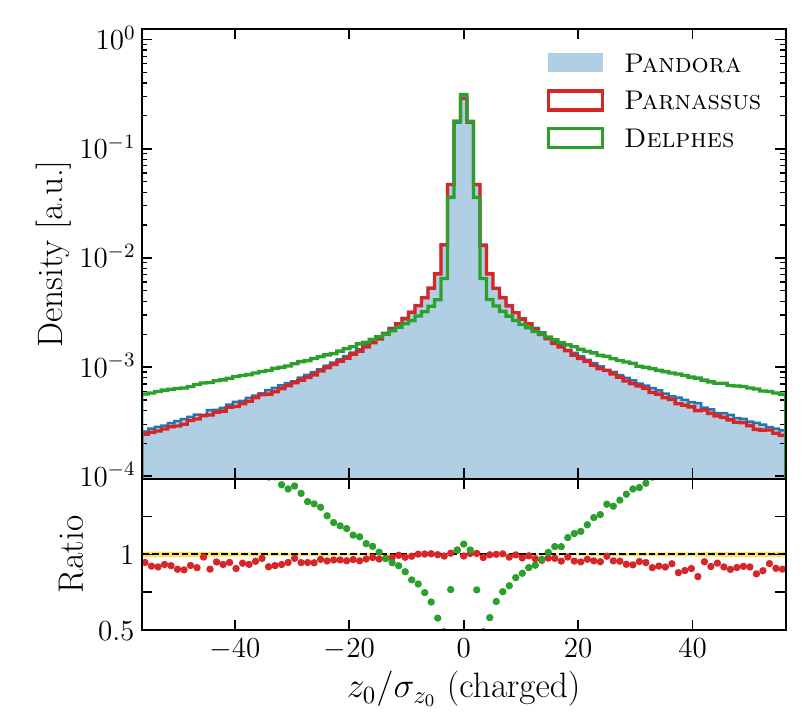}\end{subfigure}
\caption{Track impact parameters (for charged PFOs): transverse $\dzero$ and
longitudinal $\zzero$ and their significances. The displaced heavy-flavor tails
are reproduced by {\parnassus} through the truth-vertex conditioning; {\delphes}
departs from CLD, reflecting its different vertex resolution.}
\label{fig:particle_ip}
\end{figure}

Figure~\ref{fig:particle_kin} compares the single-particle kinematics and species
composition. The surrogate reproduces the transverse and longitudinal momentum
and the energy spectra over several orders of magnitude, the polar structure of $\cos\theta$
including the barrel-endcap acceptance features, the pseudorapidity, and the
five-species fractions. The {\delphes} baseline describes the inclusive $\pT$ and
$\eta$ spectra but its $\cos\theta$ distribution is virtually featureless since it uses a parameterized
acceptance rather than a physical boundary. The species composition for {\delphes} also differs
markedly. Figure~\ref{fig:particle_ip} shows the track impact parameters and
their significances for charged PFOs. The surrogate reproduces both the sharp
prompt core and the millimeter-scale displaced tails of $\dzero$ and $\zzero$,
learned from the truth-vertex conditioning of Eq.~\eqref{eq:ctxtfeat}, and the
significances $\dzero/\sigma_{\dzero}$, $\zzero/\sigma_{\zzero}$ on which the
tagger of Sec.~\ref{sec:ftag} relies. {\delphes}, departs from the baseline most strongly here.

\subsection{Jet-level observables}
\label{sec:res_jets}
To evaluate implications for the improved detector simulation at the level of jets and entire events, we cluster the reconstructed particles with the
Durham ($e^+e^-$ $k_\mathrm{T}$) algorithm~\cite{Catani:1991hj} in exclusive mode
with $n_{\mathrm{jet}}=2$ using {\fastjet}~\cite{Cacciari:2011ma} (v3.5.13). Exclusive
clustering to two jets assigns every particle to a hemisphere and transfers the
full event content into the jet observables. For each jet we compute the mass $m$, given as:
\begin{equation}
    m \;=\; \sqrt{\left(\sum_{i} E_{i}\right)^2 \;-\; \left(\sum_{i} \mathbf{p}_{i}\right)^2}\,,
\end{equation}
where the sum $i$ runs over the jet constituents, with $E_{i}$ and $\mathbf{p}_{i}$ as their energies and three-momenta respectively. Mass typically has a unimodal distribution, with a peak that scales with the jet $p_\mathrm{T}$. We also consider their transverse momentum, energy, $\cos\theta$, constituent multiplicity, and the girth $g$ defined as:
\begin{align*}
    g=\frac{1}{\pT^{\,\mathrm{jet}}}\sum_i p_{\mathrm{T},i}\,\Delta R(i,\mathrm{jet}),
\end{align*}
and the
two-point energy-correlation ratio
$e_2^{(\beta=2)}$ ~\cite{Larkoski:2013eya}, as:
\begin{align}
    e_2 =\frac{1}{p_\mathrm{T, jet}^2}\sum_{i<j}p_{\mathrm{T},i}p_{\mathrm{T},j}\Delta R_{ij}^2.
\end{align}
Dijets are formed by combining the four-momenta of the two jets and the dijet invariant mass is computed as $m_{jj}=\sqrt{E_{jj}^2 - \textbf{p}_{jj}^2}$.  Taken together, these observables exhibit a wide variety of properties. Jet mass, jet girth, $e_2$, $\cos \theta$, $\eta$ and $\pT$ are all infrared and collinear (IRC) safe, and therefore well-defined in fixed-order perturbation theory. Constituent multiplicity is IRC unsafe, making it highly sensitive to nonperturbative effects and detector distortions. Jet mass and the dijet mass probe mass-sensitive observables, while the energy correlation function $e_2$ and girth $g$ are indicative of multi-prong or angular structure. Constituent multiplicity $M$ is particularly sensitive to soft radiation. As such, these observables present a diverse set of behaviors and distribution shapes and provide a functional testbed for our baselines.

\begin{figure}
\centering
\begin{subfigure}{0.32\linewidth}\includegraphics[width=\linewidth]{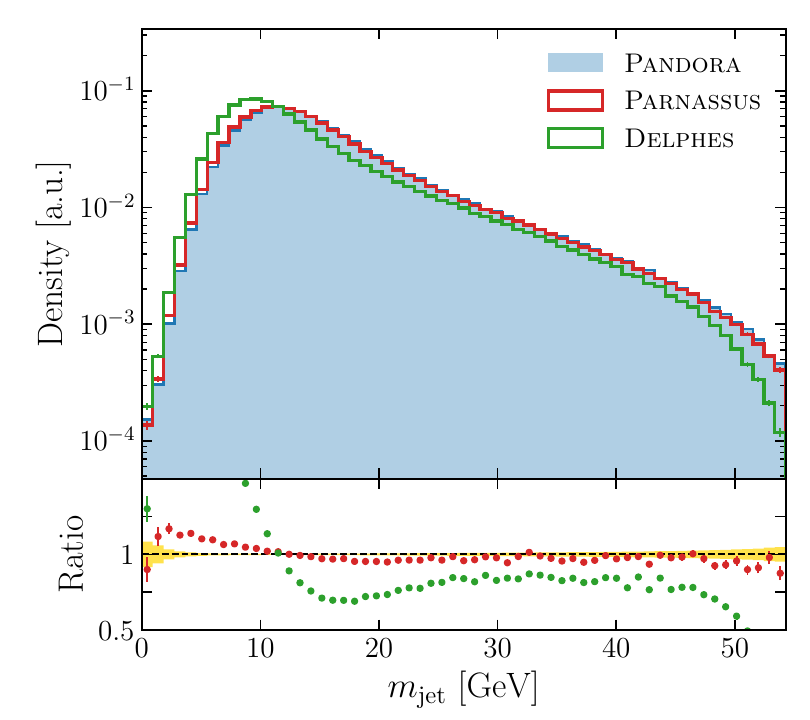}\end{subfigure}
\begin{subfigure}{0.32\linewidth}\includegraphics[width=\linewidth]{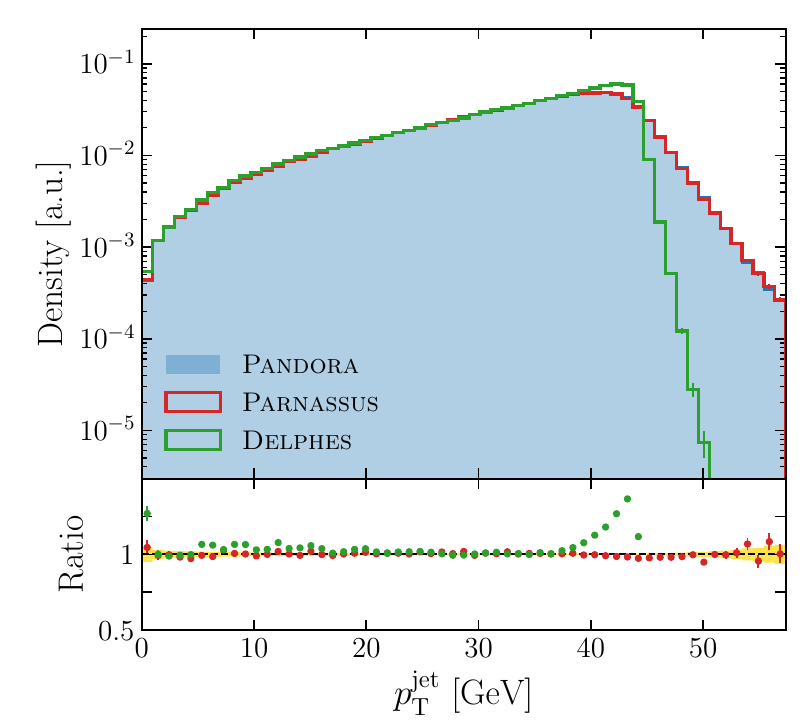}\end{subfigure}
\begin{subfigure}{0.32\linewidth}\includegraphics[width=\linewidth]{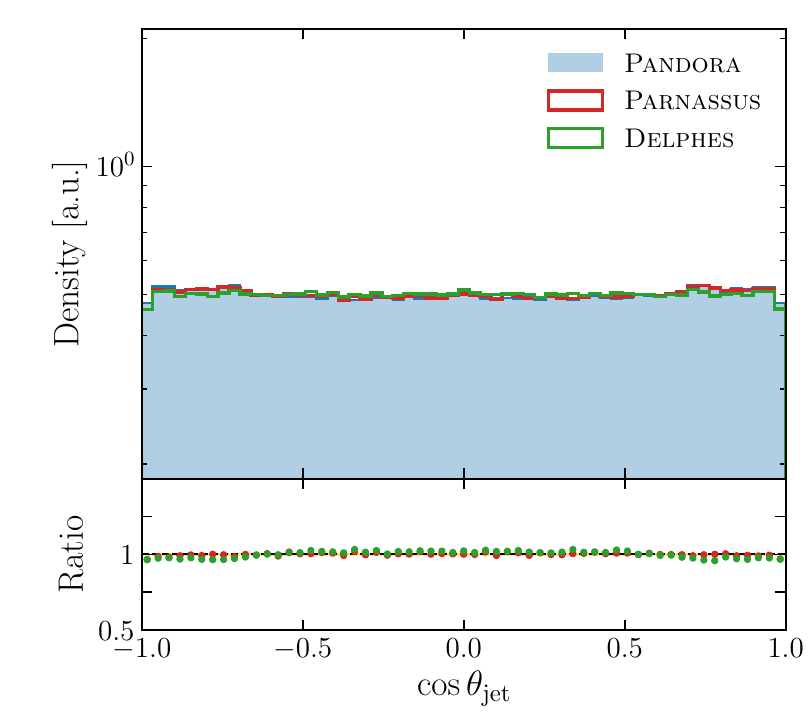}\end{subfigure}
\begin{subfigure}{0.32\linewidth}\includegraphics[width=\linewidth]{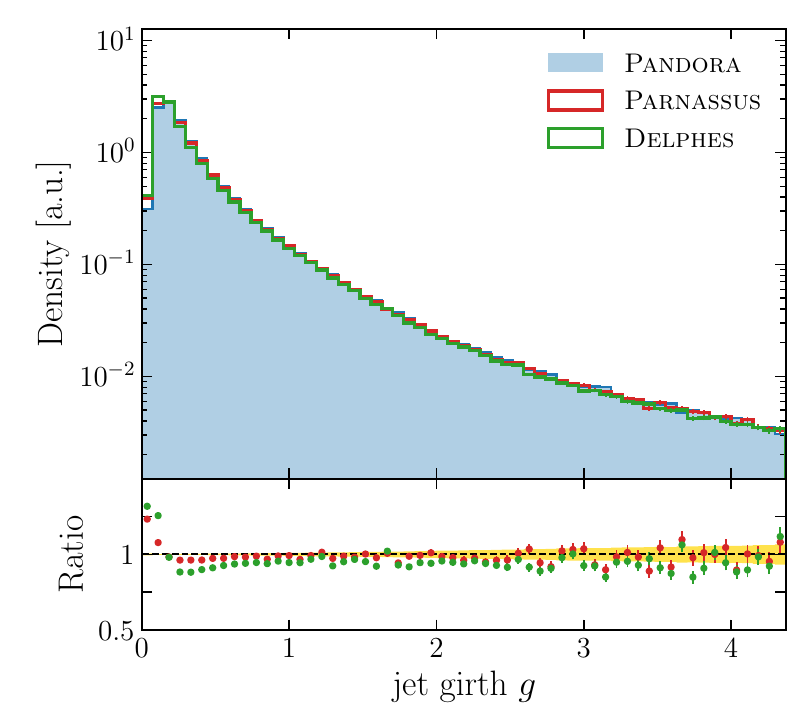}\end{subfigure}
\begin{subfigure}{0.32\linewidth}\includegraphics[width=\linewidth]{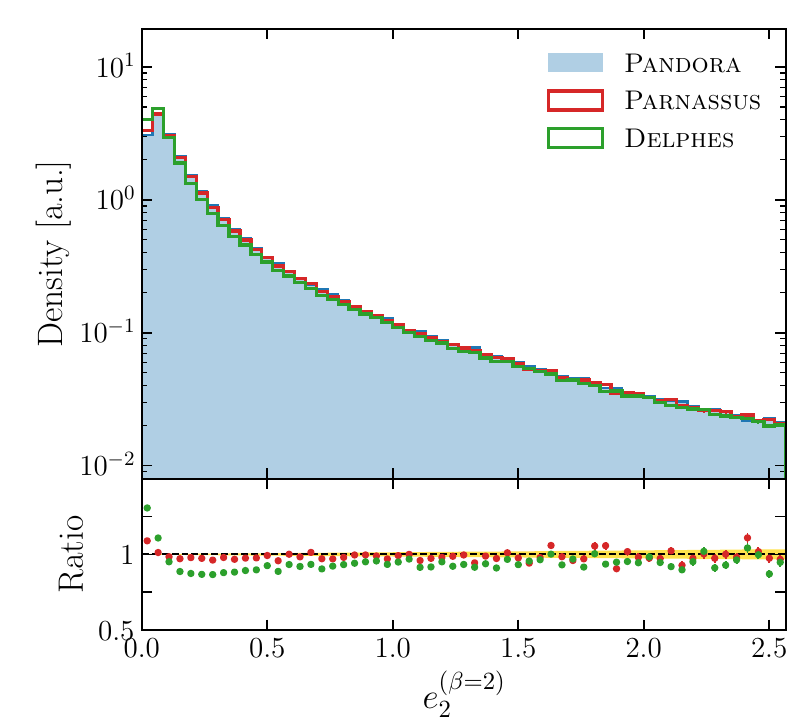}\end{subfigure}
\begin{subfigure}{0.32\linewidth}\includegraphics[width=\linewidth]{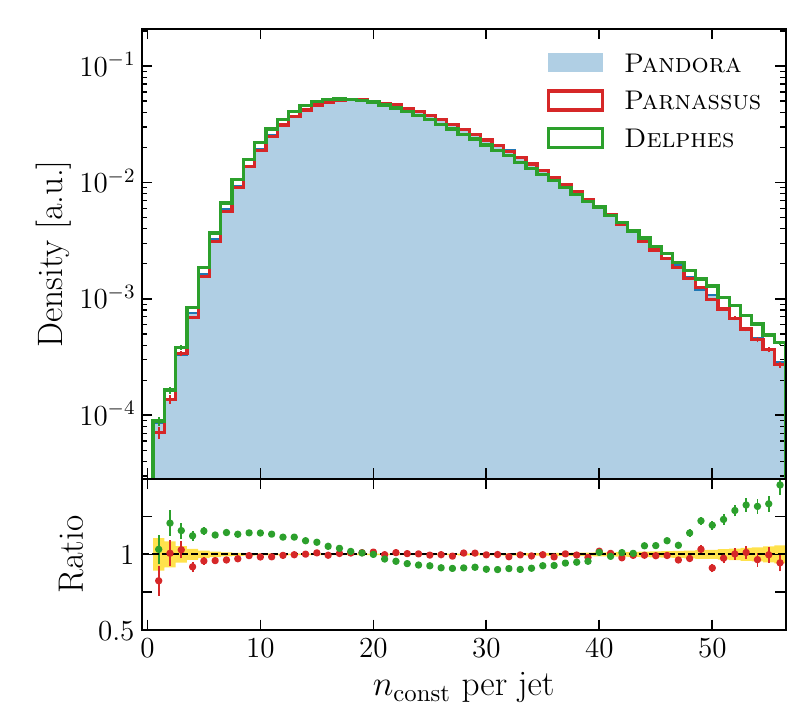}\end{subfigure}
\begin{subfigure}{0.32\linewidth}\includegraphics[width=\linewidth]{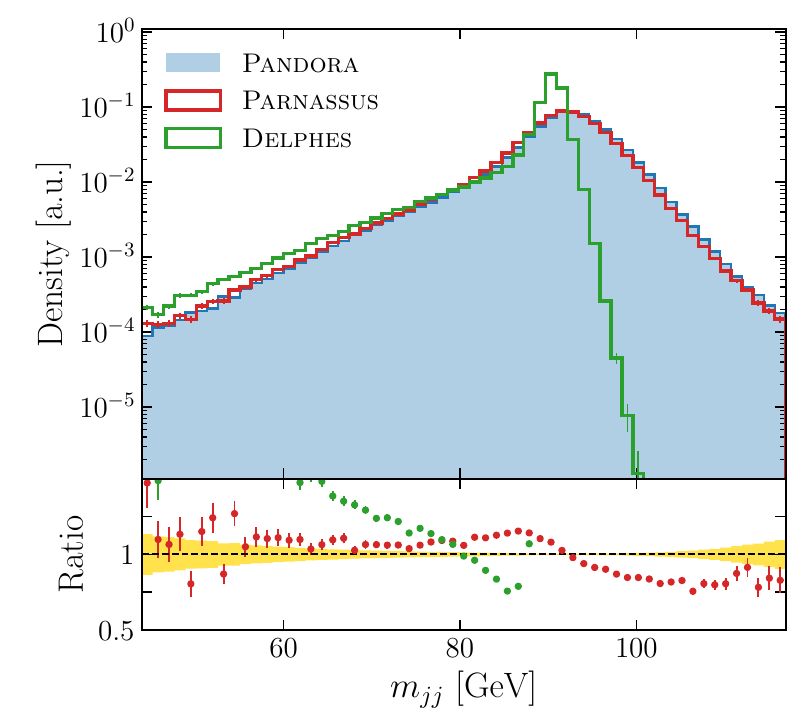}\end{subfigure}
\caption{Jet-level observables after Durham exclusive clustering with
$n_{\mathrm{jet}}=2$: jet mass, $\pT$, $\cos\theta$, girth, $e_2^{(\beta=2)}$,
constituent multiplicity, and the dijet invariant mass $m_{jj}$. Colors and ratio
panels as in Fig.~\ref{fig:particle_kin}.}
\label{fig:jets}
\end{figure}

Figure~\ref{fig:jets} shows that the
surrogate reproduces the jet mass, energy, angular, multiplicity, and
substructure distributions, and the $m_{jj}$ peak at the $Z$ mass; {\delphes} on the other hand is unable to track the jet mass, $p_\mathrm{T}$, jet multiplicity and shows a much sharper $m_{jj}$ edge. This is likely due to {\delphes} not modeling neutral fakes and duplicates, which leads it to underpredict these distributions at higher energies and masses.

\subsection{Event-level observables}
\label{sec:res_event}
\begin{figure}
\centering
\begin{subfigure}{0.32\linewidth}\includegraphics[width=\linewidth]{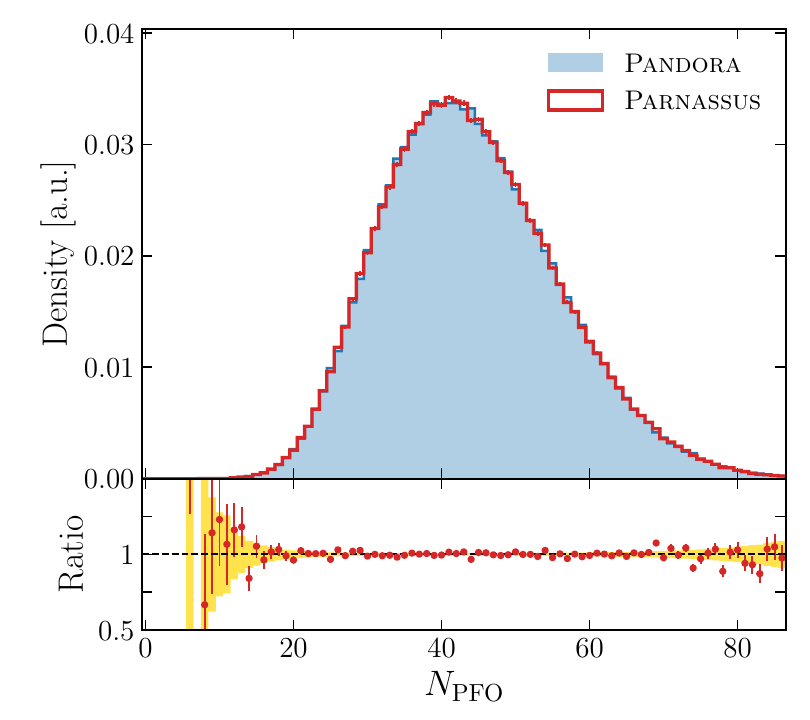}\end{subfigure}
\begin{subfigure}{0.32\linewidth}\includegraphics[width=\linewidth]{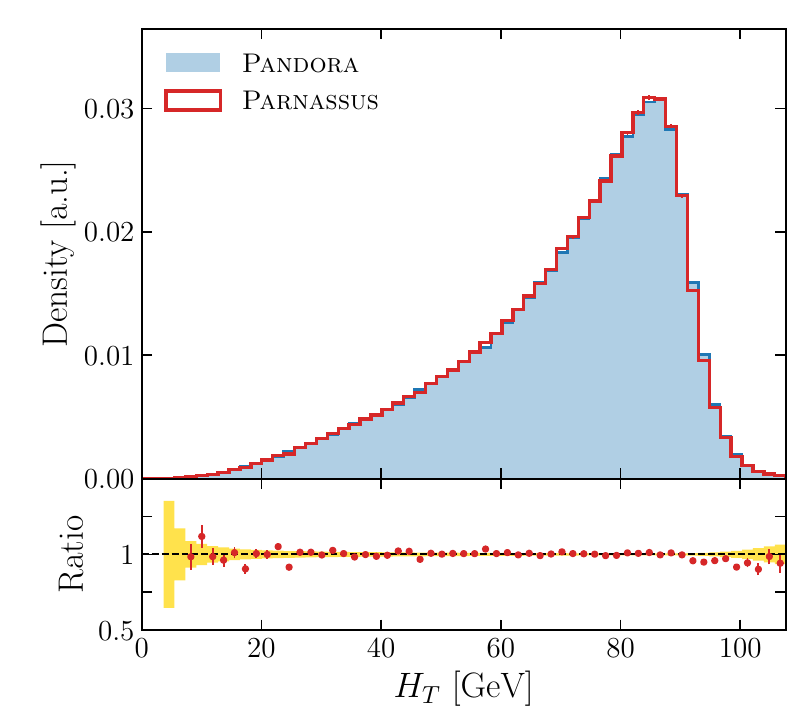}\end{subfigure}
\begin{subfigure}{0.32\linewidth}\includegraphics[width=\linewidth]{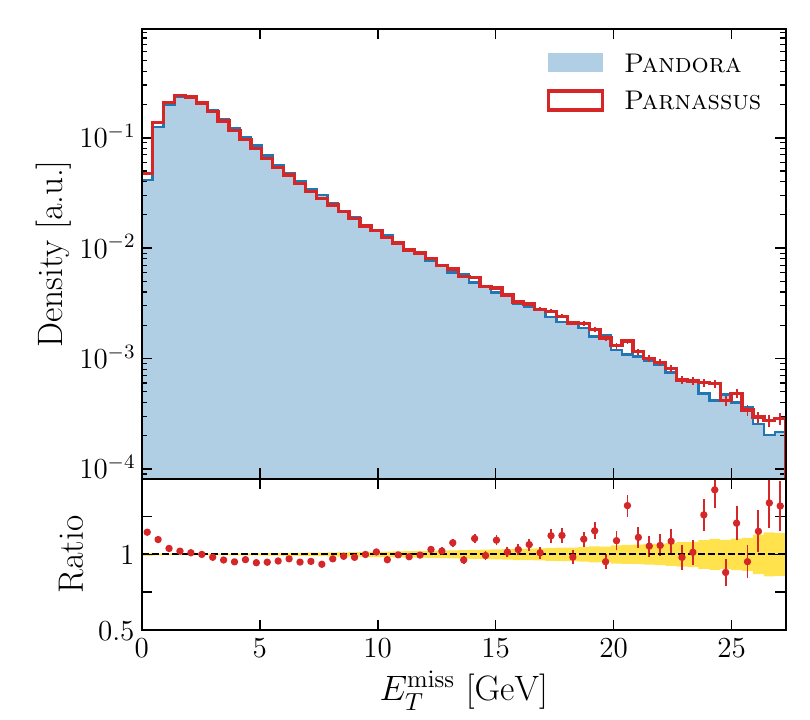}\end{subfigure}
\begin{subfigure}{0.32\linewidth}\includegraphics[width=\linewidth]{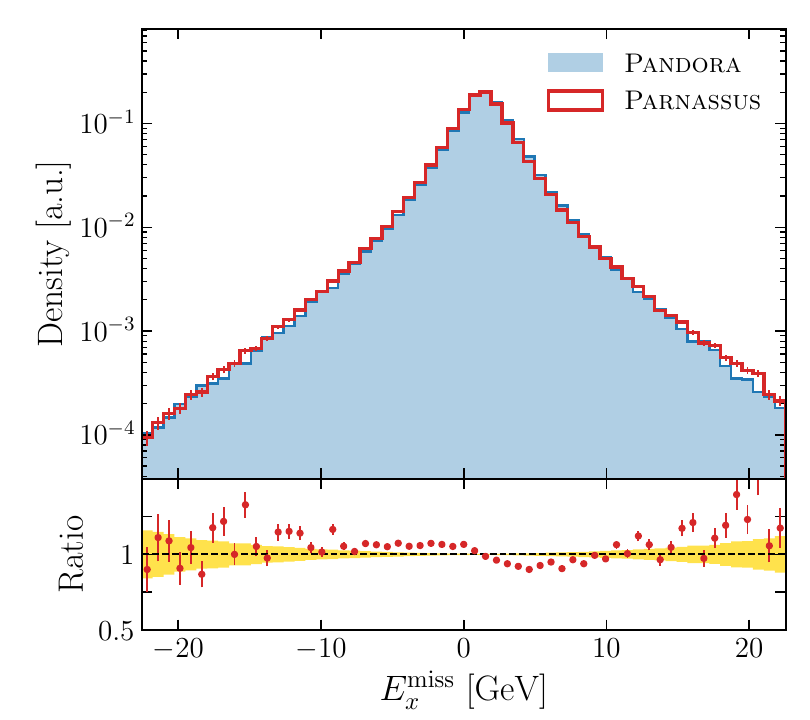}\end{subfigure}
\begin{subfigure}{0.32\linewidth}\includegraphics[width=\linewidth]{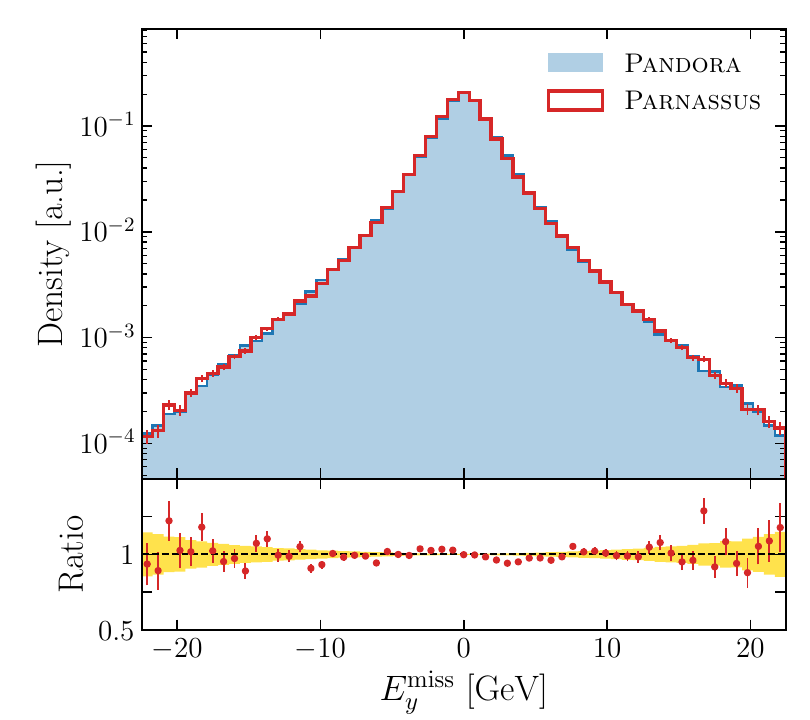}\end{subfigure}
\begin{subfigure}{0.32\linewidth}\includegraphics[width=\linewidth]{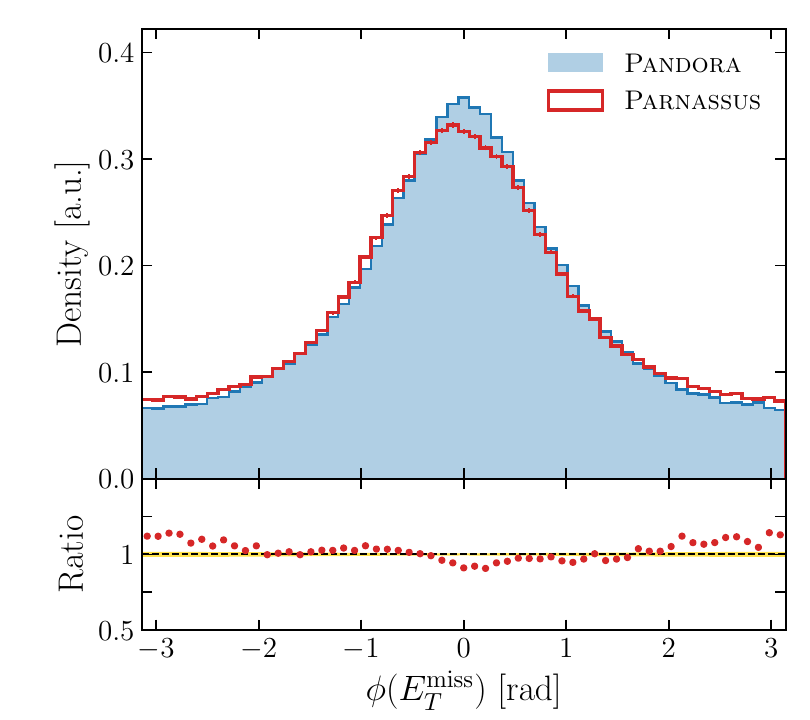}\end{subfigure}
\caption{Event-level observables from the event stage:
reconstructed multiplicity $N_{\mathrm{PFO}}$, scalar sum $\HT$, missing-energy
magnitude $E_T^{\mathrm{miss}}$ and components $E_{x,y}^{\mathrm{miss}}$, and the
missing-energy azimuth. {\pandora} (blue), {\parnassus} event model (red).}
\label{fig:event}
\end{figure}

Figure~\ref{fig:event} compares the reconstructed multiplicity $N_\bR$, the scalar sum $\HT^\bR$, the missing-energy magnitude and
components, and the missing-energy azimuth against full simulation. The event
stage reproduces the multiplicity distribution and its width, the $\HT$ peak, and
the symmetric, sharply peaked missing-momentum components.

\subsection{Distribution distances}
\label{sec:res_dist}
To move beyond visual comparison we quantify the distance between each surrogate
distribution and the {\pandora} reference with three complementary measures. For
each observable, let $p$ and $q$ be the unit-normalized histograms of the two
samples on a common binning. The
{Vincze-Le Cam} (VLC) divergence, also known as the triangular
discriminant~\cite{Topsoe:2000}, is given as:
\begin{equation}
\Delta(p,q)=\sum_i \frac{(p_i-q_i)^2}{p_i+q_i}\in[0,2],
\label{eq:vlc}
\end{equation}
is a symmetric divergence whose square root is a metric; we quote $\Delta$ scaled by
$10^{2}$. Similarly, {Jensen-Shannon} (JS) distance is the square root of the
Jensen-Shannon divergence,
$\mathrm{JS}(p,q)=[\tfrac12 D_{\mathrm{KL}}(p\Vert m)+\tfrac12 D_{\mathrm{KL}}(q\Vert m)]^{1/2}$
with $m=\tfrac12(p+q)$ and $D_{\mathrm{KL}}$ the Kullback-Leibler divergence in
base $2$, so that $\mathrm{JS}\in[0,1]$. The {$1$-Wasserstein} (earth-mover)
distance is computed on the raw (unbinned) samples $\{x\}\sim p$, $\{y\}\sim q$ as:
\begin{equation}
W_1(p,q)=\int_{0}^{1}\big|F_p^{-1}(u)-F_q^{-1}(u)\big|\,\dd u ,
\label{eq:w1}
\end{equation}
with $F^{-1}$ the quantile function; it carries the units of the observable. For every observable we report
$d(\textsc{Parnassus},\textsc{Pandora})$ and $d(\textsc{Delphes},\textsc{Pandora})$;
the smaller value i.e.~closer to the full simulation, is set in boldface.

Tables~\ref{tab:pdist} and~\ref{tab:jdist} give the particle- and jet-level
distances for all three measures. {\parnassus} is closer to {\pandora} than
{\delphes} for every observable and every metric, typically by an order of
magnitude, and the gap is largest for the species composition, the jet mass, and
the impact-parameter significances.

\begin{table}[htbp]
\centering
\caption{Particle-level distribution distances between the surrogate and {\pandora} full simulation: Vincze--Le Cam divergence ($\times10^{2}$), $1$-Wasserstein distance $W_1$, and Jensen--Shannon distance. Lower is closer to full simulation; the smaller of the two reconstructions is in boldface. Impact observables are for charged PFOs. A parenthesized digit indicates the $1\sigma$ statistical uncertainty on the last shown digit from a $1000$-replica bootstrap (omitted where it is below that).}
\label{tab:pdist}
\smallskip
\def\arraystretch{1.3}
\begin{tabular}{c | c c c | c c c}
\hline
 & \multicolumn{3}{c|}{\textbf{\textsc{Parnassus}}} & \multicolumn{3}{c}{\textbf{\textsc{Delphes}}} \\
\textbf{~Observable~} & VLC & $W_1$ & JS & VLC & $W_1$ & JS \\
\hline
$\pT$ & \textbf{0.010} & \textbf{0.017} & \textbf{0.006} & 0.148(2) & 0.042 & 0.023 \\
$p_z$ & \textbf{0.005} & \textbf{0.017(1)} & \textbf{0.004} & 0.148(2) & 0.038 & 0.023 \\
$E$ & \textbf{0.018(1)} & \textbf{0.024(1)} & \textbf{0.008} & 1.114(4) & 0.077 & 0.063 \\
$\cos\theta$ & \textbf{0.014(1)} & \textbf{0.005} & \textbf{0.007} & 0.198(2) & 0.016 & 0.027 \\
$\eta$ & \textbf{0.027(1)} & \textbf{0.009} & \textbf{0.010} & 0.309(2) & 0.032 & 0.034 \\
species & \textbf{0.025(1)} & \textbf{0.025(1)} & \textbf{0.009} & 5.830(9) & 0.153 & 0.148 \\
$\dzero$ & \textbf{0.038(1)} & \textbf{0.013} & \textbf{0.012} & 0.487(5) & 0.112 & 0.042 \\
$\zzero$ & \textbf{0.023(1)} & \textbf{0.018} & \textbf{0.009} & 0.585(5) & 0.149 & 0.046 \\
$\dzero/\sigma_{\dzero}$ & \textbf{0.037(1)} & \textbf{0.250(3)} & \textbf{0.012} & 1.512(8) & 3.376(3) & 0.074 \\
$\zzero/\sigma_{\zzero}$ & \textbf{0.023(1)} & \textbf{0.286(3)} & \textbf{0.009} & 1.807(9) & 3.422(4) & 0.081 \\
\hline
\end{tabular}
\end{table}

\begin{table}[htbp]
\centering
\caption{Jet-level distribution distances (Durham $n_{\mathrm{jet}}=2$) between the
surrogate and {\pandora}. Conventions as in table~\ref{tab:pdist}.}
\label{tab:jdist}
\smallskip
\def\arraystretch{1.3}

\begin{tabular}{c | c c c | c c c}
\hline
 & \multicolumn{3}{c|}{\textbf{\textsc{Parnassus}}} & \multicolumn{3}{c}{\textbf{\textsc{Delphes}}} \\
\textbf{~Observable~} & VLC & $W_1$ & JS & VLC & $W_1$ & JS \\
\hline
$m_{\mathrm{jet}}$ & \textbf{0.086(6)} & \textbf{0.247(12)} & \textbf{0.018(1)} & 5.488(46) & 1.886(12) & 0.141(1) \\
$\pT^{\mathrm{jet}}$ & \textbf{0.008(2)} & \textbf{0.062(12)} & \textbf{0.005(1)} & 5.375(32) & 0.720(14) & 0.154 \\
$\cos\theta_{\mathrm{jet}}$ & \textbf{0.005(2)} & \textbf{0.002} & \textbf{0.004(1)} & 0.025(3) & 0.005 & 0.010(1) \\
girth $g$ & \textbf{0.145(8)} & \textbf{0.008(1)} & \textbf{0.023(1)} & 0.673(17) & 0.022(1) & 0.049(1) \\
$e_2^{(\beta=2)}$ & \textbf{0.076(6)} & \textbf{0.007(1)} & \textbf{0.017(1)} & 0.865(19) & 0.026(1) & 0.056(1) \\
$n_{\mathrm{const}}$ & \textbf{0.007(2)} & \textbf{0.033(7)} & \textbf{0.005(1)} & 0.147(8) & 0.298(9) & 0.023(1) \\
$m_{jj}$ & \textbf{0.518(20)} & \textbf{0.541(14)} & \textbf{0.043(1)} & 58.565(137) & 3.492(14) & 0.506(1) \\
\hline
\end{tabular}
\end{table}

\subsection{Flavor dependence}
\label{sec:res_flavor}
We recompute the full set of particle-level distances of table~\ref{tab:pdist}
separately for each $Z$ decay channel;
Tables~\ref{tab:flavor_qq}-\ref{tab:flavor_bb} report all ten observables, in the
same format, for the $q\bar q$, $s\bar s$, $c\bar c$, and $b\bar b$ channels. The
kinematic and species distances are essentially flavor-independent for both
surrogates, whereas the impact-parameter distances and their significances grow
toward heavy flavor as the displaced-track population increases. Notably, however, the reconstruction performance is agnostic to whether the processes were included in training or not since $b\bar{b}$, which was included in training, exhibits worse performance than $s\bar{s}$ and $c\bar{c}$ which were not. The
{\parnassus}-{\pandora} $W_1$ of $\dzero/\sigma_{\dzero}$ rises from $0.14$ in
$q\bar q$ to $0.41$ in $b\bar b$, yet remains an order of magnitude below the
corresponding {\delphes}-{\pandora} distance ($2.06$ to $4.94$) in every channel,
with the {\delphes} gap peaking in $b\bar b$. {\parnassus} is the closer of the two surrogates to full
simulation for every observable, every metric, and every flavor. The full
per-flavor comparison plots for all observables are collected in
Appendix~\ref{app:flavor}. 

\input{flavor_tables.tex}
\section{Heavy-flavor tagging}
\label{sec:ftag}

The decisive test of a fast-simulation surrogate is whether a complex, highly correlated downstream analysis
performed on its output reaches the same conclusions as one trained on full
simulation. Jet flavor tagging naturally presents itself as such an analysis at the FCC-ee, where
the $Z/H\to b\bar b$, $c\bar c$, and $s\bar s$ programs demand accurate $b/c/s$
identification~\cite{Ntounis:2025jbl}.

\subsection{Tagger and metrics}
We train a transformer \cite{vaswani2023attentionneed} tagger on the reconstructed constituents of each Durham
jet. Each jet contributes up to $50$ highest-$\pT$ constituents, and each
constituent carries $16$ features: seven relative kinematics
($\log\pT$, $\log E$, $\log\ptrel$, $\log E^{\mathrm{rel}}$, $\Delta\eta$,
$\Delta\phi$, $\Delta R$ to the jet axis), the five-species one-hot, and the four
impact observables $(\dzero,\zzero,\dzero/\sigma_{\dzero},\zzero/\sigma_{\zzero})$, which is exactly the information the surrogate generates. A per-constituent multilayer
perceptron embeds the features into width $d=128$; a learnable class token is
prepended and the sequence is processed by an $8$-layer transformer encoder with
$16$ attention heads, feed-forward width $256$, GELU activations, pre-layer-norm,
and padding-masked attention. The class-token output feeds a two-layer head to a
four-way softmax over $\{q,s,c,b\}$, trained with the cross-entropy loss:
\begin{equation}
\mathcal{L}_{\mathrm{CE}}=-\frac1N\sum_{n=1}^{N}\sum_{f\in\{q,s,c,b\}}
   y^{(n)}_f\,\log p^{(n)}_f ,
\label{eq:ce}
\end{equation}
where $y^{(n)}$ is the one-hot flavor label, the flavor of the generating $Z$
decay, and $p^{(n)}$ the softmax output. The tagger has about 1.1M
parameters and is trained with the \textsc{AdamW} \cite{loshchilov2019decoupledweightdecayregularization} optimizer at a learning rate $10^{-3}$, batch size
$512$, and cosine annealing for $30$ epochs, with an $80/10/10$ train/validation/
test split of the jets. As before, the final model selection is based on validation set loss. A {separate}
tagger is trained on each of {\pandora}, {\parnassus}, and {\delphes}. Following Ref.~\cite{Ntounis:2025jbl}, for an ordered flavor pair $(i,j)$ we form
the binary discriminant:
\begin{equation}
D_{ij}=\frac{p_i}{p_i+p_j},
\label{eq:disc}
\end{equation}
and its receiver operating characteristic (ROC), the background-$j$ mis-tag rate
as a function of the signal-$i$ efficiency $\eps_S$; we quote the area under the
curve $\mathrm{AUC}(B)=1-\int_0^1\mathrm{MR}(B)(\eps_S)\,\dd\eps_S$ and the mis-tag
rate at signal-efficiency working points of $80\%$ and $90\%$.

\subsection{Results}
Table~\ref{tab:ftag_auc} summarizes the four-class accuracy and the one-vs-rest
AUC per flavor. The three reconstructions order as expected: {\parnassus}
reproduces {\pandora} to within a percent on every flavor, while {\delphes}
is significantly stronger. Table~\ref{tab:ftag_mistag} lists the mis-tag rates at the two
working points for the physics-relevant discriminations; the surrogate tracks
{\pandora} closely (for $b$ versus $c$ at $80\%$ $b$-efficiency the mis-tag rate is
$1.58\%$ for {\parnassus} against $1.00\%$ for {\pandora}).

\begin{figure}
\centering
\begin{subfigure}{0.32\linewidth}\includegraphics[width=\linewidth]{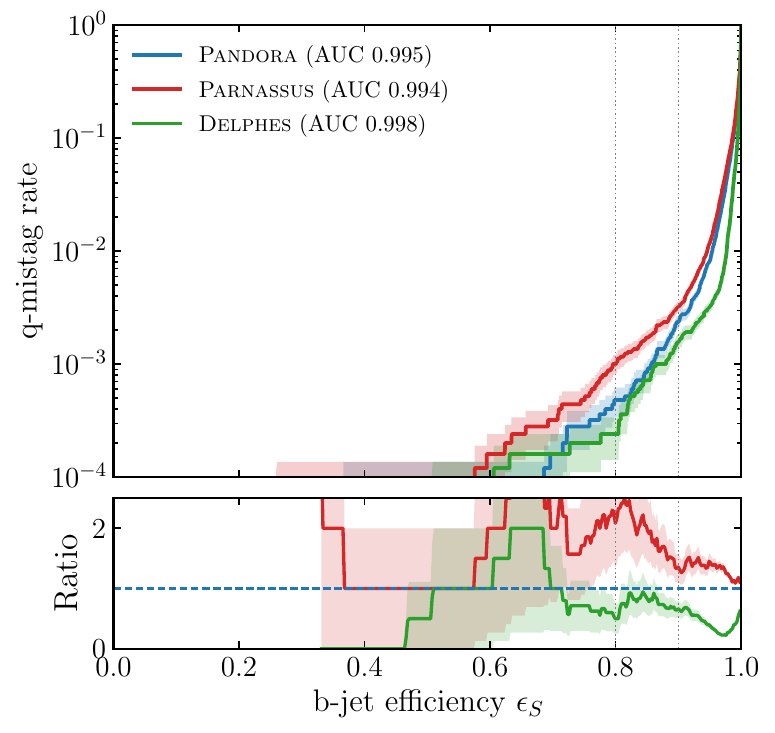}\end{subfigure}
\begin{subfigure}{0.32\linewidth}\includegraphics[width=\linewidth]{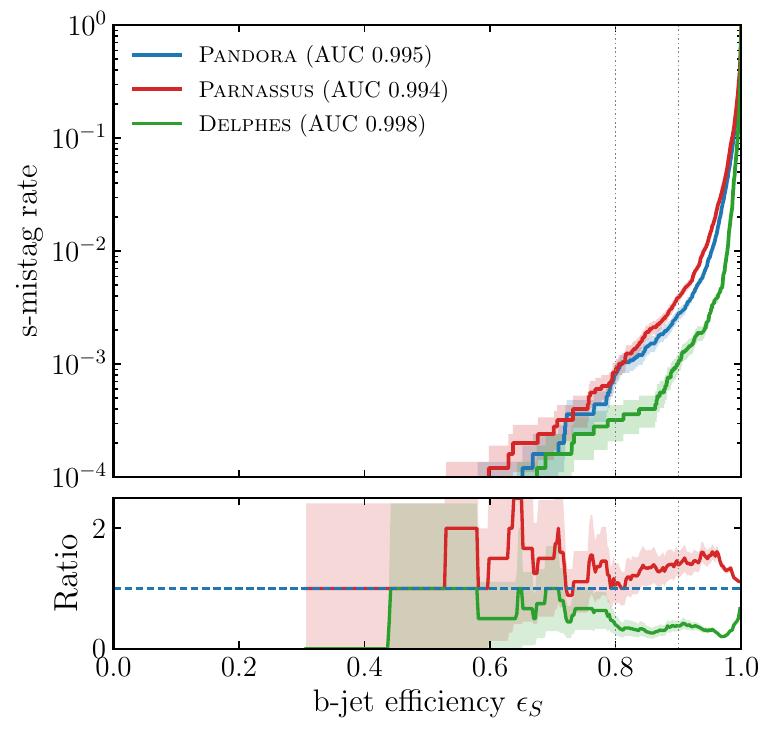}\end{subfigure}
\begin{subfigure}{0.32\linewidth}\includegraphics[width=\linewidth]{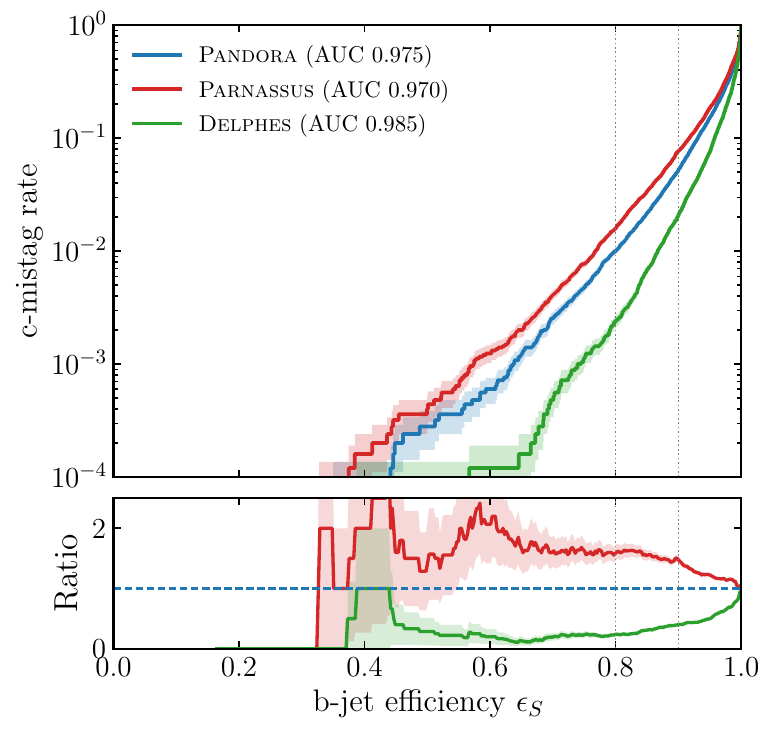}\end{subfigure}
\caption{$b$-tagging: background mis-tag rate versus $b$-jet efficiency against
light-quark, $s$, and $c$ backgrounds, via $D_{ij}=p_i/(p_i+p_j)$. Each panel
overlays {\pandora} (blue), {\parnassus} (red), {\delphes} (green) with a
ratio-to-{\pandora} sub-panel and binomial bands; dotted lines mark the $80\%$ and
$90\%$ working points.}
\label{fig:roc_b}
\end{figure}

\begin{figure}
\centering
\begin{subfigure}{0.32\linewidth}\includegraphics[width=\linewidth]{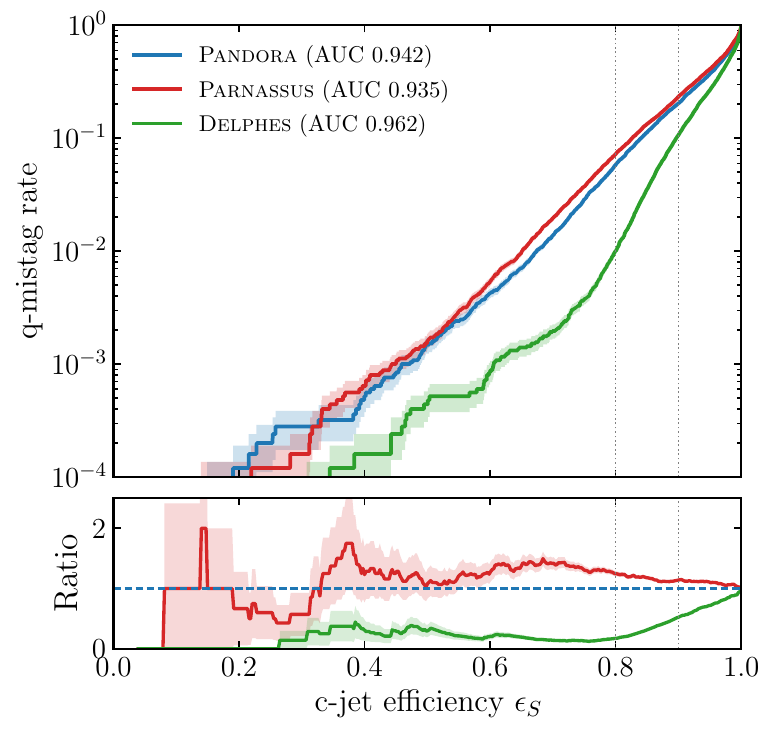}\end{subfigure}
\begin{subfigure}{0.32\linewidth}\includegraphics[width=\linewidth]{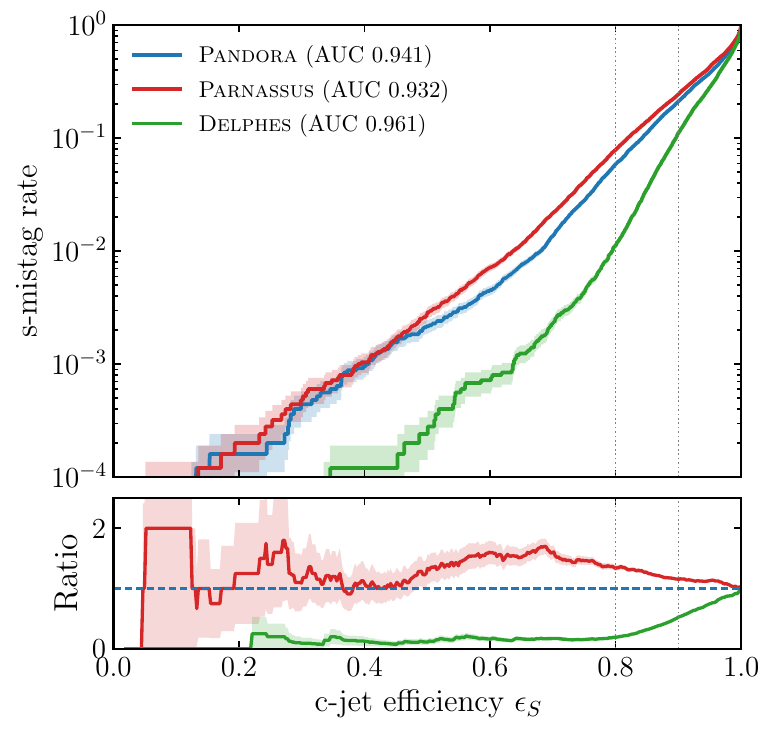}\end{subfigure}
\begin{subfigure}{0.32\linewidth}\includegraphics[width=\linewidth]{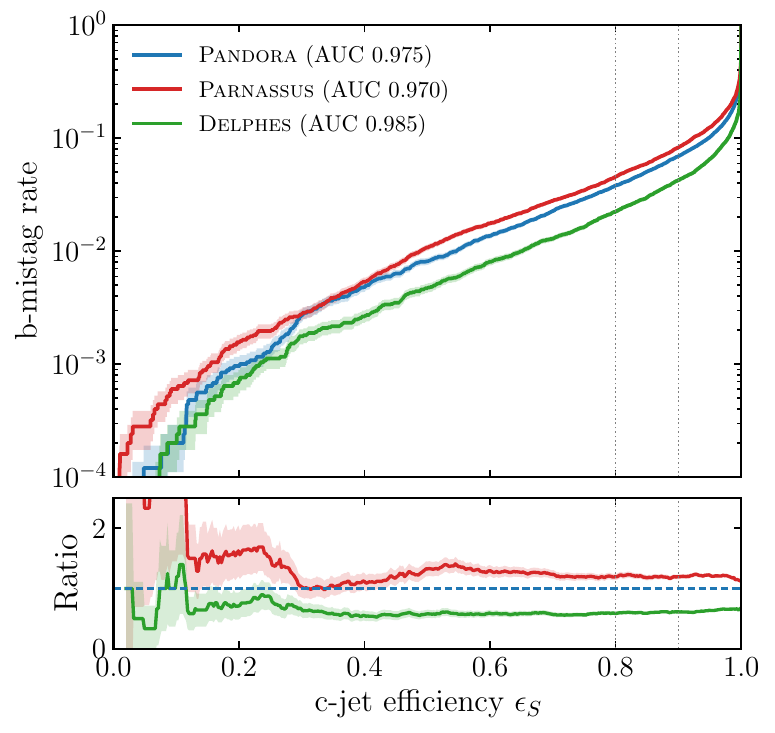}\end{subfigure}
\caption{$c$-tagging: background mis-tag rate versus $c$-jet efficiency against
light-quark, $s$, and $b$ backgrounds. Conventions as in Fig.~\ref{fig:roc_b}.}
\label{fig:roc_c}
\end{figure}

\begin{figure}
\centering
\begin{subfigure}{0.32\linewidth}\includegraphics[width=\linewidth]{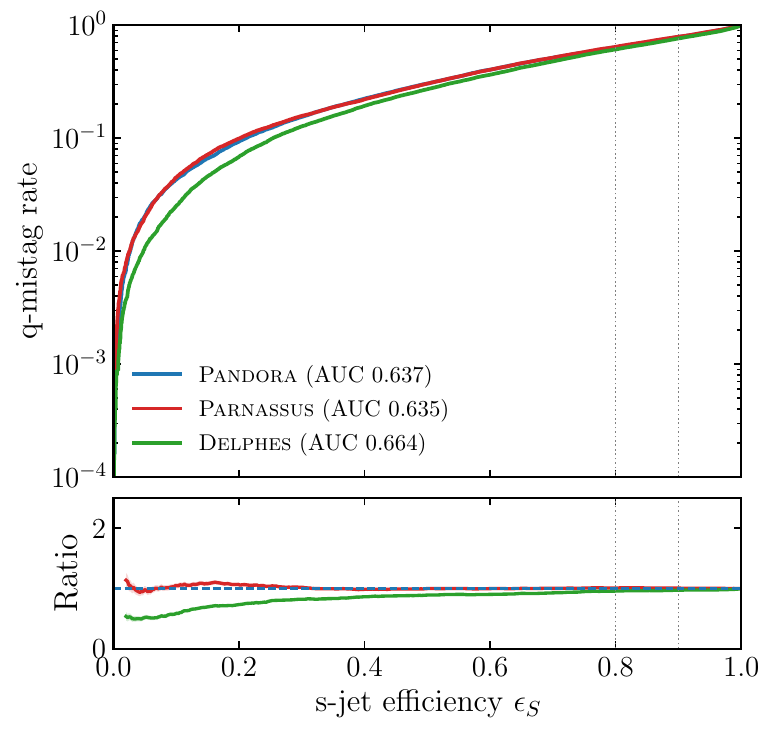}\end{subfigure}
\begin{subfigure}{0.32\linewidth}\includegraphics[width=\linewidth]{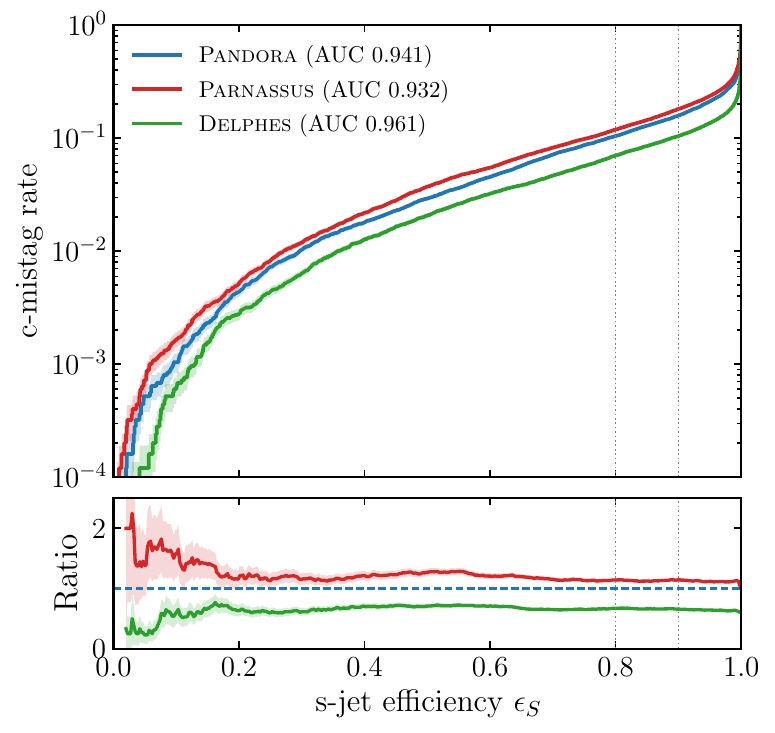}\end{subfigure}
\begin{subfigure}{0.32\linewidth}\includegraphics[width=\linewidth]{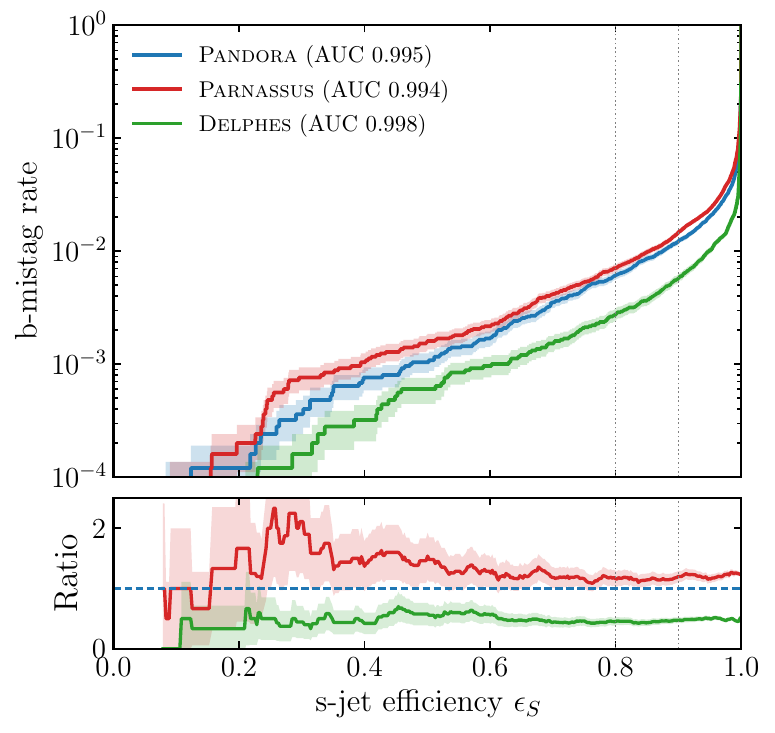}\end{subfigure}
\caption{$s$-tagging: background mis-tag rate versus $s$-jet efficiency against
light-quark, $c$, and $b$ backgrounds. Conventions as in Fig.~\ref{fig:roc_b}.}
\label{fig:roc_s}
\end{figure}

\begin{figure}
\centering
\begin{subfigure}{0.32\linewidth}\includegraphics[width=\linewidth]{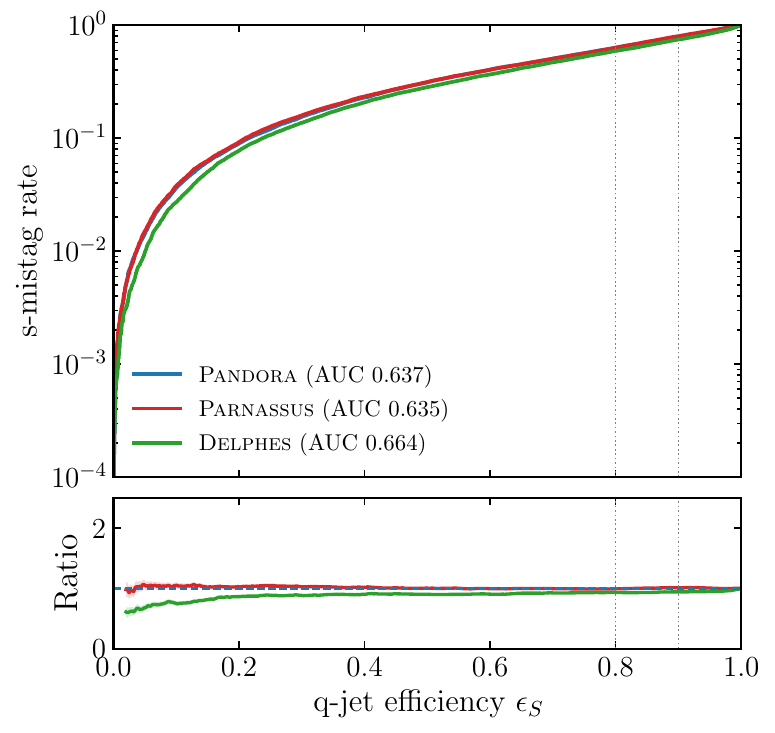}\end{subfigure}
\begin{subfigure}{0.32\linewidth}\includegraphics[width=\linewidth]{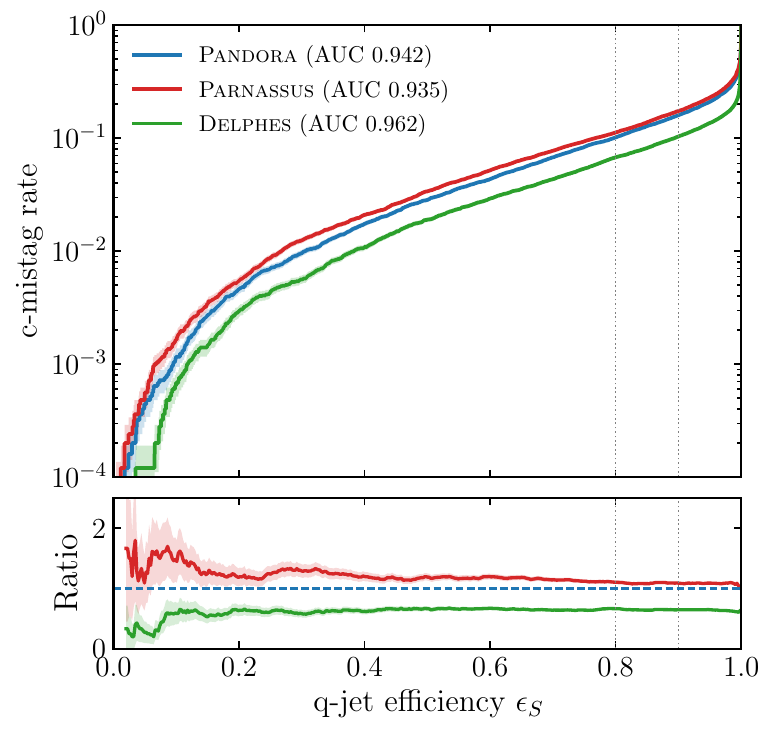}\end{subfigure}
\begin{subfigure}{0.32\linewidth}\includegraphics[width=\linewidth]{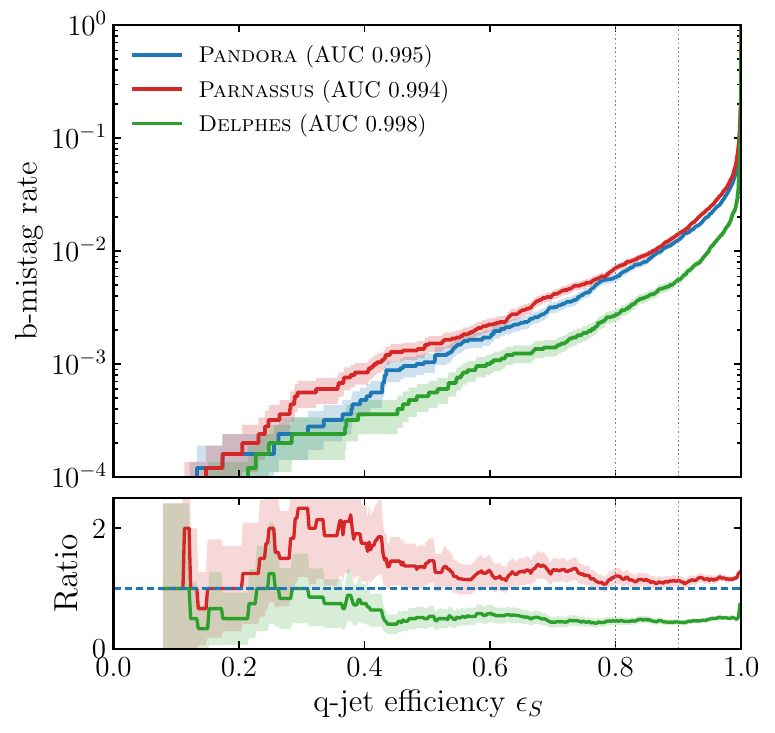}\end{subfigure}
\caption{Light-quark ($q$) tagging: background mis-tag rate versus $q$-jet
efficiency against $s$, $c$, and $b$ backgrounds. Conventions as in
Fig.~\ref{fig:roc_b}.}
\label{fig:roc_q}
\end{figure}

Figures~\ref{fig:roc_b}-\ref{fig:roc_q} show the binary ROC curves for the four
target flavors, each against its three backgrounds, with a ratio-to-{\pandora}
sub-panel. The {\parnassus} curves lie close to {\pandora} throughout, confirming
that a tagger trained on surrogate jets selects the same working points and
reaches comparable purity as one trained on full simulation; the largest residual
differences appear in $s$-tagging (Fig.~\ref{fig:roc_s}), the hardest and most
particle-identification-driven case.

\begin{table}[htbp]
\centering
\caption{Flavor-tagging performance per reconstruction: four-class accuracy and
one-vs-rest AUC for each target flavor. The baseline closest to \textsc{Pandora} is the best performing.}
\label{tab:ftag_auc}
\smallskip
\def\arraystretch{1.3}
\begin{tabular}{c c c c c c}
\hline
\textbf{Reconstruction} & \textbf{Accuracy} & $q$ & $s$ & $c$ & $b$ \\
\hline
\textsc{Pandora}   & 0.693 & 0.843 & 0.842 & 0.934 & 0.986 \\
\textsc{Parnassus} & 0.684 & 0.841 & 0.836 & 0.924 & 0.984 \\
\textsc{Delphes}   & 0.732 & 0.864 & 0.861 & 0.958 & 0.991 \\
\hline
\end{tabular}
\end{table}

\begin{table}[htbp]
\centering
\caption{Background mis-tag rate (\%) at signal efficiency $\eps_S=80\%\,/\,90\%$
for the key flavor discriminations, per reconstruction. The baseline closest to \textsc{Pandora} is the best performing.}
\label{tab:ftag_mistag}
\smallskip
\def\arraystretch{1.3}
\begin{tabular}{c c c c}
\hline
\textbf{Discrimination} & \textbf{\textsc{Pandora}} & \textbf{\textsc{Parnassus}} & \textbf{\textsc{Delphes}} \\
\hline
$b$ vs $c$ & $1.00\,/\,5.23$ & $1.58\,/\,7.68$ & $0.24\,/\,2.08$ \\
$b$ vs $q$ & $0.05\,/\,0.24$ & $0.10\,/\,0.32$ & $0.02\,/\,0.16$ \\
$c$ vs $b$ & $3.80\,/\,6.92$ & $4.56\,/\,8.25$ & $2.23\,/\,4.25$ \\
$s$ vs $q$ & $63.7\,/\,78.4$ & $64.2\,/\,79.0$ & $60.8\,/\,76.2$ \\
\hline
\end{tabular}
\end{table}

\subsection{Cross-reconstruction transfer}
\label{sec:ftag_transfer}
The taggers of the previous subsection are each trained and evaluated on their own respective samples. In this subsection, we present a complementary test of whether a tagger trained on full simulation can be applied unchanged to the
surrogate. We therefore train the tagger once on {\pandora}, freeze it, and
evaluate it on the {\pandora}, {\delphes} and {\parnassus} test sets without any retuning. Table~\ref{tab:ftag_transfer} reports the resulting
accuracy and per-flavor AUC.

The {\pandora}-trained tagger transfers to {\parnassus} with a $0.012$ drop
in accuracy from $0.694$ to $0.682$ and per-flavor AUC changes at the $10^{-3}$-$10^{-2}$
level, confirming that surrogate jets are close enough to full simulation to be
tagged correctly by the very same model. It is also worth noting that these numbers are comparable to a tagger trained on the \parnassus{} samples in Table \ref{tab:ftag_auc}.

Applied to {\delphes}, the same {\pandora}-trained tagger achieves an accuracy of $0.718$ and one-vs-rest AUCs of $0.859$, $0.854$, $0.949$, and $0.989$ for $q$, $s$, $c$, and $b$, respectively, exceeding the corresponding full-simulation values. The accuracy is $0.024$ above the {\pandora} reference, with AUC increases of $0.016$, $0.013$, $0.015$, and $0.003$, respectively. Compared with the tagger trained and evaluated on {\delphes} in Table~\ref{tab:ftag_auc}, the transferred tagger has a lower accuracy ($0.718$ versus $0.732$) and lower AUCs for every flavor, but the stronger discrimination relative to {\pandora} persists. Thus, transfer reduces the excess tagging performance of {\delphes}, but without eliminating it. In absolute differences from the {\pandora} reference, {\parnassus} is closer in accuracy and in the $q$, $s$, and $c$ AUCs, while the $b$ AUC differs by the same magnitude ($0.003$) for both models at the quoted precision.

The interpretation of this test requires care when comparing samples with different flavor separation. As shown in Table~\ref{tab:ftag_auc}, the taggers trained and evaluated on each reconstruction indicate that {\delphes} overestimates the separability of the $b/c/s/q$ flavors relative to {\pandora}, while {\parnassus} slightly underestimates it. A tagger trained exclusively on {\pandora} is not necessarily optimal for either alternative reconstruction, so transferring it will reduce its accuracy and AUC relative to a tagger trained on the corresponding samples. Such a reduction can partially compensate for the excess separation in {\delphes}, bringing its tagging metrics closer to the reference, while increasing the existing deficit in {\parnassus}. Conversely, higher absolute tagging performance on {\delphes} can reflect an easier classification task rather than a more faithful detector response. For surrogate validation, the relevant criterion is reproduction of the {\pandora} response, including its flavor ambiguities. The transfer metrics should therefore be interpreted alongside the distribution-level comparisons, since apparent agreement can arise from compensating effects.

\begin{table}[htbp]
\centering
\caption{Results of the cross-reconstruction transfer test: a single tagger trained on {\pandora} and
inference (four-class accuracy and
one-vs-rest AUC per flavor) is additionally performed on the {\parnassus} and {\delphes} samples.
1$\sigma$ uncertainties computed over a $1000$-replica jet-level bootstrap
of the $10^{5}$-jet test set are $\approx 0.001$ on every entry and are omitted.}
\label{tab:ftag_transfer}
\smallskip
\def\arraystretch{1.3}
\begin{tabular}{c c c c c c}
\hline
\textbf{Test set} & \textbf{Accuracy} & $q$ & $s$ & $c$ & $b$ \\
\hline
\textsc{Pandora} (in-domain) & 0.694 & 0.843 & 0.841 & 0.934 & 0.986 \\
\textsc{Parnassus}           & 0.682 & 0.840 & 0.836 & 0.924 & 0.983 \\
\textsc{Delphes}             & 0.718 & 0.858 & 0.854 & 0.949 & 0.989 \\
\hline
\end{tabular}
\end{table}

\section{Computational performance}
\label{sec:res_timing}

\begin{figure}
\centering
\includegraphics[width=\linewidth]{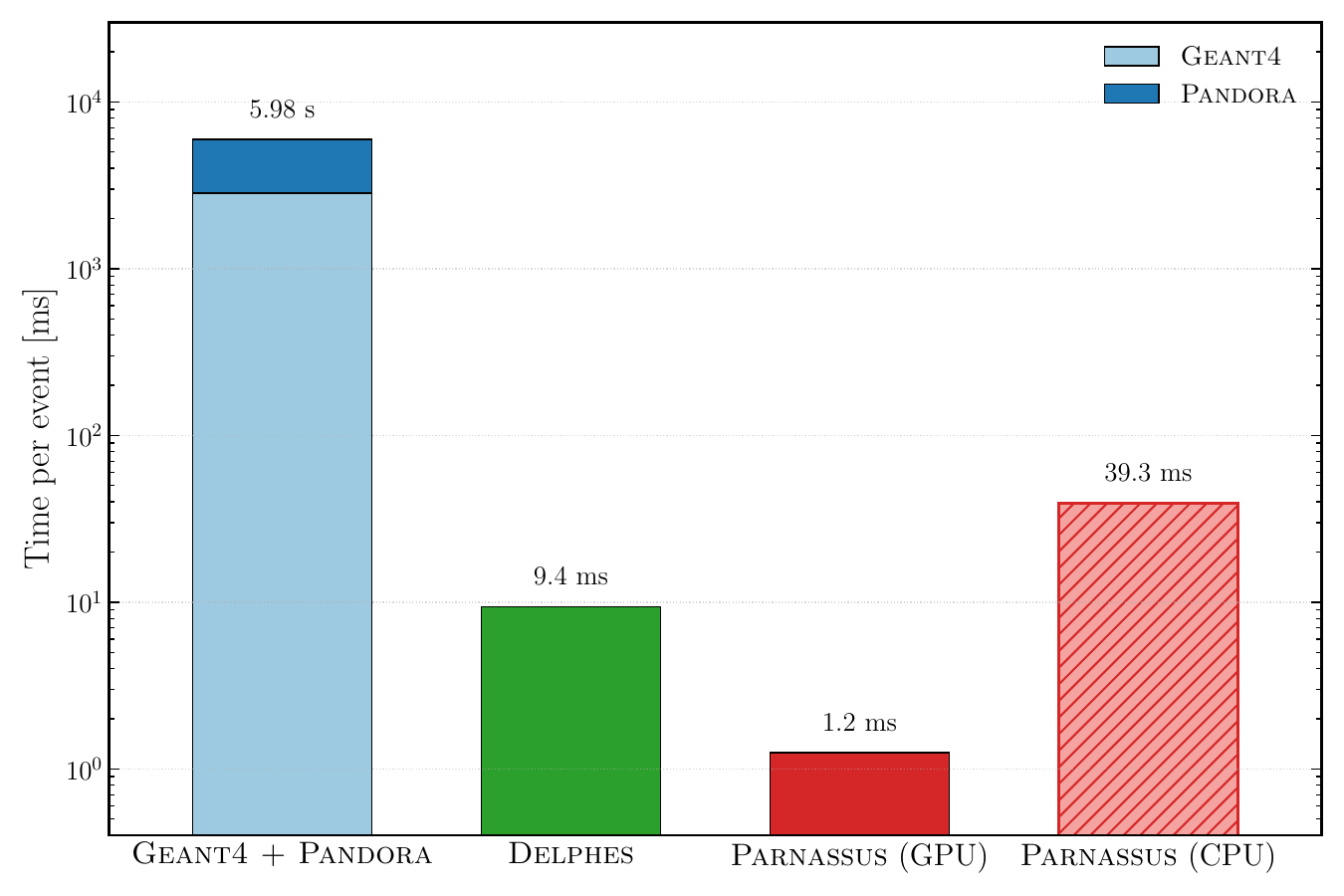}
\caption{Average per-event generation time for the full {\gfour} simulation plus
{\pandora} reconstruction (stacked into its simulation and reconstruction parts),
{\delphes}, and the {\parnassus} surrogate on a single GPU and CPU core. Full simulation, reconstruction, and {\delphes} are timed on a
single AMD EPYC~7713 (Milan) core; the {\parnassus} GPU point uses one
NVIDIA~A100.}
\label{fig:timing}
\end{figure}

We benchmark the per-event generation cost of {\parnassus} against the full
simulation and reconstruction chain and the results are collected in Fig.~\ref{fig:timing}. A full {\gfour} simulation and reconstruction of the CLD response costs about $6.0$~seconds per event on a
single AMD EPYC~7713 (Milan) core, split in roughly equal parts between the
{\gfour} detector simulation ($2.8$~s) and the {\pandora} particle-flow
reconstruction ($3.1$~s). The {\delphes} parametric fast simulation costs
$9.4$~ms per event on the same hardware.

To isolate the cost of the methods themselves, we first run {\parnassus} on the
same single Milan CPU core and it generates
one event in about $39$~ms using a batch size of 512; which is already two orders of magnitude faster than full
simulation. The GPU support of {\parnassus}
improves this substantially. On a single NVIDIA~A100 GPU using the same batch size, {\parnassus} takes about
$1.3$~ms per event, a speedup of over three orders of magnitude over the reference and faster than {\delphes} as well. The training of the model, a one-time cost that is not repeated during
sampling, completes in well under 12 hours on a single A100 and is itself far
cheaper than simulating the training sample with {\gfour}. The above quoted numbers constitute the full end-to-end chain and include I/O, reading, processing, writing of the relevant output files, and GPU transfer time if using GPU-acceleration for \textsc{Parnassus}.

\section{Extension to the \texorpdfstring{{\hitpf}}{HitPF} reconstruction}
\label{sec:hitpf}

Because the surrogate conditions only on generator-level input and targets
whatever reconstruction it is trained against, the same framework applies without
modification to alternative reconstruction algorithms. To demonstrate this we
additionally consider {\hitpf}~\cite{HitPF:2026}, a hit-level machine-learned
particle-flow algorithm implemented in {\keyfourhep}. {\hitpf} proceeds in two
learned stages: a condensation network embeds every calorimeter hit and track into
a learned space in which density-peak clustering groups constituents belonging to
the same particle, and a regression/identification network then assigns an energy
and a species to each resulting shower. Since {\hitpf} consumes the same
simulated hits and tracks as {\pandora} and emits a collection of the same EDM4hep
type, it enters our pipeline as a drop-in replacement for the target collection. As such,
the representation of Sec.~\ref{sec:dataset}, the architecture of
Sec.~\ref{sec:method}, and the training configuration are all unchanged, with the only difference being the target distribution.

\begin{figure}[htbp]
\centering
\begin{subfigure}{0.32\textwidth}\includegraphics[width=\linewidth]{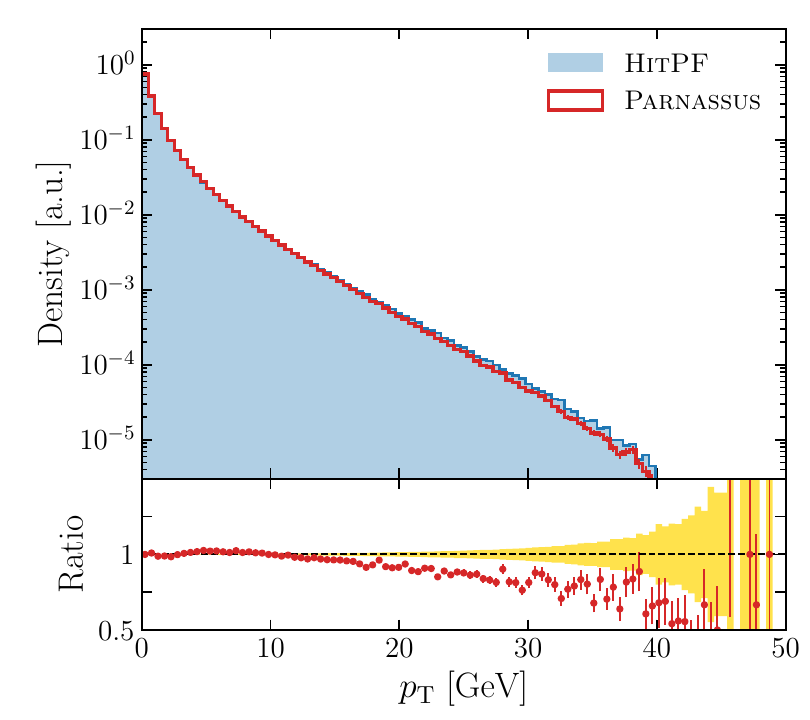}\end{subfigure}
\begin{subfigure}{0.32\textwidth}\includegraphics[width=\linewidth]{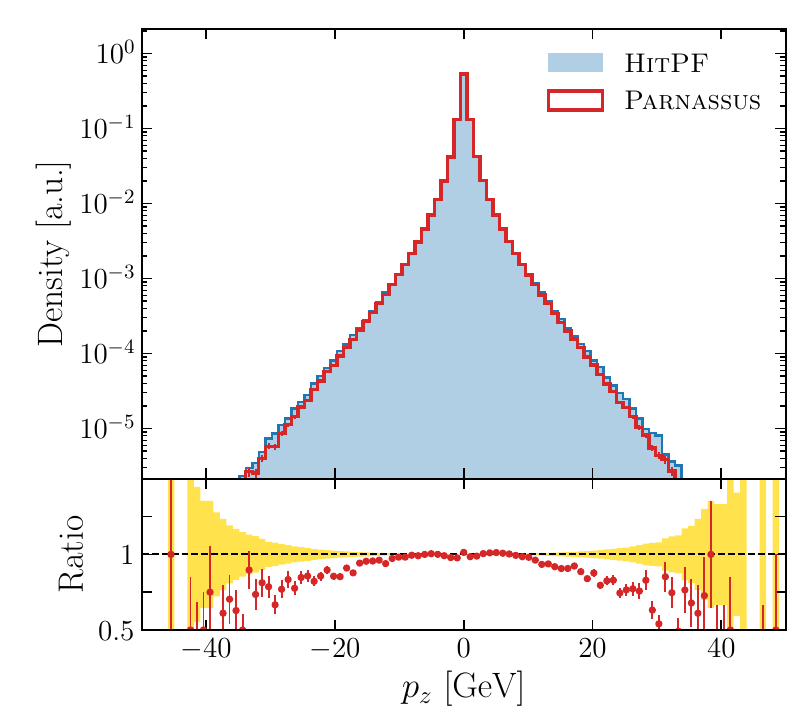}\end{subfigure}
\begin{subfigure}{0.32\textwidth}\includegraphics[width=\linewidth]{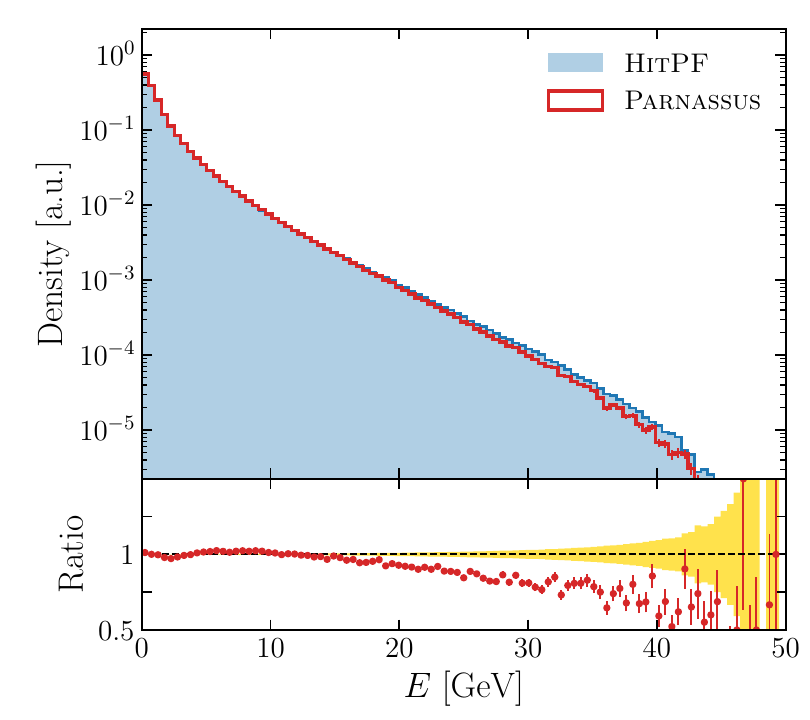}\end{subfigure}
\begin{subfigure}{0.32\textwidth}\includegraphics[width=\linewidth]{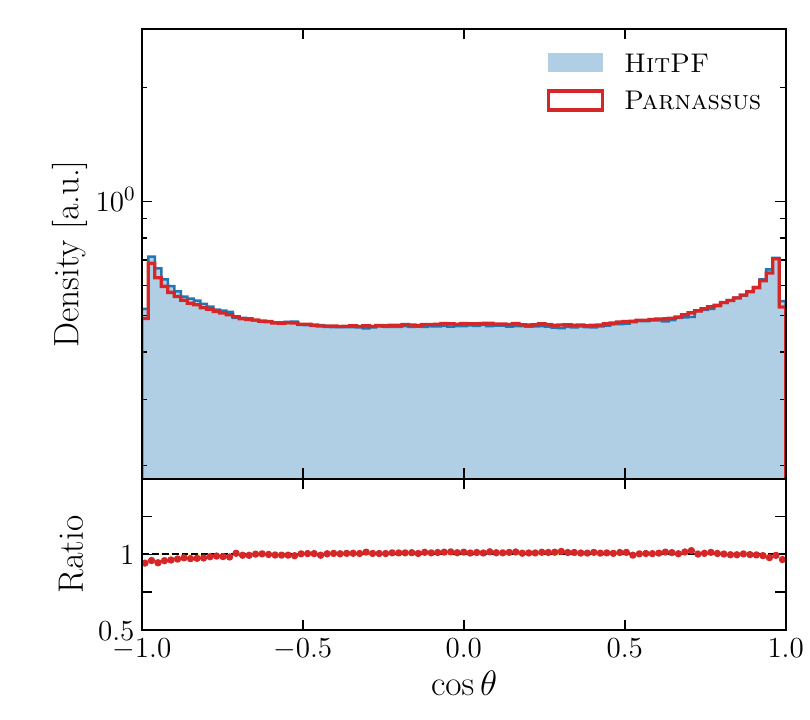}\end{subfigure}
\begin{subfigure}{0.32\textwidth}\includegraphics[width=\linewidth]{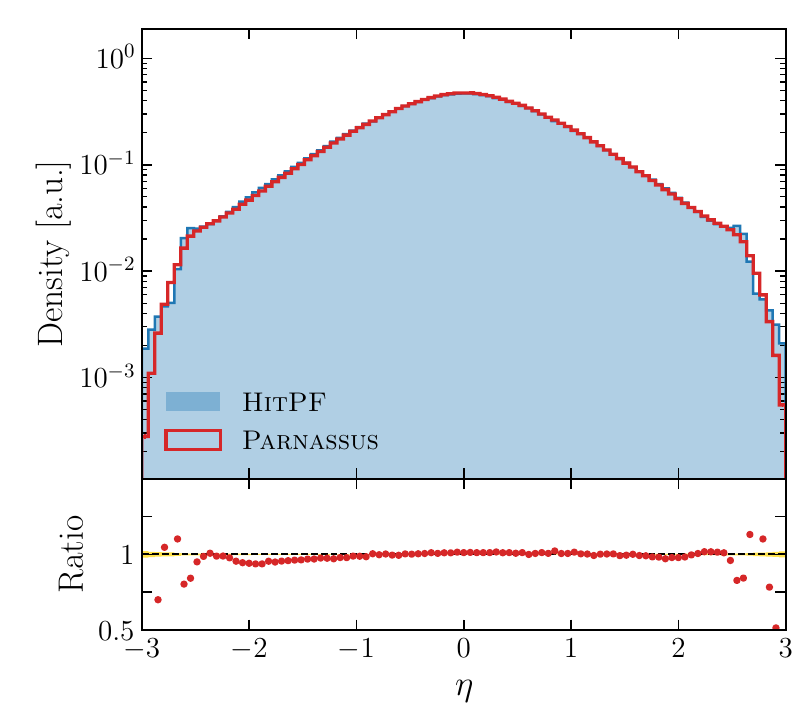}\end{subfigure}
\begin{subfigure}{0.32\textwidth}\includegraphics[width=\linewidth]{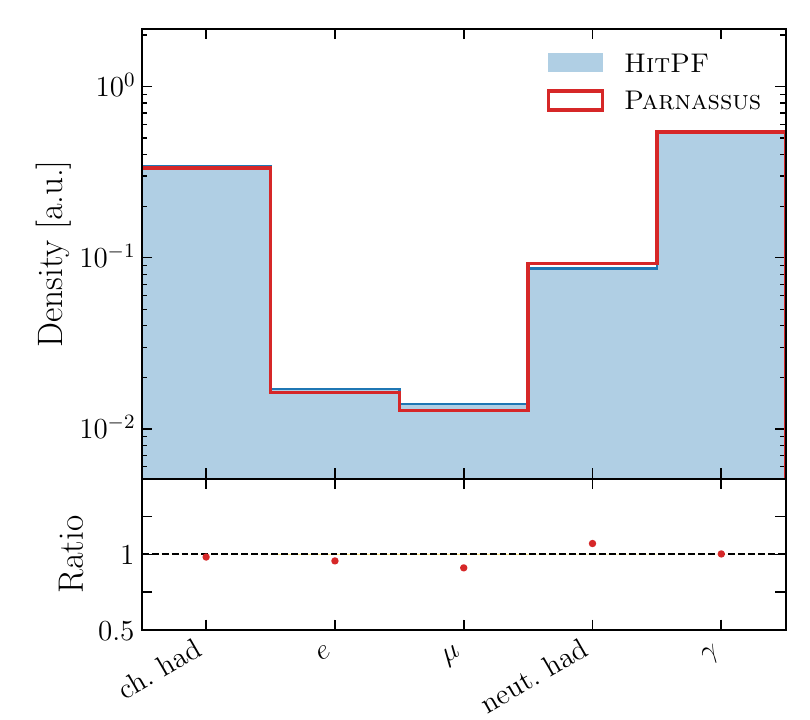}\end{subfigure}
\caption{Particle-level kinematics and species composition for the {\hitpf}
reconstruction: {\hitpf} (blue filled), {\parnassus} (red). Upper panels show
unit-normalized densities; lower panels the ratio to {\hitpf} with its
statistical uncertainty as a band.}
\label{fig:hitpf_kin}
\end{figure}

\begin{figure}[htbp]
\centering
\begin{subfigure}{0.49\textwidth}\includegraphics[width=\linewidth]{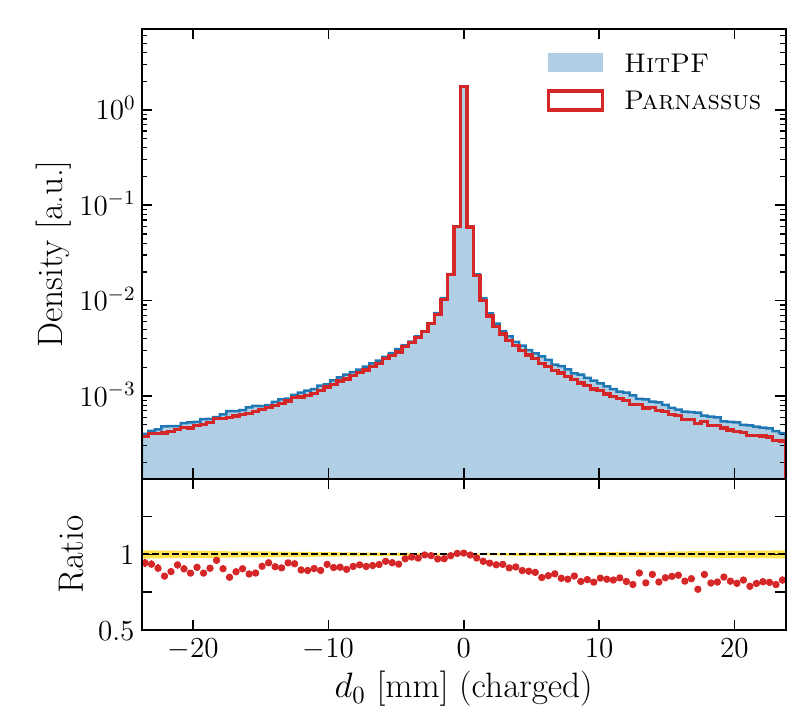}\end{subfigure}
\begin{subfigure}{0.49\textwidth}\includegraphics[width=\linewidth]{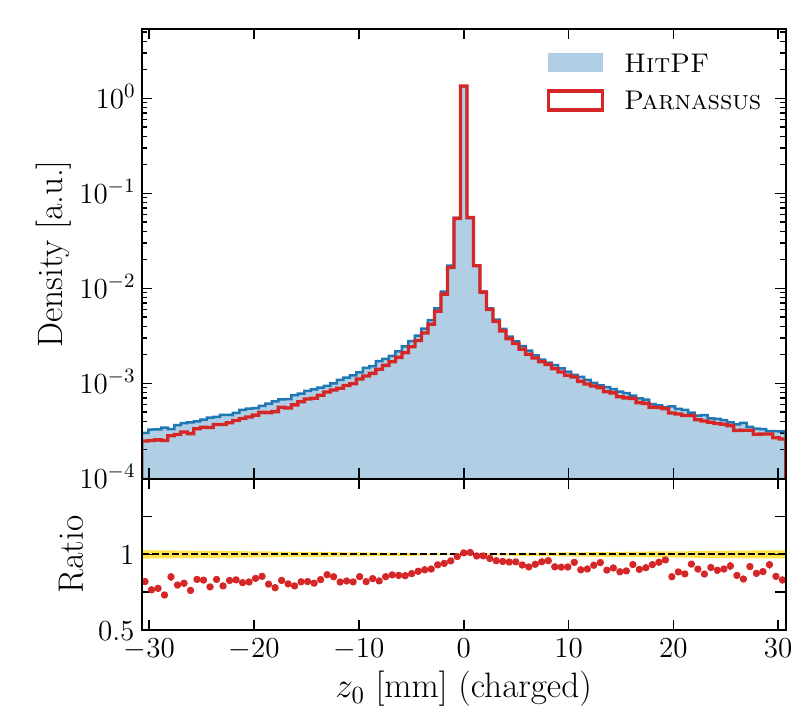}\end{subfigure}\\[2pt]
\begin{subfigure}{0.49\textwidth}\includegraphics[width=\linewidth]{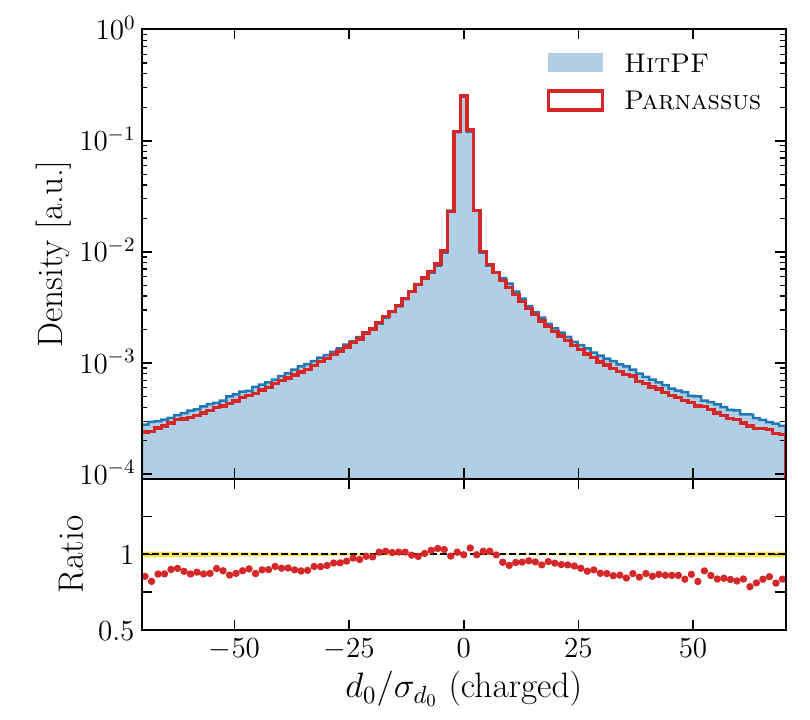}\end{subfigure}
\begin{subfigure}{0.49\textwidth}\includegraphics[width=\linewidth]{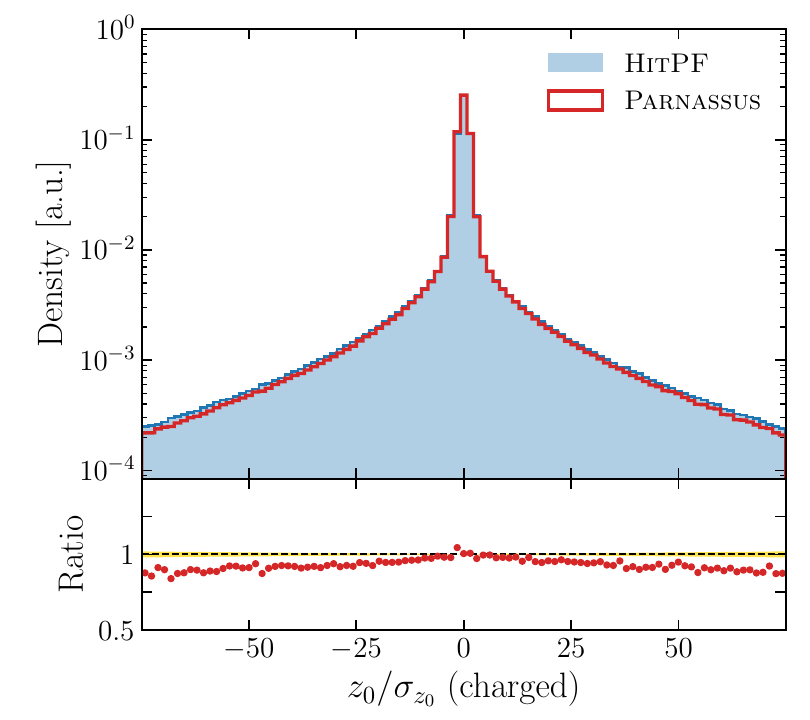}\end{subfigure}
\caption{Track impact parameters (for charged PFOs) shown for the {\hitpf} reconstruction:
transverse $\dzero$ and longitudinal $\zzero$ and their significances. Conventions as in Fig.~\ref{fig:hitpf_kin}.}
\label{fig:hitpf_ip}
\end{figure}

\begin{figure}[htbp]
\centering
\begin{subfigure}{0.32\textwidth}\includegraphics[width=\linewidth]{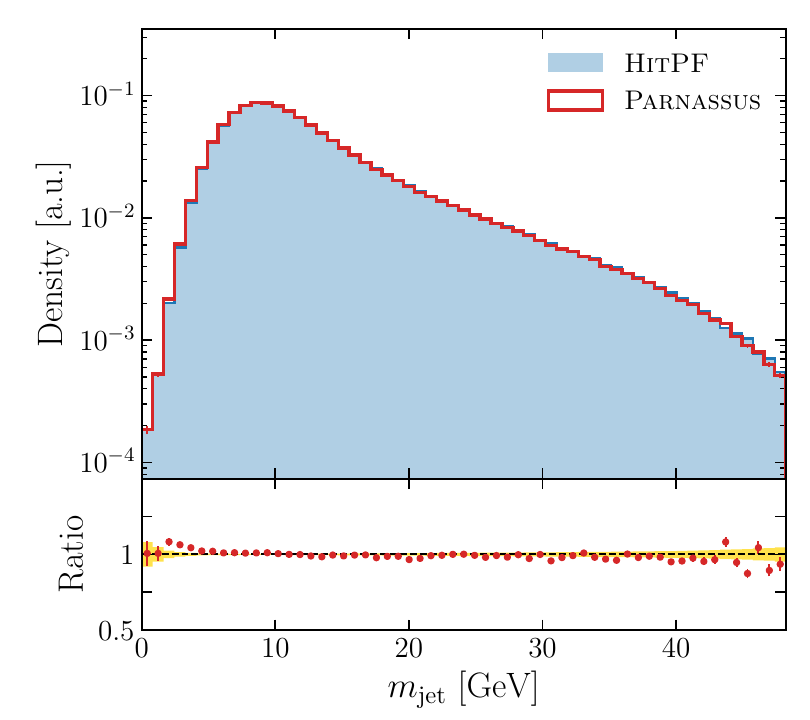}\end{subfigure}
\begin{subfigure}{0.32\textwidth}\includegraphics[width=\linewidth]{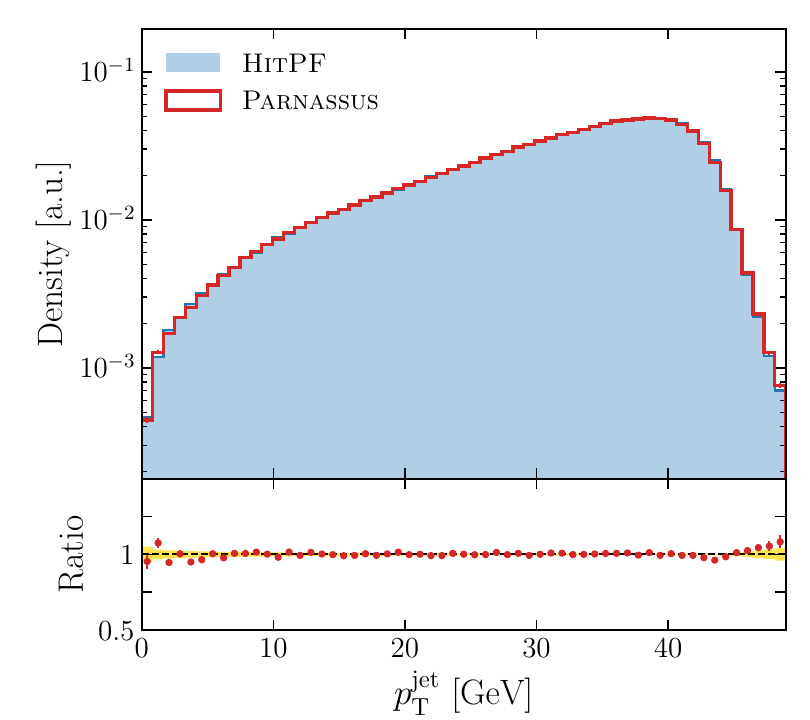}\end{subfigure}
\begin{subfigure}{0.32\textwidth}\includegraphics[width=\linewidth]{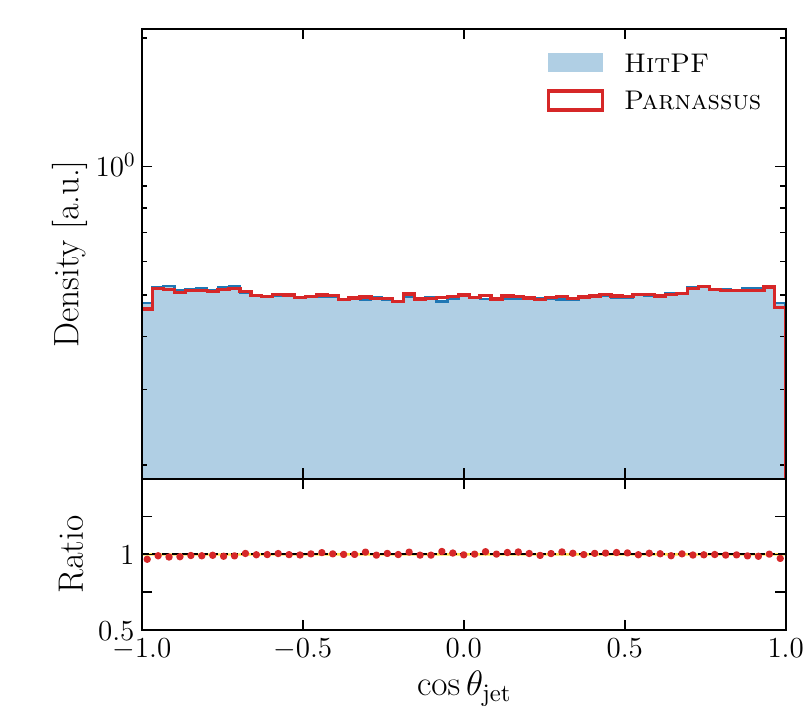}\end{subfigure}
\begin{subfigure}{0.32\textwidth}\includegraphics[width=\linewidth]{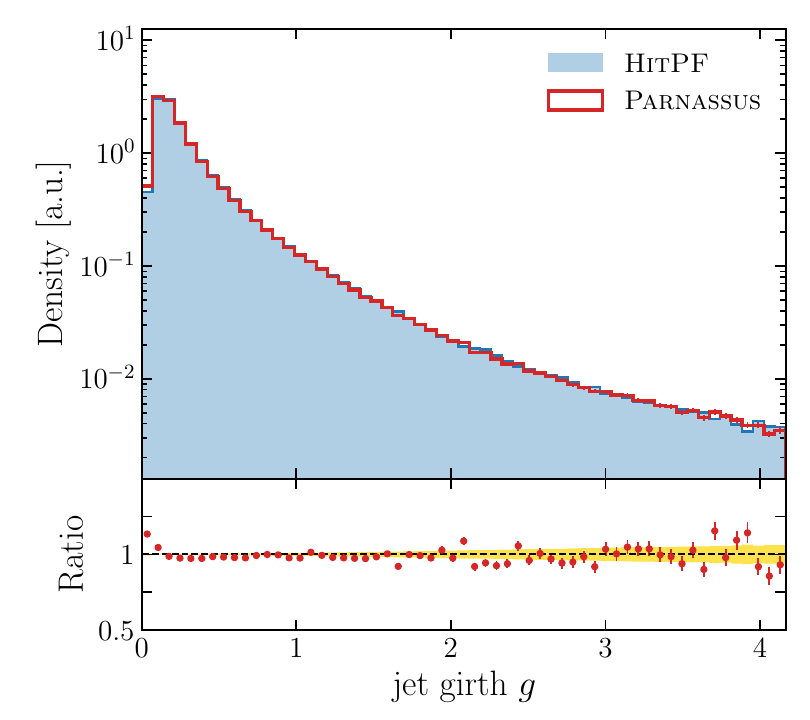}\end{subfigure}
\begin{subfigure}{0.32\textwidth}\includegraphics[width=\linewidth]{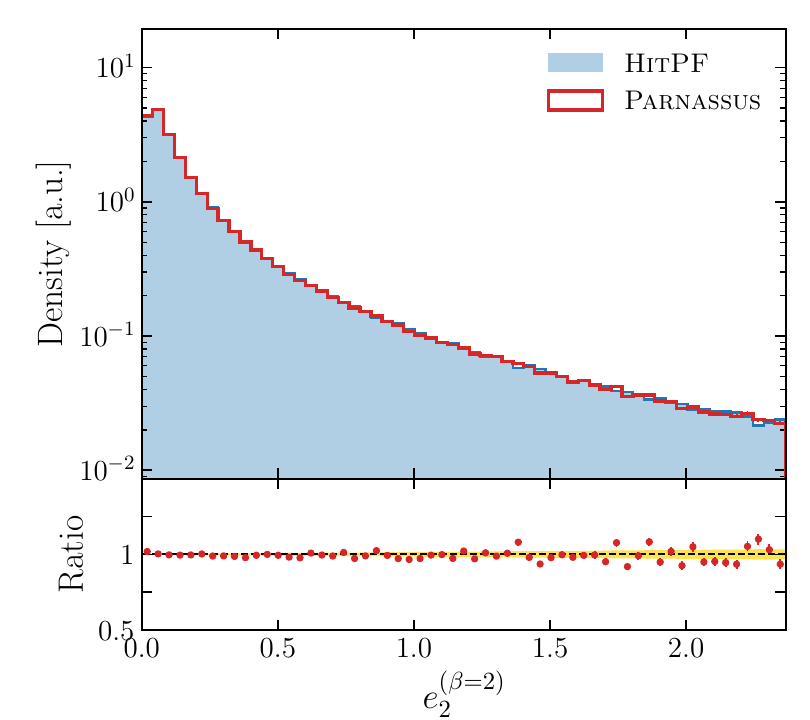}\end{subfigure}
\begin{subfigure}{0.32\textwidth}\includegraphics[width=\linewidth]{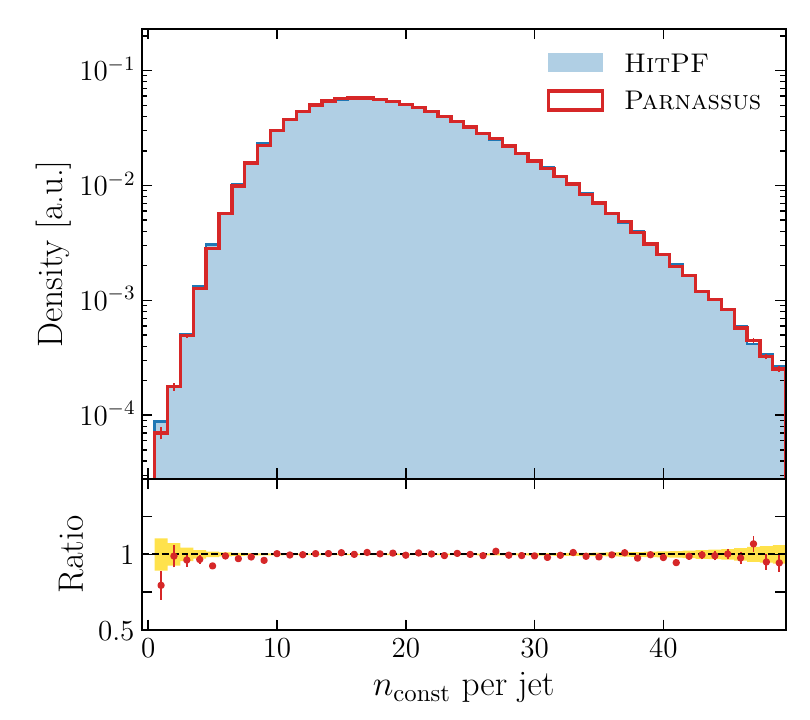}\end{subfigure}
\begin{subfigure}{0.32\textwidth}\includegraphics[width=\linewidth]{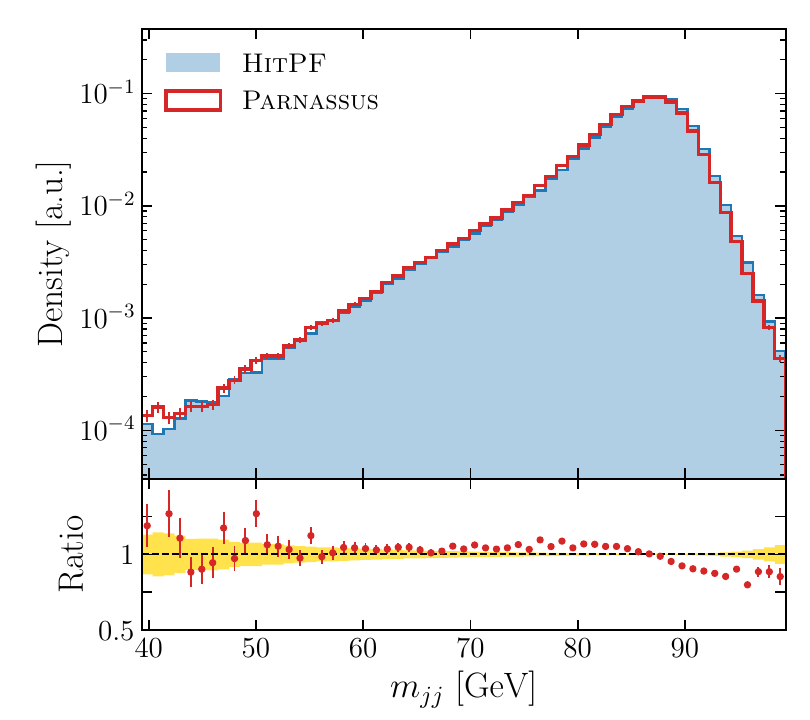}\end{subfigure}
\caption{Jet-level observables for the {\hitpf} reconstruction after Durham
exclusive clustering with $n_{\mathrm{jet}}=2$: jet mass, $\pT$, $\cos\theta$,
girth, $e_2^{(\beta=2)}$, constituent multiplicity, and the dijet invariant mass
$m_{jj}$. {\hitpf} (blue filled), {\parnassus} (red); ratio panels as in
Fig.~\ref{fig:hitpf_kin}.}
\label{fig:hitpf_jets}
\end{figure}

This setup yields the particle and jet-level distributions of Figs.~\ref{fig:hitpf_kin}- \ref{fig:hitpf_jets}. The surrogate reproduces the {\hitpf} reconstruction with roughly the same fidelity it achieves for {\pandora}. Single-particle kinematics, impact species fractions (Fig.~\ref{fig:hitpf_kin}), and after Durham jet clustering, the jet mass, substructure, constituent multiplicity, and the dijet mass (Fig.~\ref{fig:hitpf_jets}) all agree to the percent level across the bulk of every distribution, with the ratio panels flat within their statistical bands. Some deviation from the reference is observed in the charged-track impact parameters and their significances (Fig.~\ref{fig:hitpf_ip}). This serves as a proof of concept that the same architecture can be used, without modification or hyperparameter tuning, to build surrogates for both classical and machine-learned particle-flow reconstruction algorithms, requiring only a separate training run for each target.

\section{Discussion and conclusions}
\label{sec:discussion}

We have applied the {\parnassus} generative surrogate to the CLD detector concept
at the FCC-ee, learning the conditional density of {\pandora} particle-flow
objects given stable generator-level particles, with a flow-matching objective and a
cross-attention diffusion transformer. On a held-out sample of 500k
$Z\to q\bar q,s\bar s,c\bar c,b\bar b$ events the surrogate reproduces
single-particle kinematics, species fractions, and impact-parameter
distributions, and after Durham clustering, jet mass, energy, substructure, and
dijet-mass observables. Quantified
with the Vincze-Le Cam, Wasserstein, and Jensen-Shannon distances, inclusively
and by flavor, {\parnassus} is consistently an order of magnitude closer to full
simulation than the {\delphes} baseline in the detector- and algorithm-specific
observables.

Finally, a transformer flavor tagger trained on the surrogate output
preserves the $b/c/s/q$ discrimination of the full CLD reconstruction: the
per-flavor ROC curves, AUCs, and mis-tag rates track {\pandora} closely, so that a
downstream tagging analysis would reach the same physics conclusions on surrogate
and full-simulation samples. Because the surrogate conditions only on
generator-level input and targets whatever reconstruction it is trained against,
the same framework applies unchanged to alternative reconstruction algorithms,
demonstrated in outline for the hit-level {\hitpf} in Sec.~\ref{sec:hitpf}, and
to other downstream analyses. Similarly, the automated {\parnassus} workflow also makes it straightforward to update the surrogate as the CLD detector concept evolves. Changes to the detector geometry can be accommodated by producing an updated full-simulation reference sample and rerunning the training and validation pipeline to obtain a new model.

A natural extension of this work is to condition Parnassus explicitly on detector geometry and response parameters, such as layer positions, material budgets, calorimeter segmentation, and magnetic-field strength. Training on full-simulation samples spanning different configurations would allow a single surrogate to describe a family of detector designs at inference time and enable a detector-agnostic framework within the validated design space. This would enable rapid exploration of how detector choices affect reconstructed observables and downstream physics performance, supporting detector optimization studies. Furthermore, the differentiable nature of the continuous generative mapping could enable gradient-based optimization of continuous detector parameters, provided that downstream objectives are differentiable and discrete outputs are treated appropriately. Such an approach would connect detector design directly to physics objectives, including jet energy resolution and flavor discrimination, while reducing the computational cost of repeated full simulation and reconstruction.

Another interesting direction for future work is the integration of {\parnassus} into the {\keyfourhep} software stack. The trained neural network could be exported to the \textsc{Onnx} format, similar to \textsc{HitPF}. Coupling this interface to the sampling procedure and the conversion of generated particles into EDM4hep or \textsc{HepMC3} collections would make the surrogate directly accessible to analyses using {\keyfourhep}.

Together, these results establish {\parnassus} CLD as a fast and accurate surrogate for the CLD simulation and reconstruction chain, enabling larger simulated samples to support the physics program of the FCC-ee.

\appendix
\section{Flavor-split particle-level distributions}
\label{app:flavor}

Figures~\ref{fig:flavor_kin} and~\ref{fig:flavor_ip} reproduce every particle-level
observable of Sec.~\ref{sec:res_particle} separately for each $Z$ decay flavor
($q\bar q$, $s\bar s$, $c\bar c$, $b\bar b$), grouped into kinematic/species
observables (Fig.~\ref{fig:flavor_kin}) and track impact parameters
(Fig.~\ref{fig:flavor_ip}); each is quantified inclusively in
table~\ref{tab:pdist} and by flavor in
Tables~\ref{tab:flavor_qq}-\ref{tab:flavor_bb}. The kinematic and species
distributions are reproduced by the surrogate in every channel, while the
impact-parameter observables and their significances develop progressively longer
displaced tails from $q\bar q$ to $b\bar b$; {\parnassus} tracks this evolution,
whereas {\delphes} departs from CLD by an amount that grows with flavor.

\input{flavor_appendix.tex}

\acknowledgments
We thank Dmitrii Kobylianskii and Gregor Krzmanc for insightful discussions. This work is supported by the U.S. Department of Energy (DOE) under Contract No. DE-AC02-76SF00515 and by the CompHEP Center for Computational Excellence. This research used resources of the National Energy Research Scientific Computing Center (NERSC), a DOE Office of Science User Facility supported by the Office of Science of the U.S. Department of Energy under Contract No.~DE-AC02-05CH11231, under NERSC award HEP-ERCAP0035546. This work used the resources of the SLAC Shared Science Data Facility (S3DF) at SLAC National Accelerator Laboratory. SLAC is operated by Stanford University for the U.S. Department of Energy's Office of Science. USQ is additionally supported by a Stanford Knight-Hennessy Fellowship.


\section*{Data and Code Availability}
All \textsc{Parnassus} source code, including the trained model weights, is openly available on \href{https://github.com/parnassus-hep/parnassus}{GitHub}. \href{https://parnassus-hep.github.io/parnassus/}{Detailed documentation} is also provided to facilitate reproducibility and community use. The dataset used to conduct this study is also made publicly available on \href{https://huggingface.co/datasets/umarsqur/cld-fcc-ee-fullsim-rec}{Hugging Face} \cite{umar_sohail_qureshi_2026}.

\section*{AI Usage Statement}
We acknowledge the use of Claude Code (Anthropic) for generating and orchestrating a portion of the code used
for this study, and for help editing the manuscript. We have verified and take full
responsibility for the content of this publication.


\bibliographystyle{JHEP}
\bibliography{references}

\end{document}

%% file: flavor_tables.tex
\begin{table}[htbp]
\centering
\caption{Particle-level distribution distances between the surrogate and {\pandora} for the $q\bar q$ channel: Vincze-Le Cam divergence ($\times10^{2}$), $W_1$, and Jensen--Shannon distance. Conventions as in Table~\ref{tab:pdist}.}
\label{tab:flavor_qq}
\smallskip
\def\arraystretch{1.3}
\begin{tabular}{c | c c c | c c c}
\hline
 & \multicolumn{3}{c|}{\textbf{\textsc{Parnassus}}} & \multicolumn{3}{c}{\textbf{\textsc{Delphes}}} \\
\textbf{~Observable~} & VLC & $W_1$ & JS & VLC & $W_1$ & JS \\
\hline
$\pT$ & \textbf{0.011(1)} & \textbf{0.018(1)} & \textbf{0.006} & 0.126(3) & 0.042(1) & 0.021 \\
$p_z$ & \textbf{0.005(1)} & \textbf{0.017(1)} & \textbf{0.004} & 0.137(3) & 0.038(1) & 0.022 \\
$E$ & \textbf{0.016(1)} & \textbf{0.024(2)} & \textbf{0.008} & 1.074(9) & 0.077(1) & 0.062 \\
$\cos\theta$ & \textbf{0.016(1)} & \textbf{0.005} & \textbf{0.008} & 0.207(4) & 0.016 & 0.027 \\
$\eta$ & \textbf{0.028(1)} & \textbf{0.009} & \textbf{0.010} & 0.318(5) & 0.033 & 0.034 \\
species & \textbf{0.012(1)} & \textbf{0.020(1)} & \textbf{0.007} & 5.978(19) & 0.144(1) & 0.151 \\
$\dzero$ & \textbf{0.038(3)} & \textbf{0.011} & \textbf{0.012} & 1.438(15) & 0.090 & 0.073 \\
$\zzero$ & \textbf{0.023(2)} & \textbf{0.015} & \textbf{0.009} & 1.217(15) & 0.121 & 0.067 \\
$\dzero/\sigma_{\dzero}$ & \textbf{0.054(3)} & \textbf{0.142(3)} & \textbf{0.014} & 1.899(17) & 2.058(3) & 0.084 \\
$\zzero/\sigma_{\zzero}$ & \textbf{0.029(2)} & \textbf{0.182(4)} & \textbf{0.010} & 2.358(18) & 2.182(4) & 0.094 \\
\hline
\end{tabular}
\end{table}

\begin{table}[htbp]
\centering
\caption{Particle-level distribution distances between the surrogate and {\pandora} for the $s\bar s$ channel: Vincze--Le Cam divergence ($\times10^{2}$), $W_1$, and Jensen--Shannon distance. Conventions as in table~\ref{tab:pdist}.}
\label{tab:flavor_ss}
\smallskip
\def\arraystretch{1.3}
\begin{tabular}{c | c c c | c c c}
\hline
 & \multicolumn{3}{c|}{\textbf{\textsc{Parnassus}}} & \multicolumn{3}{c}{\textbf{\textsc{Delphes}}} \\
\textbf{~Observable~} & VLC & $W_1$ & JS & VLC & $W_1$ & JS \\
\hline
$\pT$ & \textbf{0.013(1)} & \textbf{0.022(1)} & \textbf{0.007} & 0.159(3) & 0.058(1) & 0.024 \\
$p_z$ & \textbf{0.005(1)} & \textbf{0.019(1)} & \textbf{0.004} & 0.143(3) & 0.043(1) & 0.023 \\
$E$ & \textbf{0.018(1)} & \textbf{0.028(2)} & \textbf{0.008} & 1.131(9) & 0.097(1) & 0.064 \\
$\cos\theta$ & \textbf{0.017(1)} & \textbf{0.005} & \textbf{0.008} & 0.216(4) & 0.017 & 0.028 \\
$\eta$ & \textbf{0.031(2)} & \textbf{0.009} & \textbf{0.011} & 0.323(5) & 0.034 & 0.034 \\
species & \textbf{0.017(1)} & \textbf{0.022(1)} & \textbf{0.008} & 6.431(20) & 0.184(1) & 0.156 \\
$\dzero$ & \textbf{0.038(3)} & \textbf{0.012} & \textbf{0.012} & 1.147(14) & 0.149 & 0.065 \\
$\zzero$ & \textbf{0.025(2)} & \textbf{0.017} & \textbf{0.009} & 0.988(13) & 0.198 & 0.060 \\
$\dzero/\sigma_{\dzero}$ & \textbf{0.049(3)} & \textbf{0.164(3)} & \textbf{0.013} & 2.008(18) & 3.145(3) & 0.087 \\
$\zzero/\sigma_{\zzero}$ & \textbf{0.031(2)} & \textbf{0.208(4)} & \textbf{0.011} & 2.992(21) & 3.110(5) & 0.106 \\
\hline
\end{tabular}
\end{table}

\begin{table}[htbp]
\centering
\caption{Particle-level distribution distances between the surrogate and {\pandora} for the $c\bar c$ channel: Vincze--Le Cam divergence ($\times10^{2}$), $W_1$, and Jensen--Shannon distance. Conventions as in table~\ref{tab:pdist}.}
\label{tab:flavor_cc}
\smallskip
\def\arraystretch{1.3}
\begin{tabular}{c | c c c | c c c}
\hline
 & \multicolumn{3}{c|}{\textbf{\textsc{Parnassus}}} & \multicolumn{3}{c}{\textbf{\textsc{Delphes}}} \\
\textbf{~Observable~} & VLC & $W_1$ & JS & VLC & $W_1$ & JS \\
\hline
$\pT$ & \textbf{0.012(1)} & \textbf{0.017(1)} & \textbf{0.007} & 0.124(3) & 0.037(1) & 0.021 \\
$p_z$ & \textbf{0.006(1)} & \textbf{0.017(1)} & \textbf{0.005} & 0.129(3) & 0.034(1) & 0.022 \\
$E$ & \textbf{0.021(1)} & \textbf{0.025(2)} & \textbf{0.009} & 1.034(9) & 0.069(1) & 0.061 \\
$\cos\theta$ & \textbf{0.015(1)} & \textbf{0.005} & \textbf{0.007} & 0.202(4) & 0.016 & 0.027 \\
$\eta$ & \textbf{0.026(1)} & \textbf{0.009} & \textbf{0.010} & 0.313(5) & 0.033 & 0.034 \\
species & \textbf{0.023(1)} & \textbf{0.023(1)} & \textbf{0.009} & 5.714(19) & 0.155(1) & 0.146 \\
$\dzero$ & \textbf{0.035(3)} & \textbf{0.012} & \textbf{0.011} & 0.734(11) & 0.114 & 0.052 \\
$\zzero$ & \textbf{0.023(2)} & \textbf{0.016} & \textbf{0.009} & 0.740(11) & 0.154 & 0.052 \\
$\dzero/\sigma_{\dzero}$ & \textbf{0.041(3)} & \textbf{0.223(5)} & \textbf{0.012} & 1.197(13) & 3.079(4) & 0.066 \\
$\zzero/\sigma_{\zzero}$ & \textbf{0.024(2)} & \textbf{0.258(6)} & \textbf{0.009} & 1.616(15) & 3.077(6) & 0.077 \\
\hline
\end{tabular}
\end{table}

\begin{table}[htbp]
\centering
\caption{Particle-level distribution distances between the surrogate and {\pandora} for the $b\bar b$ channel: Vincze--Le Cam divergence ($\times10^{2}$), $W_1$, and Jensen--Shannon distance. Conventions as in table~\ref{tab:pdist}.}
\label{tab:flavor_bb}
\smallskip
\def\arraystretch{1.3}
\begin{tabular}{c | c c c | c c c}
\hline
 & \multicolumn{3}{c|}{\textbf{\textsc{Parnassus}}} & \multicolumn{3}{c}{\textbf{\textsc{Delphes}}} \\
\textbf{~Observable~} & VLC & $W_1$ & JS & VLC & $W_1$ & JS \\
\hline
$\pT$ & \textbf{0.011(1)} & \textbf{0.015(1)} & \textbf{0.006} & 0.205(4) & 0.054(1) & 0.027 \\
$p_z$ & \textbf{0.007(1)} & \textbf{0.015(1)} & \textbf{0.005} & 0.192(4) & 0.046(1) & 0.026 \\
$E$ & \textbf{0.022(1)} & \textbf{0.022(1)} & \textbf{0.009} & 1.230(9) & 0.090(1) & 0.067 \\
$\cos\theta$ & \textbf{0.013(1)} & \textbf{0.005} & \textbf{0.007} & 0.178(3) & 0.015 & 0.025 \\
$\eta$ & \textbf{0.025(1)} & \textbf{0.008} & \textbf{0.010} & 0.292(4) & 0.030 & 0.033 \\
species & \textbf{0.050(2)} & \textbf{0.033(1)} & \textbf{0.013} & 5.982(18) & 0.133(1) & 0.150 \\
$\dzero$ & \textbf{0.053(3)} & \textbf{0.016} & \textbf{0.014} & 0.174(5) & 0.098 & 0.025 \\
$\zzero$ & \textbf{0.034(2)} & \textbf{0.023} & \textbf{0.011} & 0.296(7) & 0.128 & 0.033 \\
$\dzero/\sigma_{\dzero}$ & \textbf{0.044(3)} & \textbf{0.413(7)} & \textbf{0.013} & 1.858(18) & 4.936(7) & 0.082 \\
$\zzero/\sigma_{\zzero}$ & \textbf{0.029(2)} & \textbf{0.425(9)} & \textbf{0.010} & 1.849(17) & 5.042(10) & 0.082 \\
\hline
\end{tabular}
\end{table}

%% file: flavor_appendix.tex
\begin{figure}[p]
\centering
\makebox[0.25\linewidth][c]{$q\bar q$}\makebox[0.25\linewidth][c]{$s\bar s$}\makebox[0.25\linewidth][c]{$c\bar c$}\makebox[0.25\linewidth][c]{$b\bar b$}\\
\includegraphics[width=0.245\textwidth]{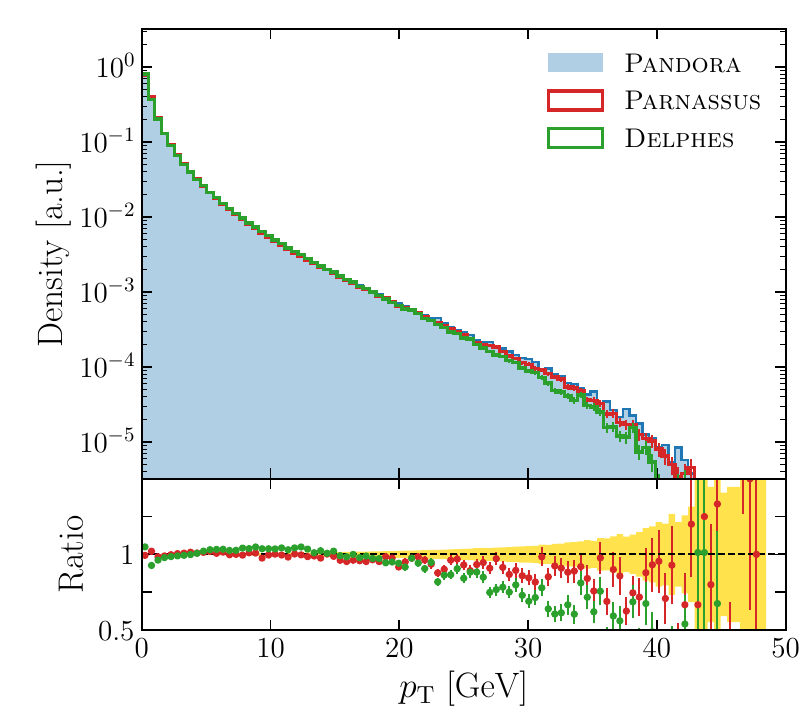}\includegraphics[width=0.245\textwidth]{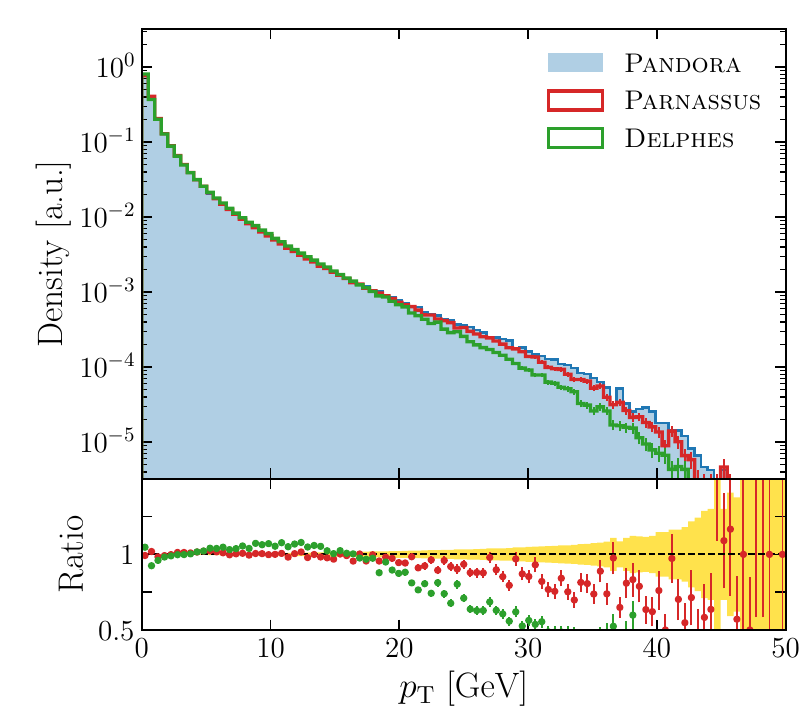}\includegraphics[width=0.245\textwidth]{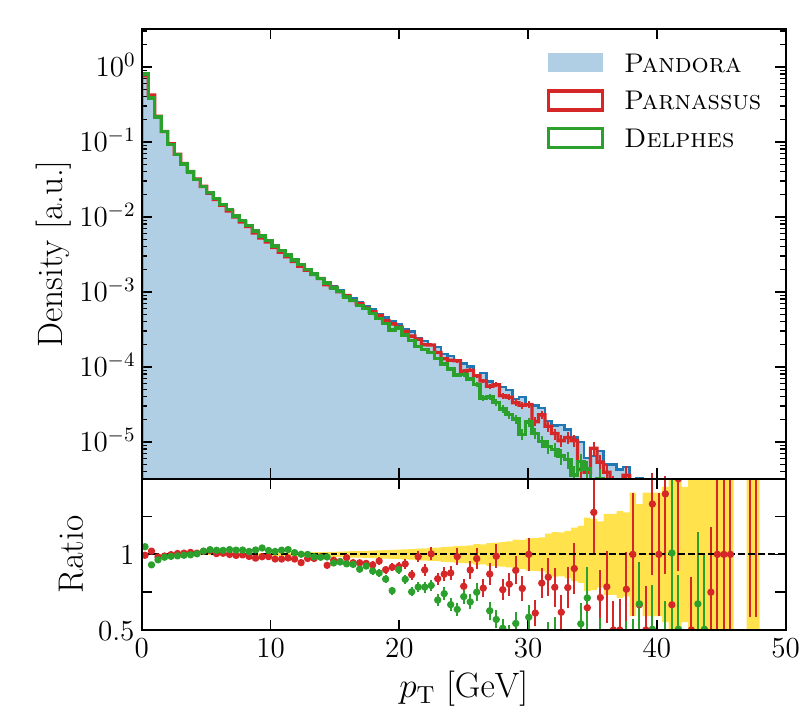}\includegraphics[width=0.245\textwidth]{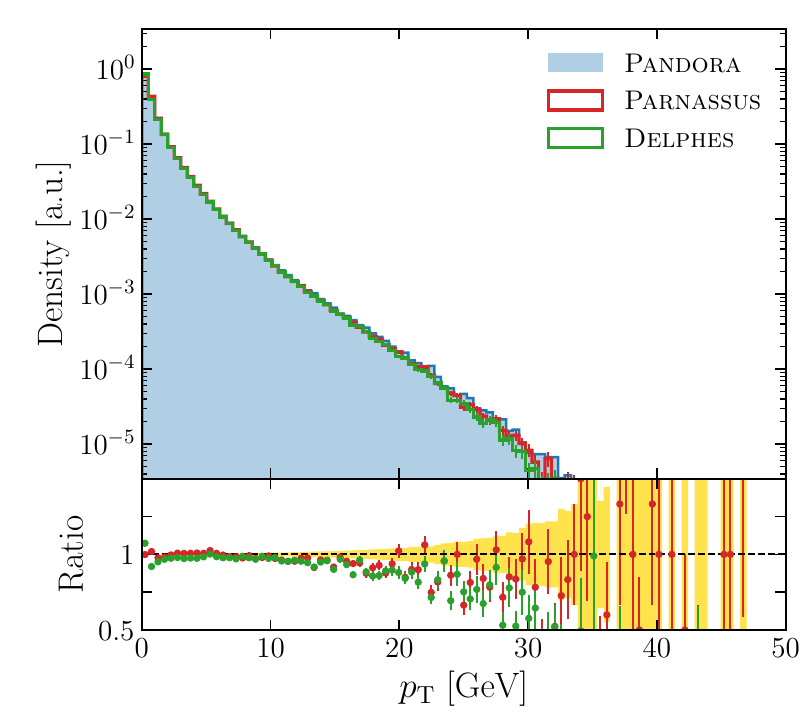}\\
\includegraphics[width=0.245\textwidth]{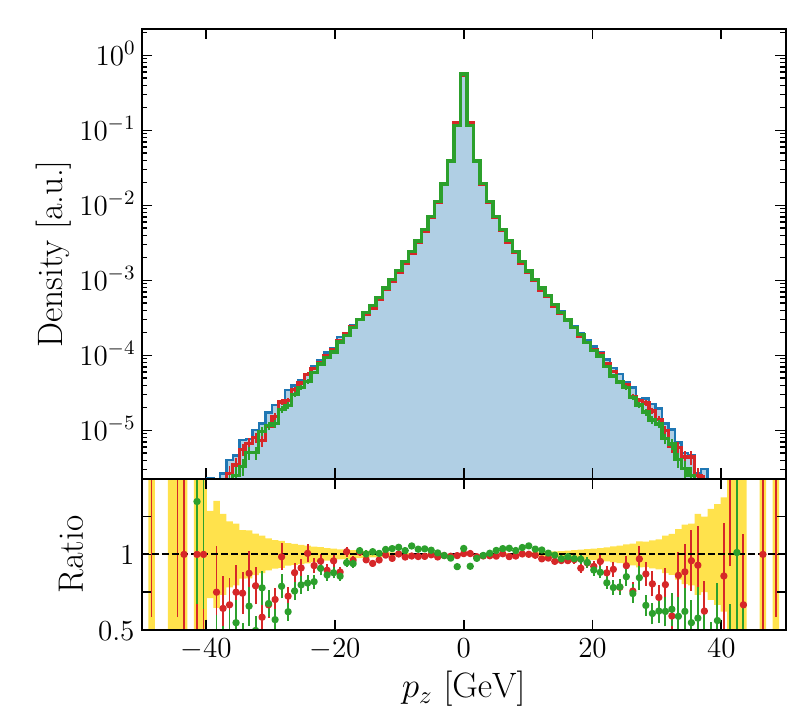}\includegraphics[width=0.245\textwidth]{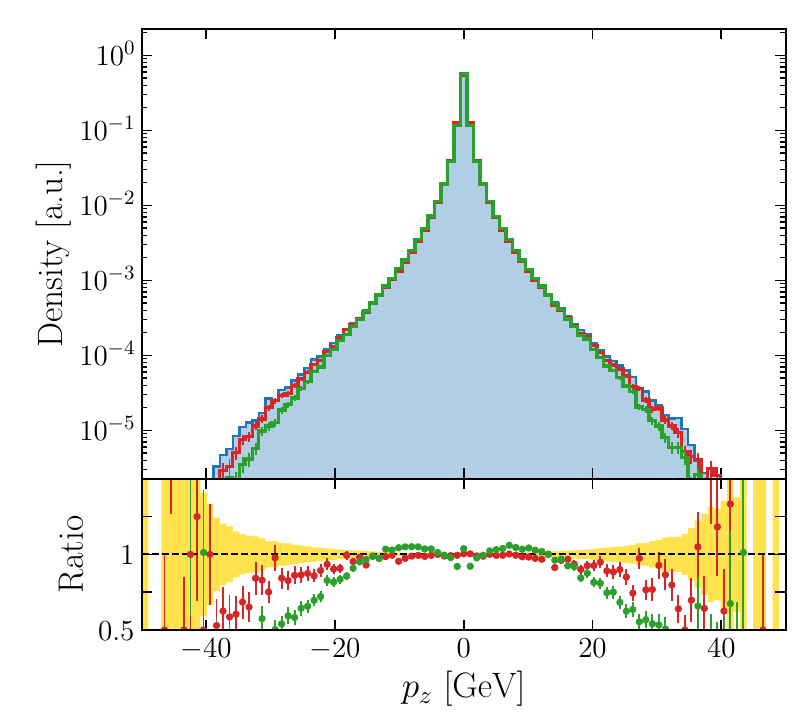}\includegraphics[width=0.245\textwidth]{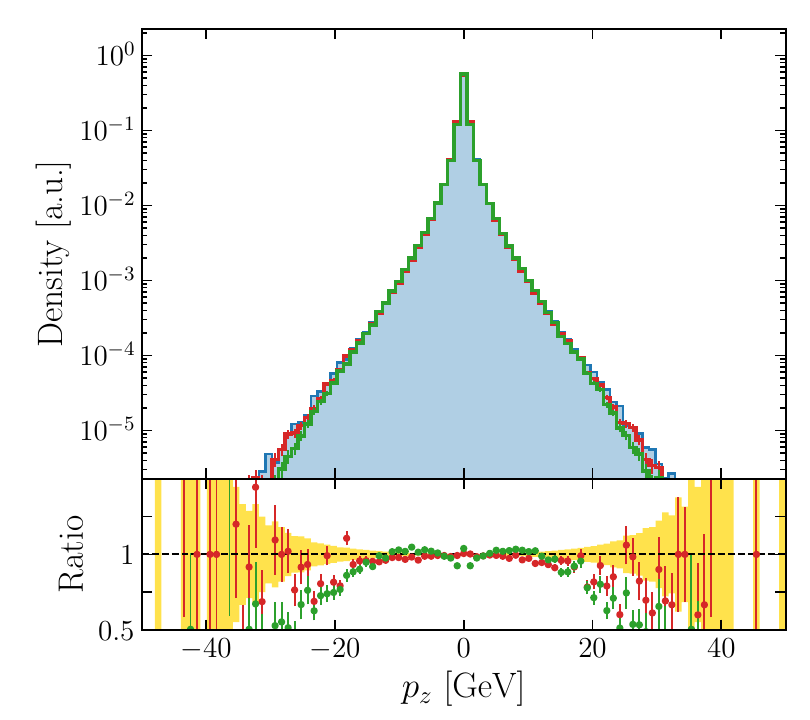}\includegraphics[width=0.245\textwidth]{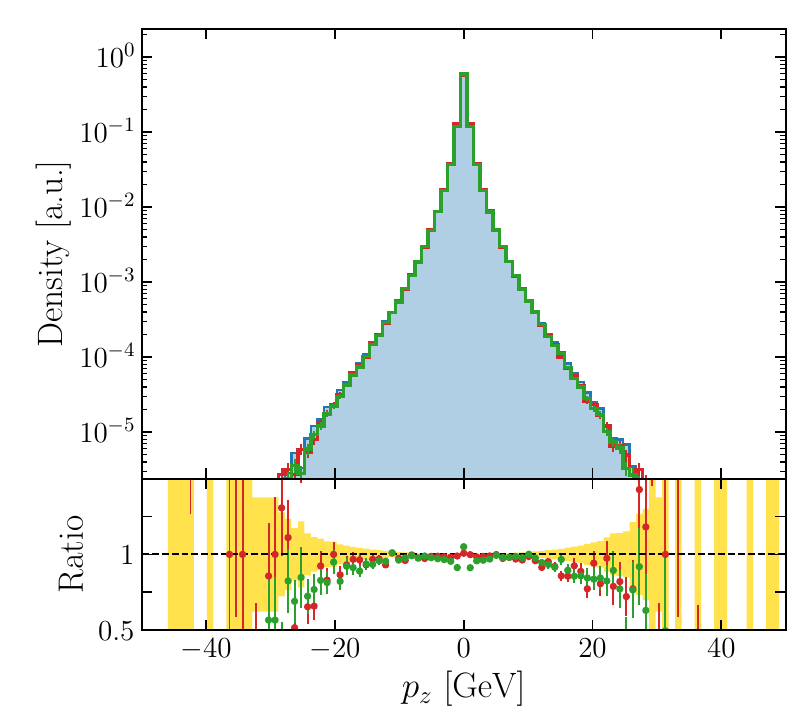}\\
\includegraphics[width=0.245\textwidth]{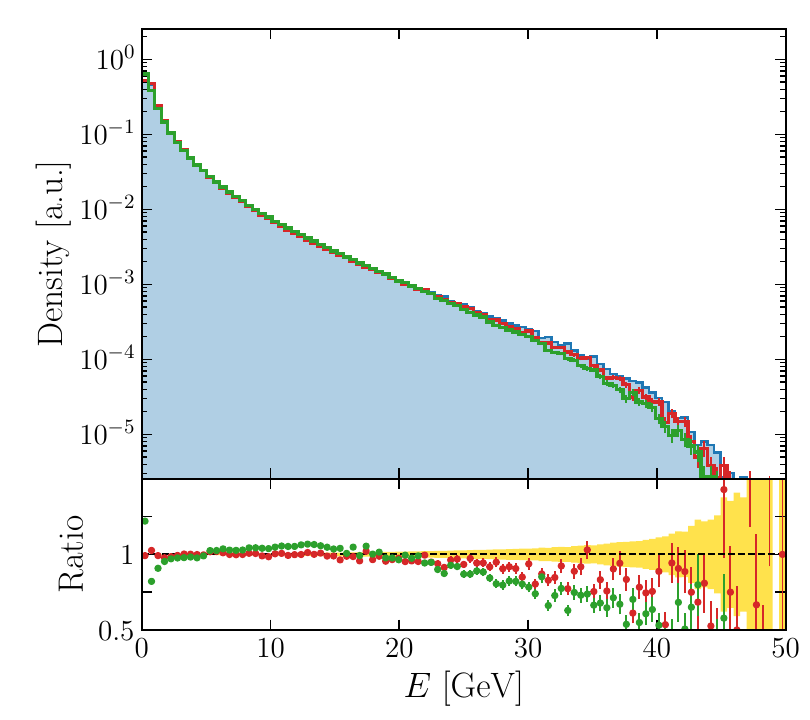}\includegraphics[width=0.245\textwidth]{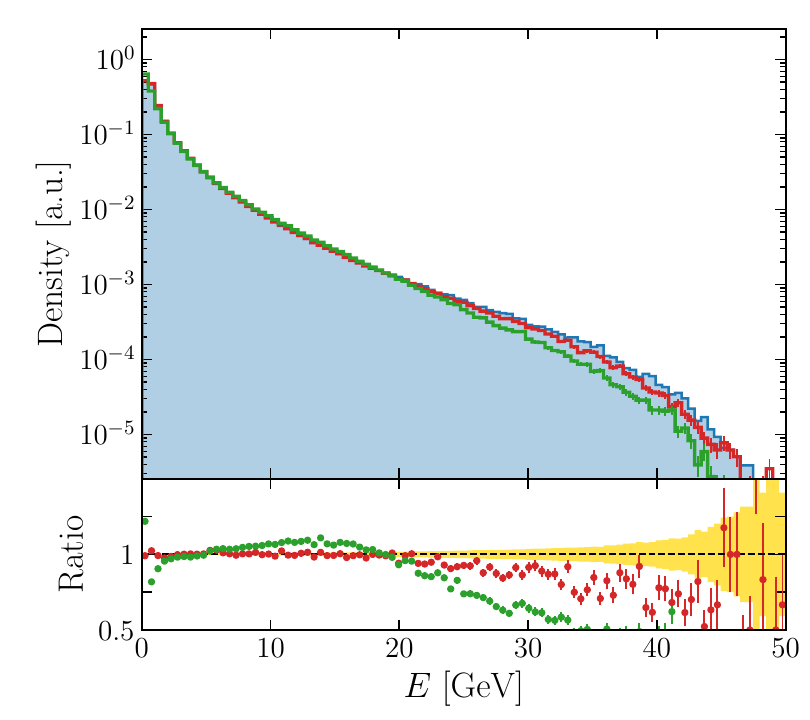}\includegraphics[width=0.245\textwidth]{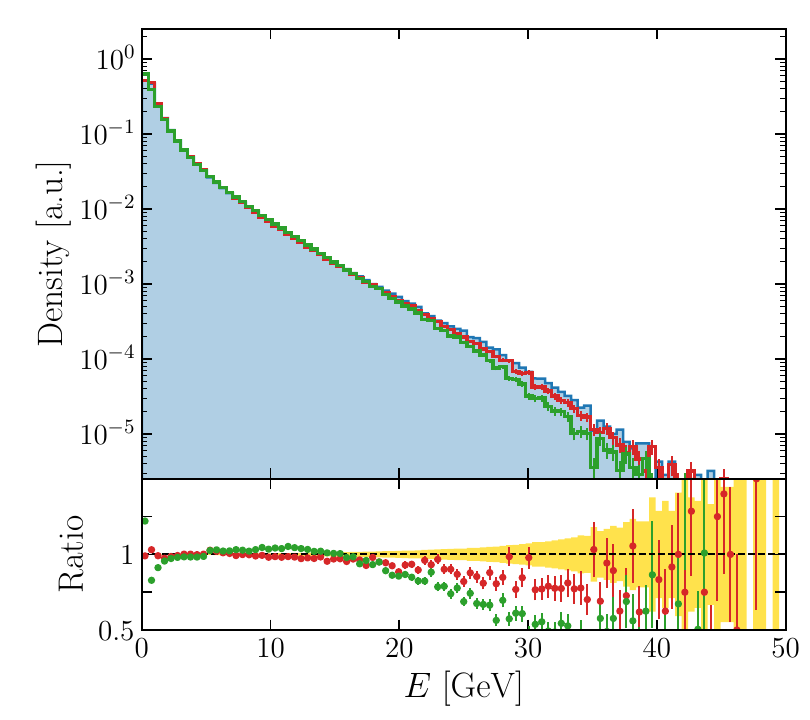}\includegraphics[width=0.245\textwidth]{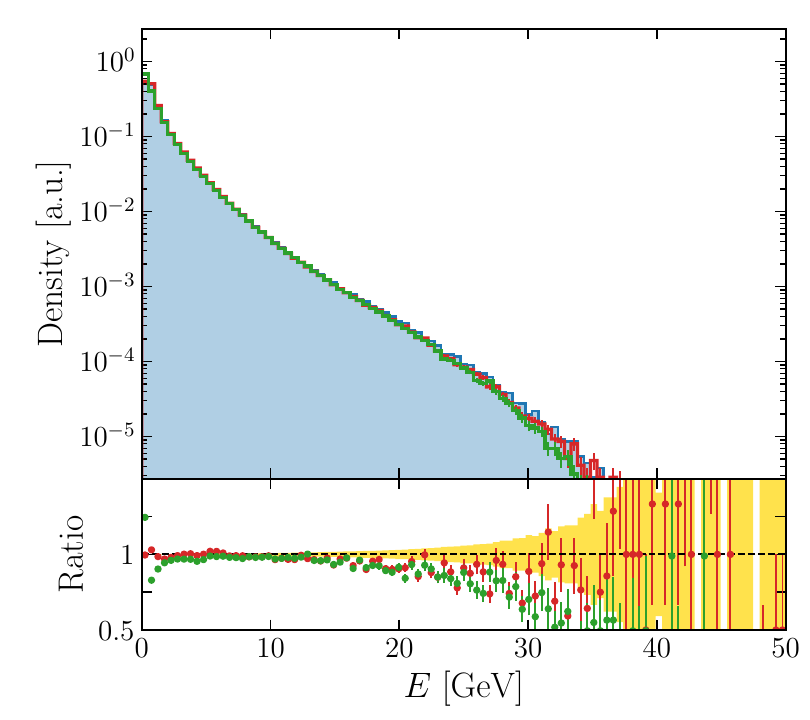}\\
\includegraphics[width=0.245\textwidth]{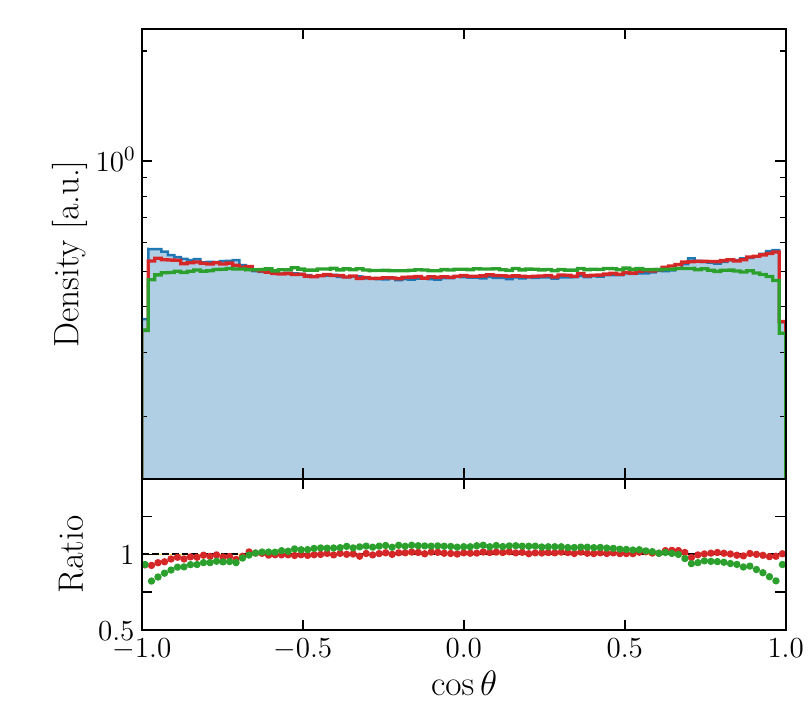}\includegraphics[width=0.245\textwidth]{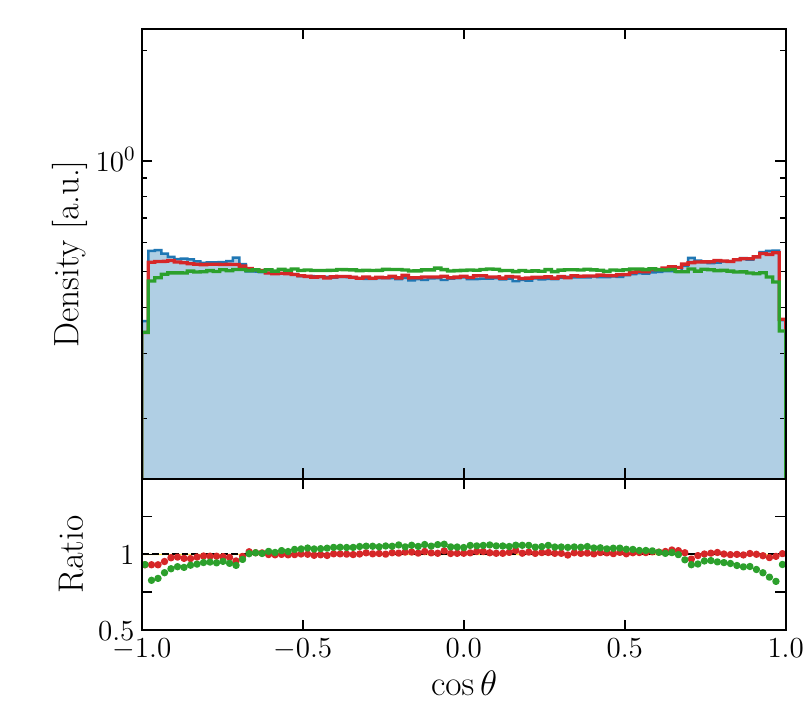}\includegraphics[width=0.245\textwidth]{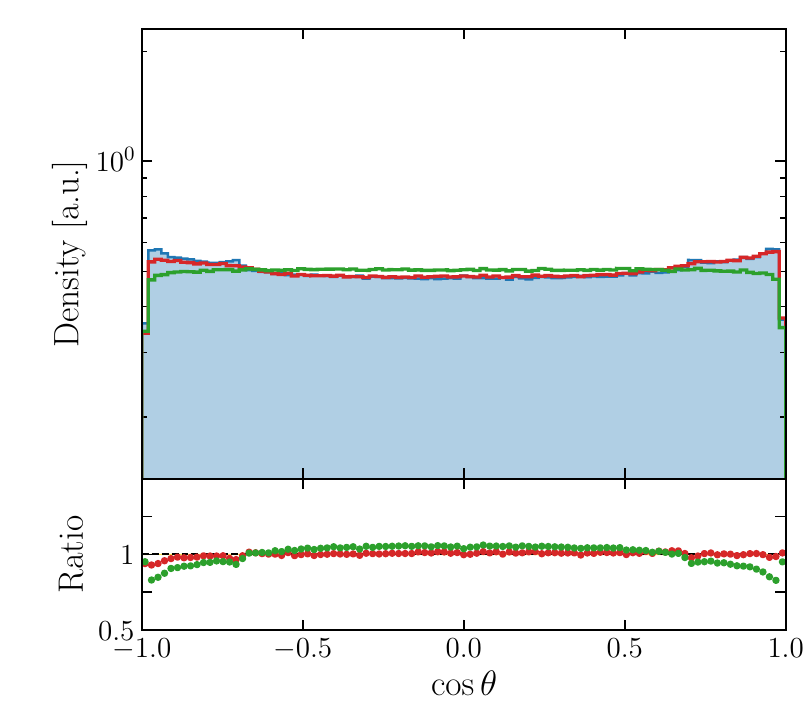}\includegraphics[width=0.245\textwidth]{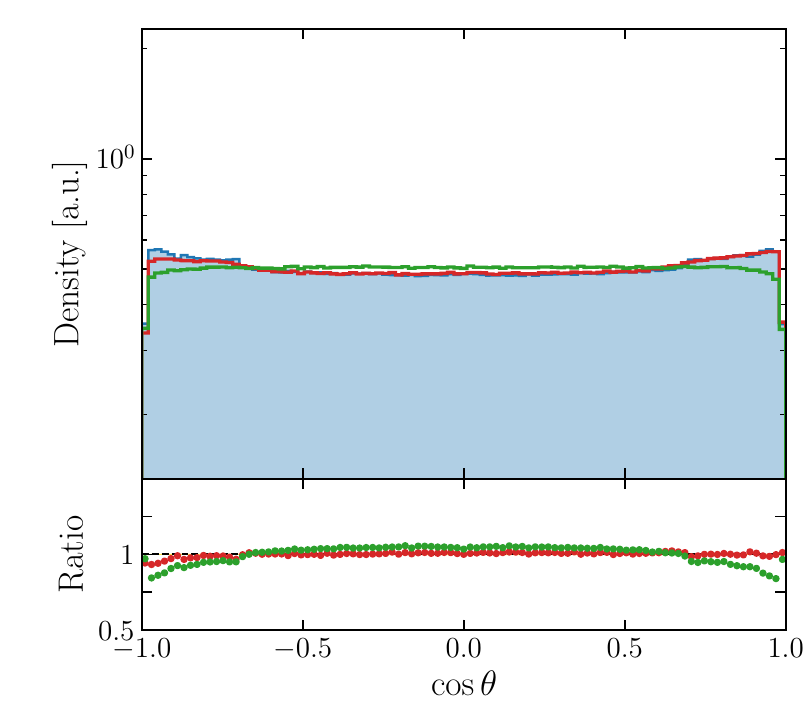}\\
\includegraphics[width=0.245\textwidth]{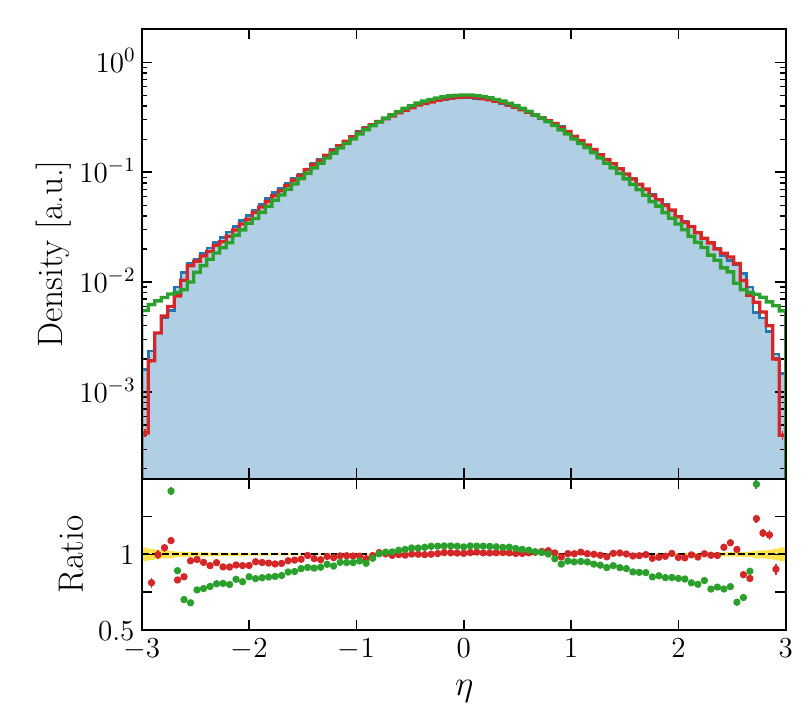}\includegraphics[width=0.245\textwidth]{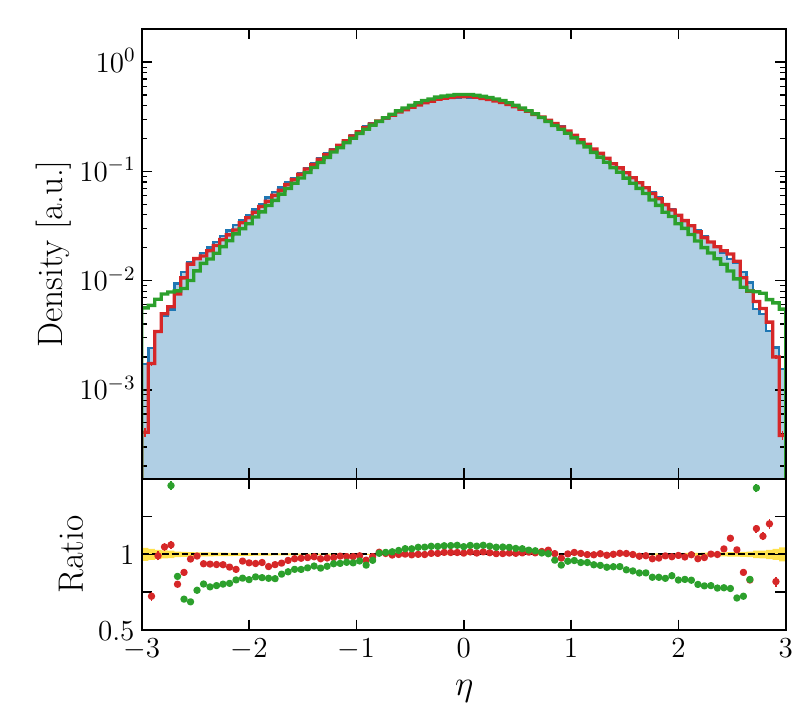}\includegraphics[width=0.245\textwidth]{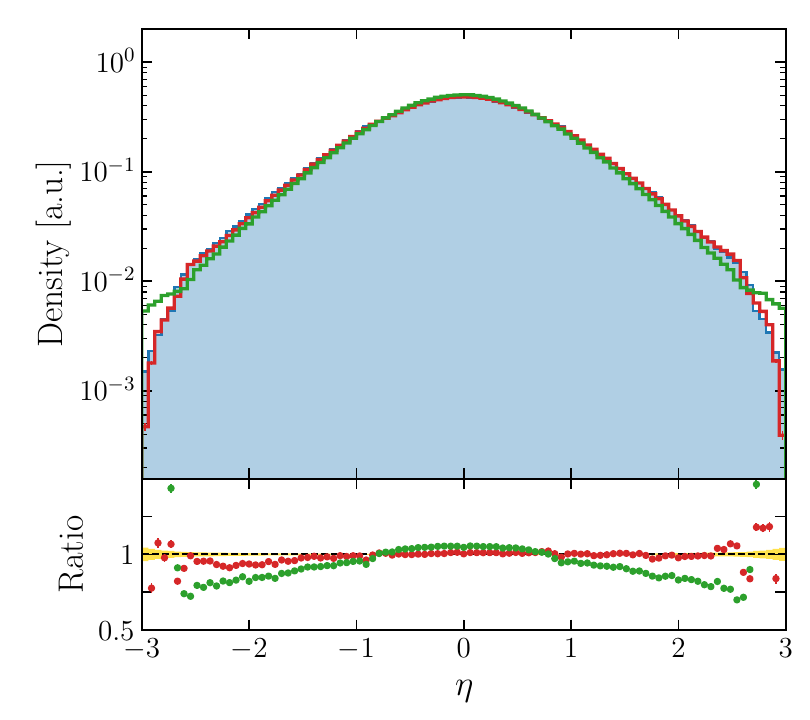}\includegraphics[width=0.245\textwidth]{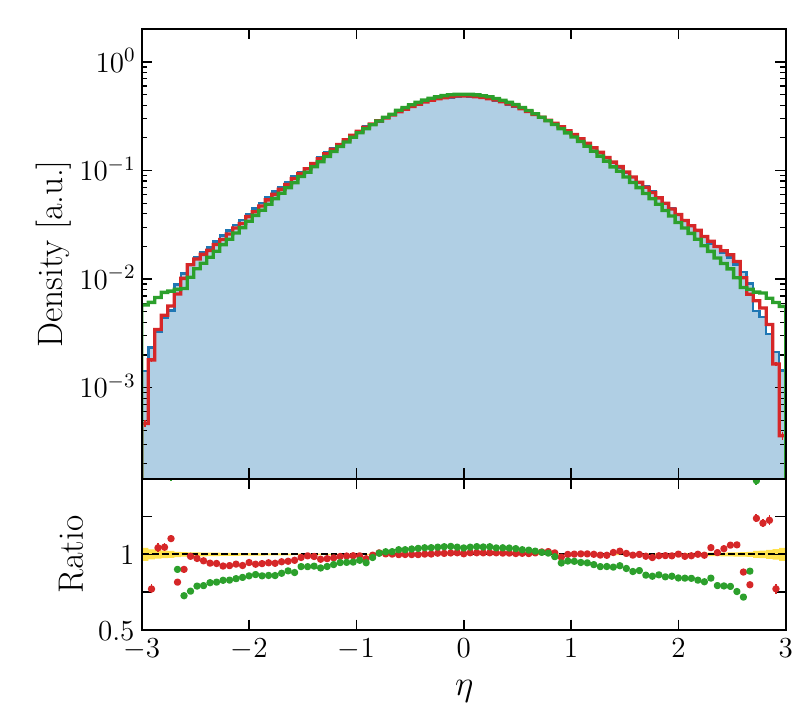}\\
\caption{Particle-level kinematics, split by $Z$ decay flavor (columns: $q\bar q$, $s\bar s$, $c\bar c$, $b\bar b$; rows, top to bottom: $\pT$, $p_z$, $E$, $\cos\theta$, $\eta$, species). {\pandora} (blue), {\parnassus} (red), {\delphes} (green).}
\label{fig:flavor_kin}
\end{figure}

\begin{figure}[p]
\centering
\makebox[0.245\textwidth][c]{$q\bar q$}\makebox[0.245\textwidth][c]{$s\bar s$}\makebox[0.245\textwidth][c]{$c\bar c$}\makebox[0.245\textwidth][c]{$b\bar b$}\\
\includegraphics[width=0.245\textwidth]{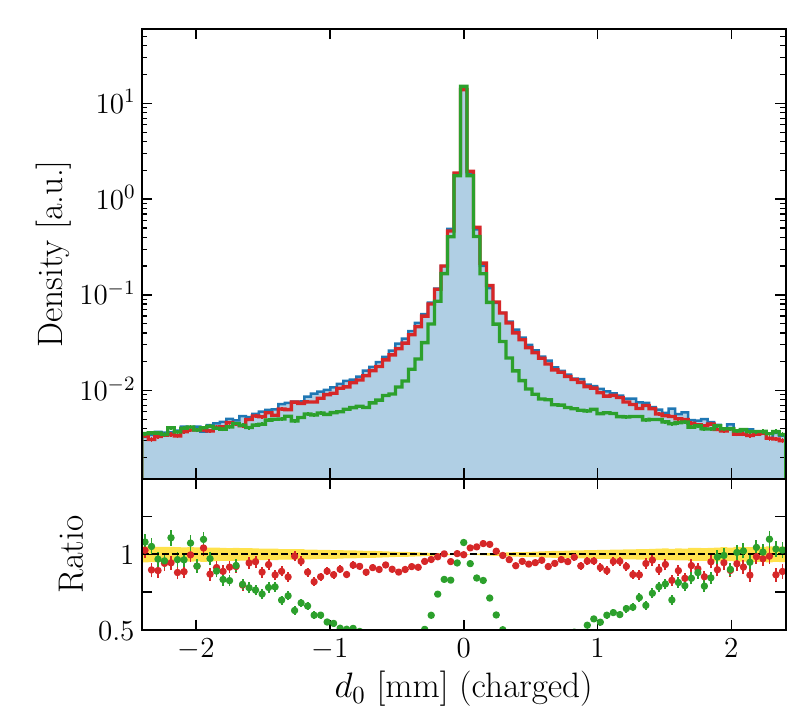}\includegraphics[width=0.245\textwidth]{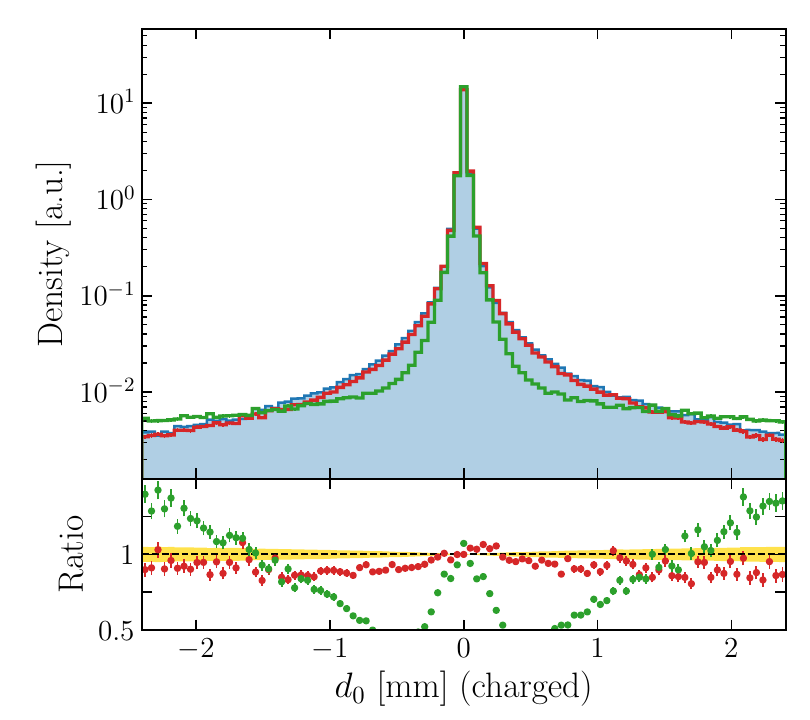}\includegraphics[width=0.245\textwidth]{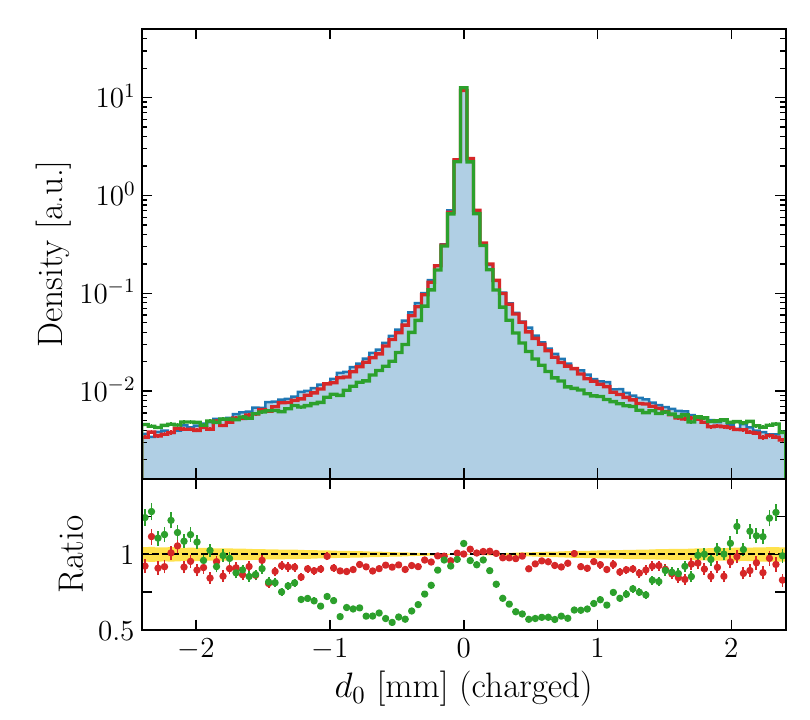}\includegraphics[width=0.245\textwidth]{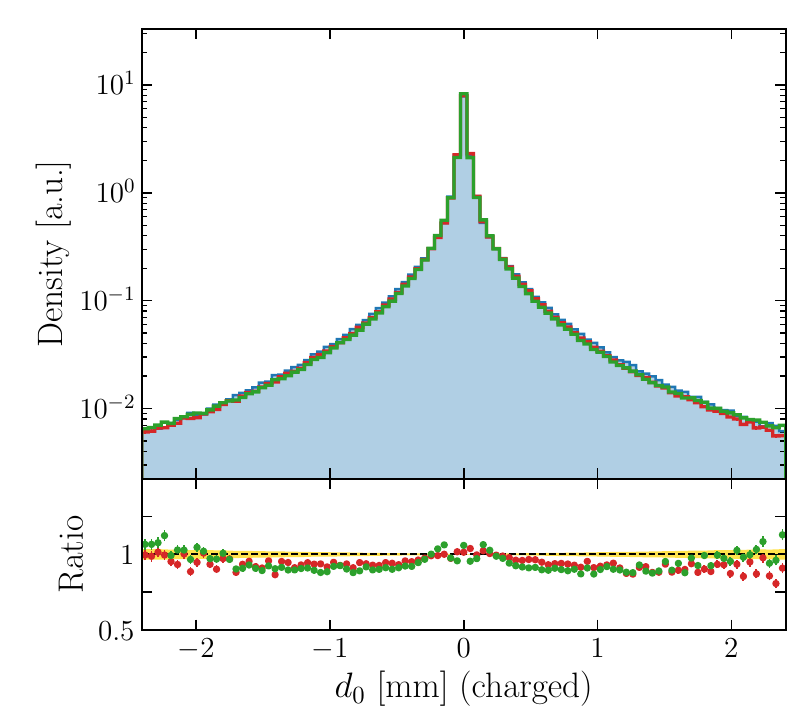}\\
\includegraphics[width=0.245\textwidth]{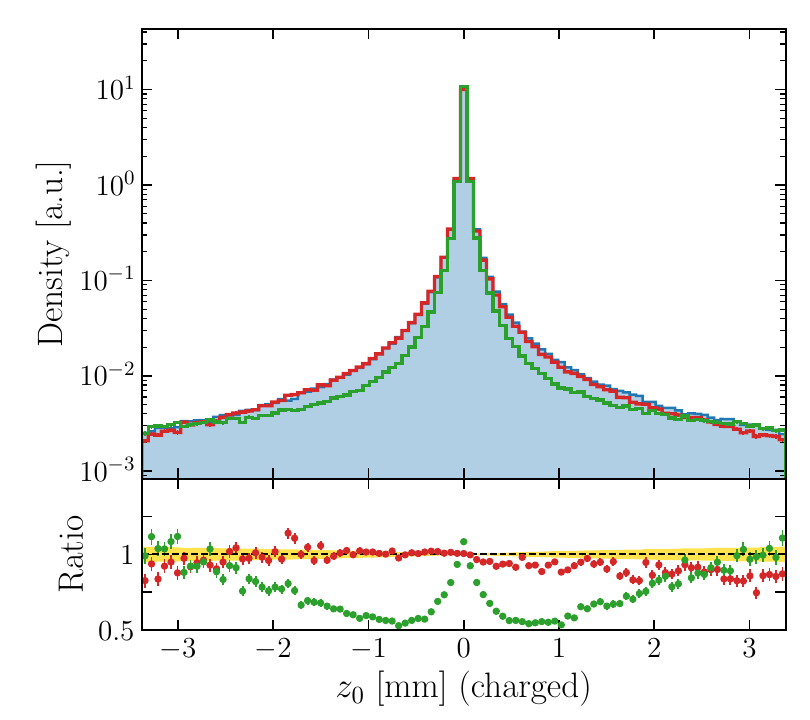}\includegraphics[width=0.245\textwidth]{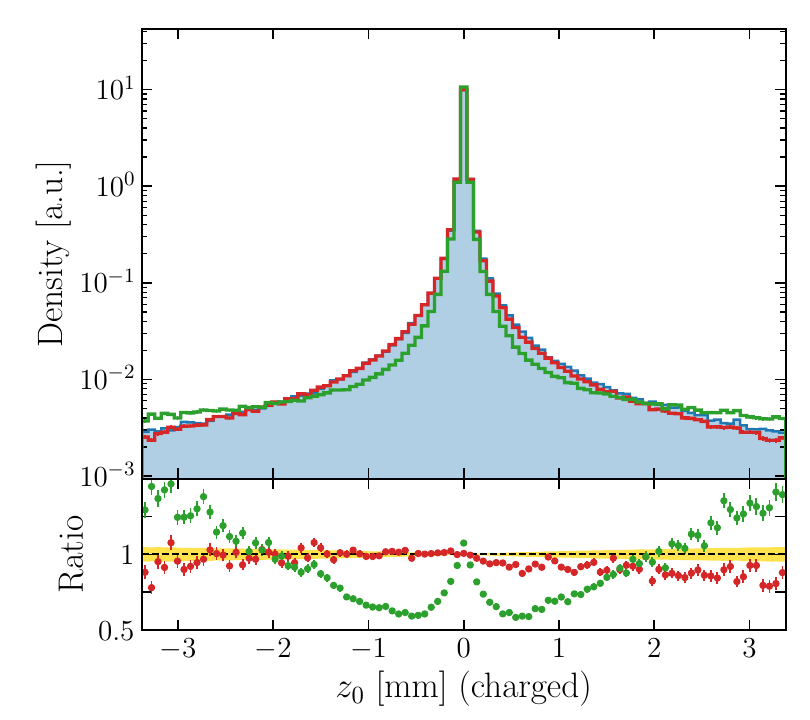}\includegraphics[width=0.245\textwidth]{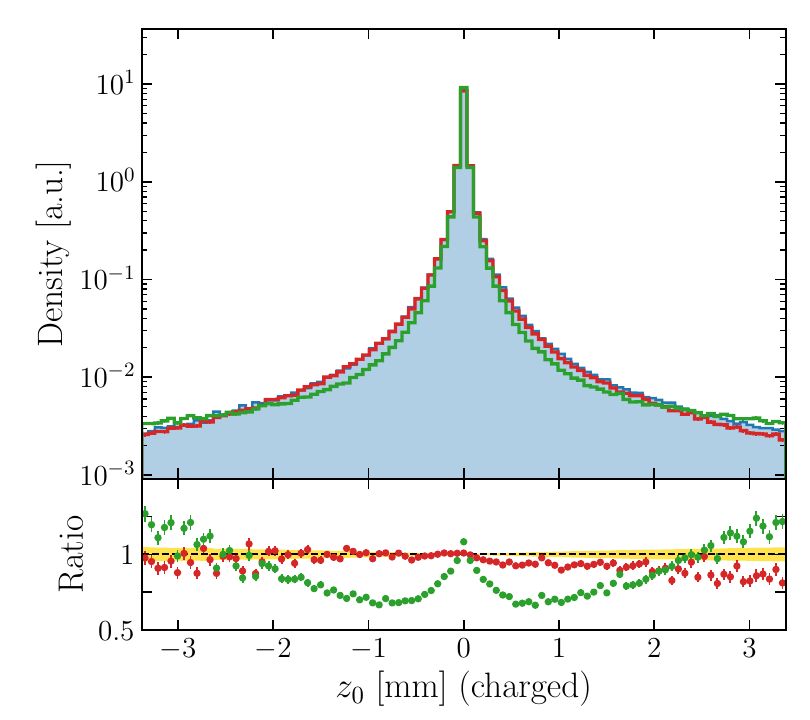}\includegraphics[width=0.245\textwidth]{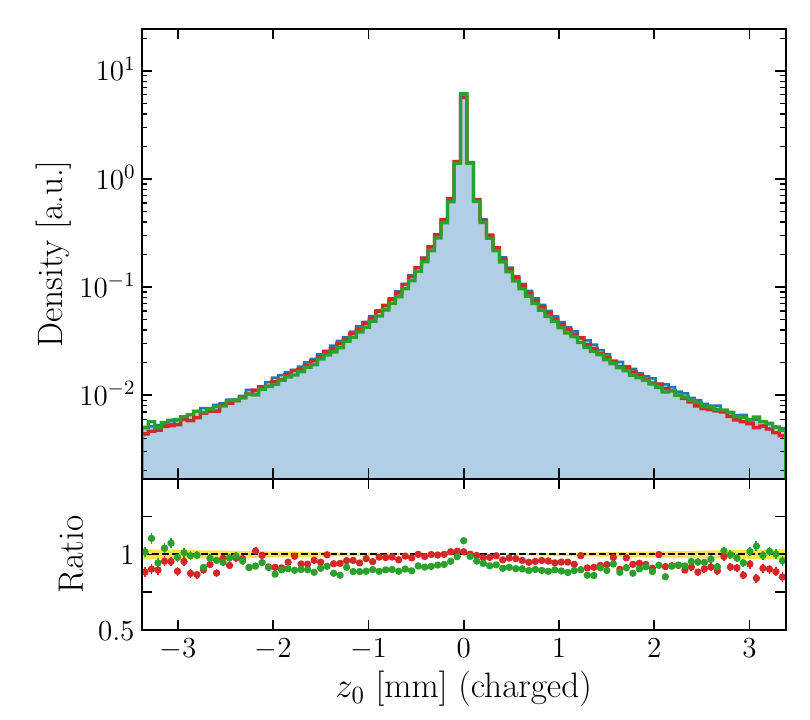}\\
\includegraphics[width=0.245\textwidth]{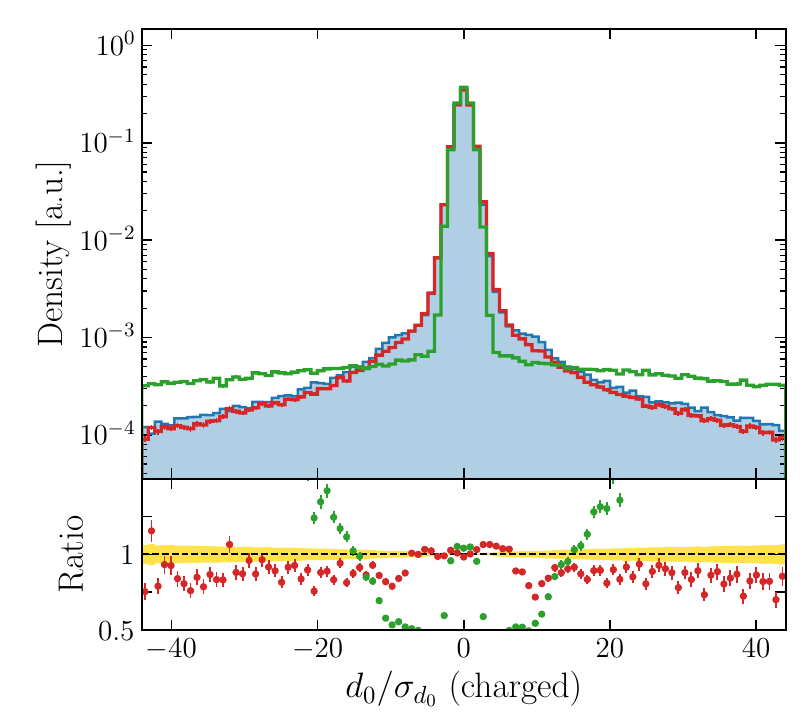}\includegraphics[width=0.245\textwidth]{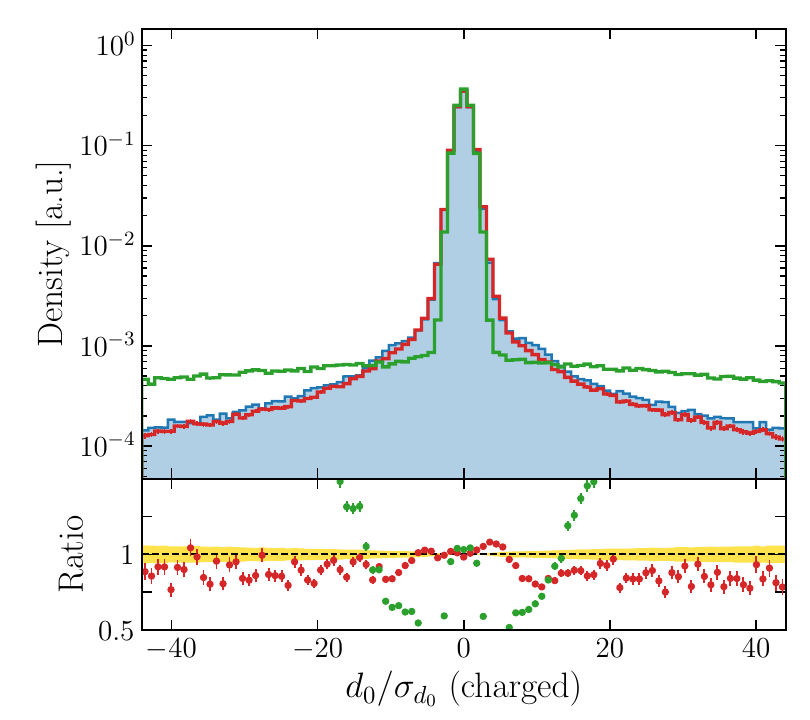}\includegraphics[width=0.245\textwidth]{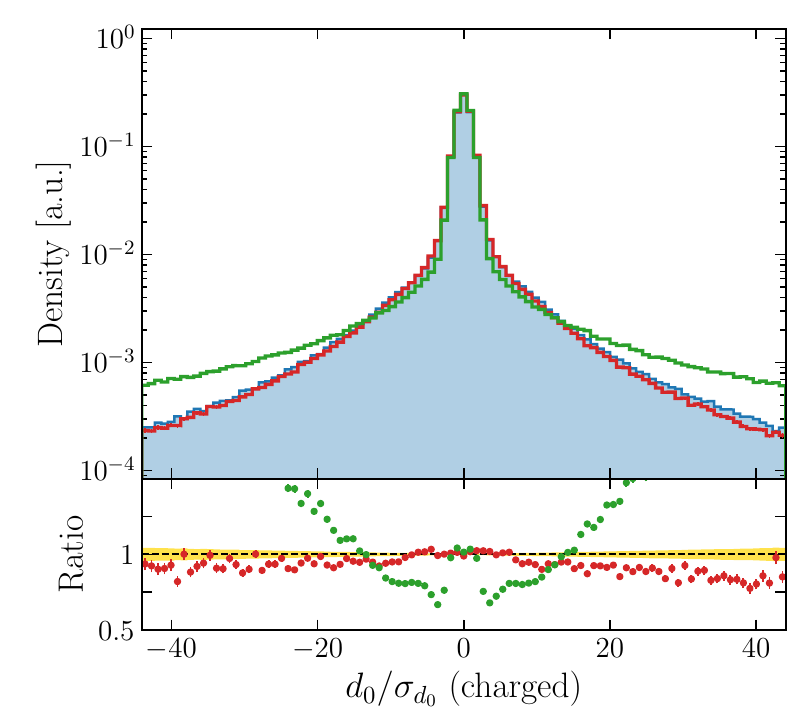}\includegraphics[width=0.245\textwidth]{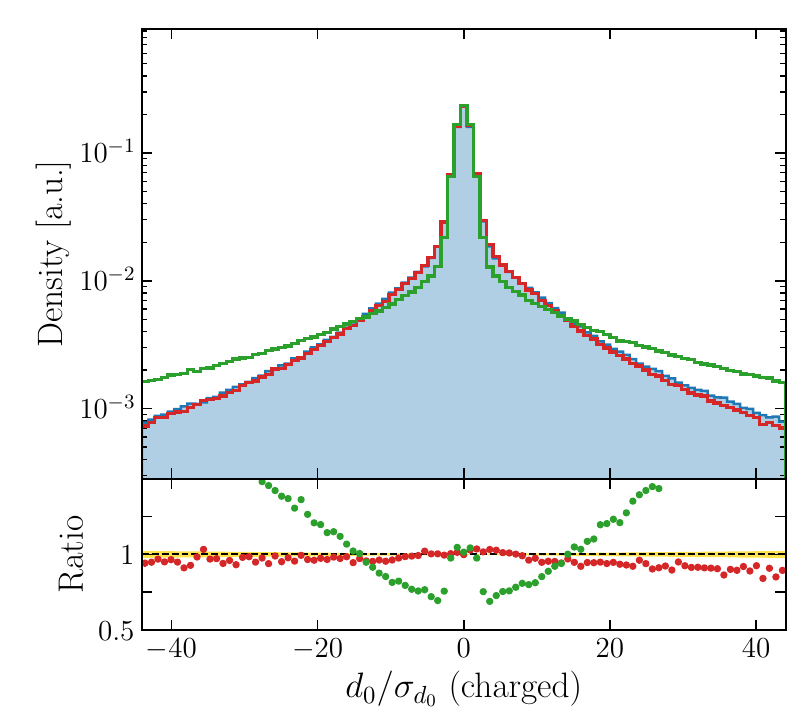}\\
\includegraphics[width=0.245\textwidth]{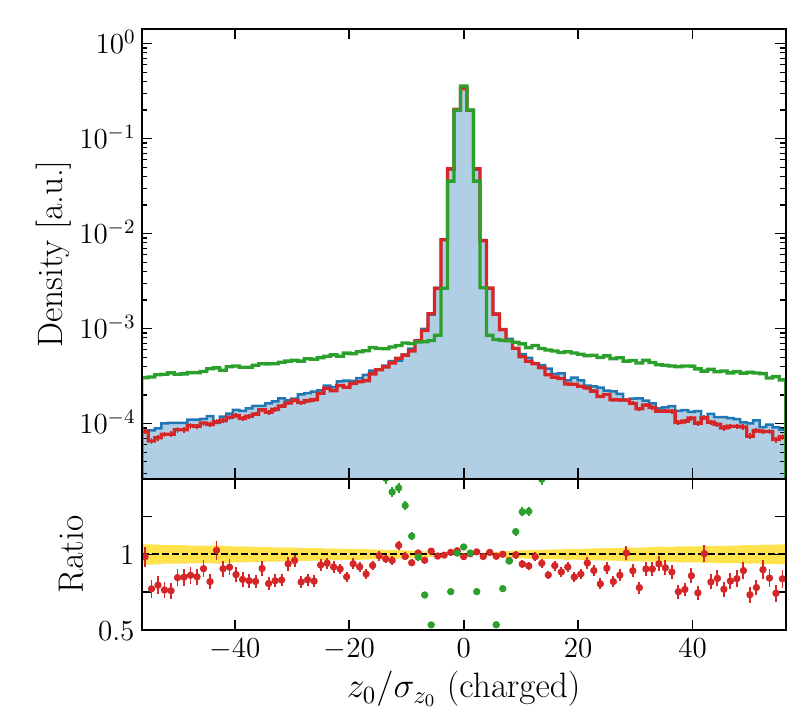}\includegraphics[width=0.245\textwidth]{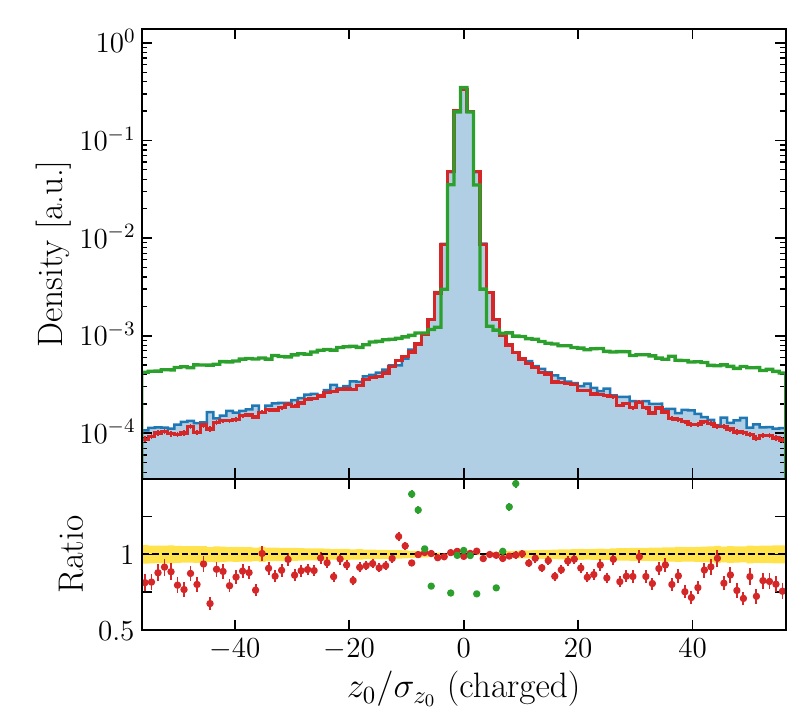}\includegraphics[width=0.245\textwidth]{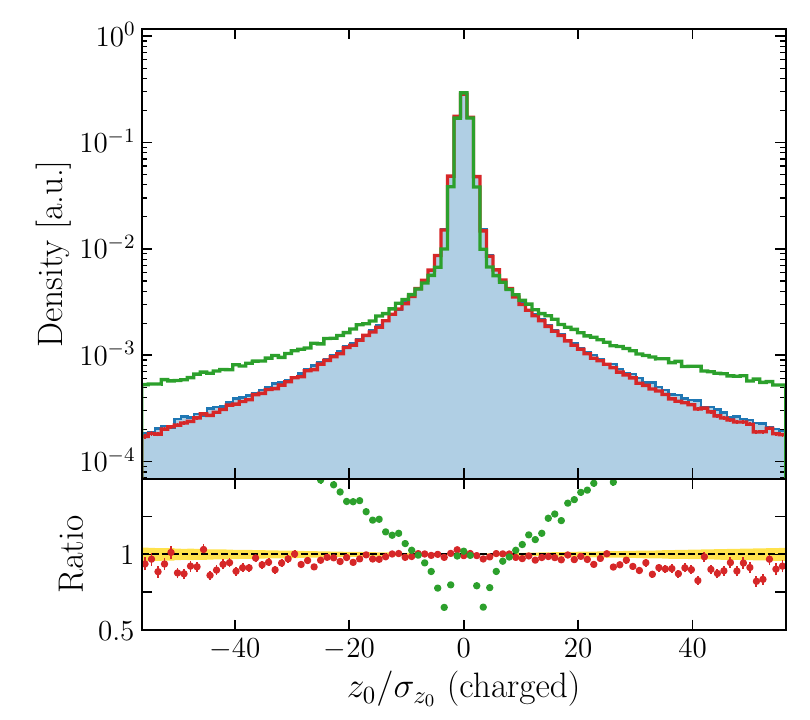}\includegraphics[width=0.245\textwidth]{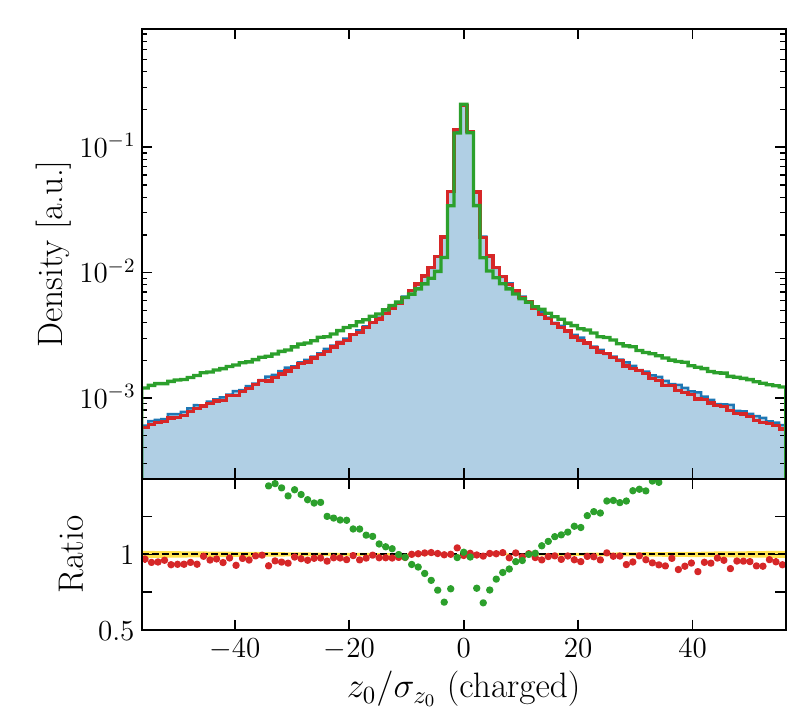}\\
\caption{Particle-level track impact-parameter distributions, split by $Z$ decay flavor (columns: $q\bar q$, $s\bar s$, $c\bar c$, $b\bar b$; rows, top to bottom: $\dzero$, $\zzero$, $\dzero/\sigma_{\dzero}$, $\zzero/\sigma_{\zzero}$). {\pandora} (blue), {\parnassus} (red), {\delphes} (green).}
\label{fig:flavor_ip}
\end{figure}